\documentclass[sigplan,10pt]{acmart}
\usepackage[english]{babel}
\usepackage{blindtext}

\usepackage{filecontents}
\usepackage{adjustbox}
\usepackage{babel}
\usepackage{ifoddpage}
\usepackage{pifont}
\usepackage{siunitx}
\usepackage{multirow}
\usepackage{adjustbox}
\usepackage{tikz}

\newcommand*\circled[1]{\tikz[baseline=(char.base)]{
            \node[shape=circle,draw,inner sep=0pt] (char) {#1};}}
\usepackage{amsmath}
\usepackage{enumitem}
\usepackage[most]{tcolorbox}
\usepackage{multicol,lipsum}
\usepackage[utf8]{inputenc}
\usepackage{algorithmicx}
\usepackage{algorithm}
\usepackage{tabularx,booktabs}
\usepackage{tabularray}
\usepackage[noend]{algpseudocode}
\usepackage{soul}

\usepackage{appendix}
\usepackage{balance}
\usepackage{makecell}
\usepackage[skip=4pt]{caption}
\usepackage{diagbox}
\usepackage{float}
\usepackage{graphicx}
\usepackage{hyperref}
\usepackage{siunitx}
\usepackage{subcaption}
\usepackage{threeparttable}
\usepackage{url}

\usepackage{verbatim} 
\usepackage{xspace}
\usepackage{xcolor}
\usepackage{totpages}
\usepackage{mathtools}
\usepackage{physics}
\usepackage{array}
\usepackage{pdfpages}
\usepackage{adjustbox}
\usepackage{float}
\usepackage{environ}
\usepackage[normalem]{ulem}
\usepackage{hyperref}

\usepackage{titlesec}
            
\usepackage{tcolorbox}
\definecolor{mycolor}{rgb}{0.122, 0.435, 0.698}
\usepackage{xparse}

\def\noeditingmarks{}

\ifx\noeditingmarks\undefined
  
\else
  
\fi

\NewEnviron{commentbox}{%
  \vspace{0.2cm}
  \begin{tcolorbox}[
    width=.48\textwidth,
    colframe=mycolor,
    coltext=mycolor,
    colback=white!10!white,
    boxrule=0.5pt,
    arc=4pt,
    boxsep=0pt,
    left=6pt, right=6pt,
    top=6pt, bottom=6pt
  ]
  \BODY
  \end{tcolorbox}
}

\newcommand{\comments}[1]{}

\newenvironment{parafont}{\fontfamily{ptm}\selectfont}{}
\newcommand{\sysname}[0]{Tiga\xspace}

\newcommand{\Para}[1]{\vspace{2pt}\noindent\begin{parafont}\textbf{\textit{#1}}\end{parafont}}

\newcommand{\ag}[1]{}

\usepackage{caption}
\renewcommand\footnotetextcopyrightpermission[1]{} % removes footnote with conference info
\algdef{SE}[EVENT]{Event}{EndEvent}[1]{\textbf{upon}\ \textsc{\small #1}\ \algorithmicdo}{\algorithmicend\ \textbf{event}}%
\algtext*{EndEvent}
\algdef{SE}[DOWHILE]{Do}{doWhile}{\algorithmicdo}[1]{\algorithmicwhile\ #1}%

\makeatletter
\NewDocumentCommand{\LeftComment}{s m}{%
  \Statex \IfBooleanF{#1}{\hspace*{\ALG@thistlm}}\(\triangleright\) #2}
\makeatother

\begin{document}
\title[Tiga]{\LARGE \bf  Tiga: Accelerating Geo-Distributed Transactions with Synchronized Clocks [Technical Report] }

% \title{\sysname: Accelerating Geo-Distributed Transactions with Synchronized Clocks [Technical Report] }

% \title{\LARGE \bf \sysname: Accelerating Geo-Distributed Transactions with Synchronized Clocks }

\makeatletter
\renewcommand{\@authornotemark}{}
\makeatother

\author{{Jinkun Geng{$^\star$*}, Shuai Mu{$^\star$}, Anirudh Sivaraman{$^\dag$}, Balaji Prabhakar{$^\ddagger$}}\\
\textit{{$^\star$}Stony Brook University}, \textit{{$^\dag$}New York University}, \textit{{$^\ddagger$}Stanford University}
}
\authornote{Part of the work was done at Stanford University.}

% \acmYear{2025}\copyrightyear{2025}
% \acmConference[SOSP '25]{ACM SIGOPS 31st Symposium on Operating Systems Principles}{October 13--16, 2025}{Seoul, Republic of Korea}
% \acmBooktitle{ACM SIGOPS 31st Symposium on Operating Systems Principles (SOSP '25), October 13--16, 2025, Seoul, Republic of Korea}
% \acmDOI{10.1145/3731569.3764854}
% \acmISBN{979-8-4007-1870-0/25/10}

\sloppy

\begin{abstract}
This paper presents \sysname, a new design for geo-replicated and scalable transactional databases such as Google Spanner. 
\sysname aims to commit transactions within 1 wide-area roundtrip time, or 1 WRTT, for a wide range of scenarios, while maintaining high throughput with minimal computational overhead. \sysname consolidates concurrency control and consensus, completing both strictly serializable execution and consistent replication in a single round. It uses synchronized clocks to proactively order transactions by assigning each a \emph{future} timestamp at submission. 
In most cases, transactions arrive at servers before their future timestamps and are serialized according to the designated timestamp, requiring 1 WRTT to commit. 
In rare cases, transactions are delayed and proactive ordering fails, in which case \sysname falls back to a slow path, committing in 1.5--2 WRTTs. Compared to state-of-the-art solutions, \sysname can commit more transactions at 1-WRTT latency, and incurs much less throughput overhead. 
Evaluation results show that 
\sysname outperforms all baselines, achieving 1.3--7.2$\times$ higher throughput and 1.4--4.6$\times$ lower latency. \sysname is open-sourced at \href{https://github.com/New-Consensus-Concurrency-Control/Tiga}{https://github.com/New-Consensus-Concurrency-Control/Tiga}.
\end{abstract}

\maketitle

% \titlespacing\section{0pt}{2pt}{1pt}
% \titlespacing\subsection{0pt}{4pt}{2pt}

\section{Introduction}
\label{sec:introduction}

Distributed online transaction processing (OLTP) systems~\cite{spanner, fast20-cad, zippydb, sigmod12-calvin,sigmod20-cockroachdb,vldb19-ocean-vista,vldb20-tidb,cosmosdb,sigmod15-helios,sigmod18-carousel,sigmod22-natto, sigmod21-foundationdb,vldb08-hstore, osdi14-silo, sigmod23-detock,vldb23-caerus} are fundamental to cloud infrastructures and online services. 
These systems partition data to scale and replicate data across different datacenters to tolerate server and datacenter outages. 
Data accesses are guaranteed to be strongly consistent for easy usage. Replication is \emph{linearizable}, and operations are wrapped in transactions with \emph{strict serializability}---together providing users with the illusion of having a single-copy, single-threaded storage with unlimited capacity. 
To provide this fault-tolerant transaction guarantee, the system uses a concurrency control protocol (e.g., two-phase locking/commit~\cite{acm76-2pl}) to isolate transactions from each other, and a consensus protocol (e.g., Multi-Paxos~\cite{paxos}) to replicate data.

Both concurrency control and consensus protocols are inherently complex and impose significant performance overhead. The overhead includes extra computation (e.g., locking) and additional message roundtrips on the critical path to commit transactions. 
These additional message roundtrips are especially costly in geo-replicated settings. 
It may require multiple wide-area roundtrip times (WRTT) to commit a transaction, which significantly impacts latency. 
Prior work has attempted to reduce the latency overhead, but they often require techniques with substantial computational overhead (e.g., dependency tracking and cycle detection~\cite{osdi16-janus,sigmod23-detock}), thus reducing throughput. 
Additionally, the optimal latency is typically achievable only for a subset of cases, such as when transactions are commutative.

We ask this question in the paper: \emph{Can we design a fault-tolerant transaction protocol that commits more transactions in 1 WRTT with less overhead?} Our answer is \sysname, a lightweight and low-latency protocol based on synchronized clocks. 
For a wide range of workloads and deployment settings (e.g., server co-location), \sysname can commit transactions in 1 WRTT. \sysname uses an efficient timestamp ordering approach to achieve strong consistency, which yields 2.0--4.5$\times$ higher throughput than the other protocols (e.g., Janus/Detock) that rely on intensive graph algorithms for the same consistency guarantee. \sysname achieves this goal through three key design decisions:

\Para{Consolidating consensus and concurrency control.} 
Both concurrency control and consensus protocols aim to establish a consistent order across servers, with one handling replicas and the other managing shards. 
When a system stacks two protocols together, it essentially overpays for achieving the same goal twice. 
To achieve 1-WRTT latency, as pointed out by previous works~\cite{osdi16-janus,sosp15-tapir}, it is necessary to consolidate these two layers of protocols. 
Given this insight, \sysname is designed as a consolidated protocol that unifies concurrency control and consensus.

\Para{Proactive ordering with synchronized clocks.} 
\sysname uses timestamps to order transactions, which is a classic lightweight approach in concurrency control~\cite{tods79-timestamp-ordering}. 
Using synchronized clocks, \sysname measures the one-way delay (OWD) from the transaction's sender (i.e., coordinator) to every participating server and assigns the transaction a future timestamp at submission. The transaction is expected to arrive at all participating servers before the future timestamp. This timestamp includes a \emph{headroom}—an estimate of transmission delay to reach every participating server, which is calculated based on the measured OWDs (\S\ref{sec:timestamp-initilization}).
The headroom effectively masks the heterogeneous latency from the sender to each server: Even if the transaction arrives early at some servers before its timestamp, the servers will hold the transactions until their local clocks pass the timestamp, and then process the transactions based on their timestamp order. 
This ordering approach reduces the occurrence of inconsistent arrival orders at different servers, making \sysname's fast path more stable and allowing more transactions to be committed with low latency. \sysname leverages the significant improvements in clock synchronization accuracy over recent decades. Today's clock synchronization algorithms (e.g., Huygens~\cite{huygens}) can achieve microsecond accuracy across data centers and scale robustly~\cite{clockwork-accurate-owd,clockwork-clock-cmp}. 
This enables \sysname to extract more performance, as the clock inaccuracy (a few microseconds) is often negligible compared to the headroom (10s of milliseconds).

\Para{Minimizing server coordination overhead.} \sysname's proactive ordering is best-effort: Transactions can still arrive later than the predicted timestamps---e.g., due to packet drop and retransmission---thereby violating the consistency requirements. Such violations can be subtle and may go undetected by clients or coordinators (Figure~\protect\ref{fig:timestamp-inversion-undetect}), unless they communicate with \emph{all} shards during every transaction commit, which is undesirable in practice. To guarantee correctness, \sysname carefully designates one leader per shard, and coordinates these leaders to agree on a timestamp for every transaction. (1) In common deployments with full replication~\cite{sigmod15-helios,osdi16-janus,sosp11-cops,aws-shard,tikv-multi-region-deployment,azure-cosmos-partition,eurosys21-dast,vldb23-caerus,vldb08-pnuts}, the leaders can be co-located in one geographic region and incur negligible LAN overhead. This setup enables a 1-WRTT fast path to commit most transactions. (2) In more general deployments with partial replication, leaders cannot be co-located in a single geographic region and their coordination introduces WAN overhead. \sysname still allows commutative transactions to be committed via the 1-WRTT fast path. The extra WAN overhead arises only when multiple conflicting transactions are submitted concurrently: the later transactions will be blocked until the leaders reach timestamp agreement for the earlier ones. If the blocking occurs, it typically costs 0.5 WRTT, leading to a graceful performance degradation compared with the full-replication setup.

\Para{Correctness challenge.} Building a consolidated protocol like \sysname is non-trivial, as both consensus and concurrency control protocols are inherently complex. 
A major challenge is that it is error-prone. 
While timestamp-based consensus protocols can achieve linearizability, timestamp-based concurrency control protocols usually only achieve serializability, not \emph{strict} serializability---i.e., they drop the external consistency guarantee provided by linearizability. 
This is summarized as ``timestamp inversion'' in recent work~\cite{osdi23-ncc}. 
We carefully designed \sysname to avoid timestamp inversion. 
We have formally proved the correctness of \sysname in Appendix~\ref{appendix:correctness-proof} and included the TLA+ specification in~\cite{tiga-tla}.

\Para{Evaluation.} We implement \sysname as a complete protocol that includes both normal processing and failure recovery. We compare \sysname to layered protocols (2PL/OCC+Paxos, NCC, Calvin+ and Detock) and consolidated protocols (Tapir and Janus) in Google Cloud, using a micro benchmark (MicroBench) and an industry-standard benchmark (TPC-C). We find:

(1) Under MicroBench, when contention is low (skew factor=0.5), \sysname outperforms all baselines by 1.3--7.2$\times$ in throughput and by 1.4--4.6$\times$ in median latency at close to their respective saturation throughputs. As we fix the load and increase the skew factor, all baselines degrade except Calvin+, but Calvin+ incurs 33\% higher latency than \sysname.

(2) Under TPC-C, which contains both one-shot and multi-shot (interactive) transactions, 2PL/OCC+Paxos, Tapir and NCC yield very low throughput (1K--2K txns/s). \sysname outperforms the remaining baselines by 1.6-3.5$\times$ in throughput and 1.5-3.7$\times$ in median latency.

(3) \sysname's performance varies when using different clocks. \sysname can achieve high performance by using physical clocks if their synchronization error is negligible compared to the cross-region message delay (tens to hundreds of \SI{}{\milli\second}). Based on our evaluation, off-the-shelf clock synchronization services (e.g., \texttt{chrony} in Google Cloud) can already satisfy this requirement (with $<\SI{5}{\milli\second}$ synchronization error).

\section{Background and Intuition}
\label{sec:background}

\begin{figure*}[!t]
    \centering
\begin{minipage}{0.43\textwidth}
    \centering
    \includegraphics[width=0.9\linewidth]{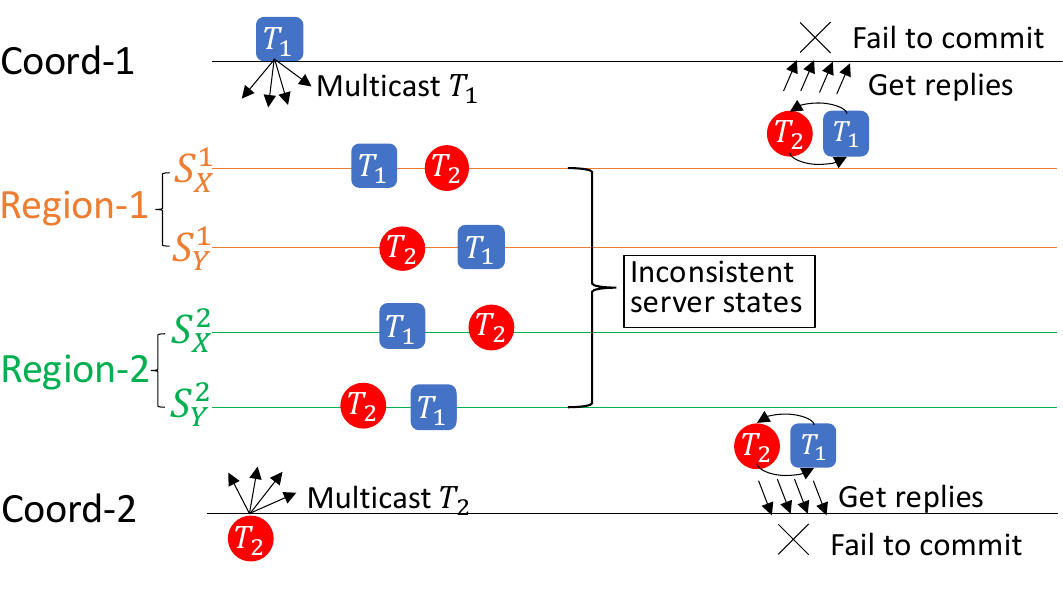}
    \caption{Tapir fails to commit transactions in the fast path because $T_1$ and $T_2$ arrive at servers in different orders, causing inconsistent server states.}
    \label{fig:janus-reorder}
\end{minipage}\hspace{1.5cm}
\begin{minipage}{0.43\textwidth}
    \centering
    \includegraphics[width=0.9\linewidth]{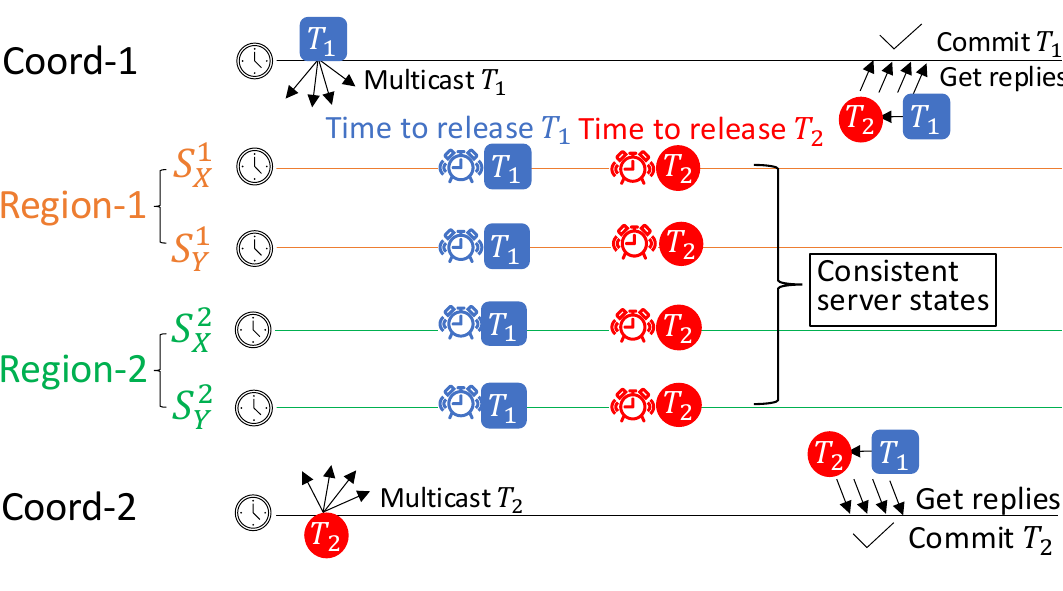}
    \caption{\sysname rectifies inconsistent arrival order based on synchronized clocks so that servers process $T_1$ and $T_2$ in the same order, and commit both in the fast path.}
    \label{fig:tiga-rectify}
\end{minipage}
\end{figure*}

\Para{Common setup.} Distributed OLTP systems are typically modeled as a sharded key-value store. Each shard is replicated across multiple geographic regions. 
We assume a partial replication setup: different shards can be replicated to different sets of regions. 
For simplicity, we often use examples with a full replication setup: each region contains a replica from every shard, but this is not necessary.

The system has three roles in transaction processing: \emph{client}, \emph{coordinator}, and \emph{server}. A client sends the transaction request to a coordinator. The coordinator communicates with servers to commit the transaction and returns the execution result to the client. We primarily focus on one-shot transactions, which are written as a stored procedure to be executed on servers. In addition, we incorporate the decomposition technique~\cite{vldb10-txn-determinisim} to support interactive transactions (details explained in Appendix~\ref{appendix:decomposition}).

\Para{Necessity of strict serializability}. Strict serializability requires that transactions' executions are equivalent to a serial execution on a single-copy system, and the transactions' executions also reflect the real-time ordering. While some systems~\cite{voltdb,yugabytedb} choose to sacrifice real-time ordering for performance and only provide serializability, we find that non-strict serializability is insufficient for many practical cases. For example, (1) Banking systems: When transaction processing does not obey real-time ordering, account balances may appear inconsistent to clients; a withdrawal might not be reflected immediately, potentially leading to overdrafts or business errors. (2) Booking/ticket systems: Late booking orders may succeed over earlier ones, thereby causing unfairness among users. (3) Locking service: Clients may observe outdated lock states and perform unsafe operations under the false assumption of ownership.

Therefore, we target strict serializability. Meanwhile, we aim to achieve fault tolerance, which guarantees strict serializability for all committed transactions in the presence of a minority of server/datacenter failures of any shard.

\Para{Consolidated concurrency control and consensus 
 and 1-WRTT fast path.} 
To achieve strict serializability and fault tolerance, we need concurrency control and consensus protocols to coordinate the servers. 
Although each category of protocols is usually studied separately, it is recognized that they share the same goal---to achieve consistent ordering among all servers~\cite{tods06-consensus-commit}. 
Prior work has proposed consolidating the two types of protocols into a single layer~\cite{sosp15-tapir,osdi16-janus}, to reduce redundant coordination overhead. 
In particular, the commit latency in geo-replicated systems can be reduced from several WRTTs to 1--2 WRTTs by the consolidation.

Existing work has proposed having a 1-WRTT fast path for commutative transactions, i.e., transactions that do not conflict with each other. If transactions conflict, the fast path will fail, and more WRTTs are required to commit the transaction. 
Consider Figure~\ref{fig:janus-reorder}, which shows the timestamp-based protocol Tapir~\cite{sosp15-tapir} and illustrates the problem. 
The example has 2 coordinators and 2 shards, $X$ and $Y$. 
Both shards are replicated in 2 regions, Region-1 and Region-2 (technically a 3rd region is needed to tolerate failures; this is omitted to simplify discussions). 
Each coordinator multicasts a transaction ($T_1$/$T_2$) to all servers. 
$T_1$ and $T_2$ arrive at the 4 servers in different orders. 
This inconsistent ordering will form a cyclic dependency that is propagated across servers. Neither $T_1$ nor $T_2$ can commit in 1 WRTT (fast path).
Thus, Tapir needs extra RTTs, which are also WRTTs if coordinators and servers are in different regions, to resolve the cycle.

We realize that the fast paths of Tapir and the others (e.g., Janus, Detock, and so on) fail under conflicts because these protocols \emph{optimistically} process transactions based on their arrival order, but transactions' arrival order on different servers can often be different in geo-replicated settings~\cite{vldb23-nezha}.

\Para{Intuition: proactive ordering with timestamps.}
Instead of being optimistic about arrival orders, \sysname chooses the \emph{proactive ordering} approach---\sysname lets the coordinators predict a timestamp for transactions to arrive before multicast. Servers serialize transactions in their timestamp order. The timestamps are generated with synchronized clocks and indicate an approximate serialization point in the global ordering. 

When clocks are used to serialize transactions, their accuracy directly impacts system performance. Classic clock synchronization techniques (e.g., NTP~\cite{ntp}) often suffer high synchronization errors (10s of milliseconds)~\cite{sigmod95-clocc}, which limited the effectiveness of protocols relying on synchronized clocks.

In recent years, however, clock synchronization accuracy has been improved substantially in practice~\cite{huygens,osdi20-sundial,aws-clock-sync,nsdi22-graham}. 
For example, Huygens~\cite{huygens} has become generally deployable in the public cloud and can yield microsecond- or even nanosecond-level synchronization errors~\cite{clockwork-accurate-owd,clockwork-clock-cmp}. 
In our evaluation, even the default NTP service of Google Cloud (i.e., \texttt{chrony})~\cite{gcp-chrony} can steadily converge to \SI{4.54}{\milli\second} error thanks to well-built cloud infrastructure. 
This has opened up new opportunities for using physical clocks to serialize transactions: Multiple servers can share a common timeline given the high accuracy of the synchronized clocks. When receiving the transaction, these servers can take actions simultaneously according to its timestamp. However, it is worth noting that even though the clock synchronization accuracy has been greatly improved, there is no guarantee of a \emph{deterministic} error bound. 
Even for advanced schemes like Huygens, the worst-case error can still go arbitrarily large in theory. 
Therefore, a desirable protocol should still assume the clock is loosely synchronized, following Liskov's design principle to ``depend on clock synchronization for performance but not for correctness''~\cite{liskov_clock_sync}.

Figure~\ref{fig:tiga-rectify} illustrates \sysname's idea with the same example. 
When the coordinator multicasts $T_1$ (or $T_2$), it proactively assigns a future timestamp to its transaction, and the timestamp is decided by summing up the sending time and the expected delay from the coordinator to enough (i.e., a super quorum, \S\ref{sec:timestamp-initilization}) participating servers to commit the transaction. 
Servers receiving the transactions will hold them until the local clocks pass the transactions' timestamps. 
Thus, all participating servers can consistently process $T_1$ and $T_2$, and commit both transactions.

\Para{Challenges.}
While having synchronized clocks can create favorable conditions for \sysname's fast path performance, using the simple rationale we just demonstrated itself is not sufficient to guarantee system correctness. 
The reason is three-fold. 
First, timestamps are not assumed to be always accurate. 
Clocks can be out of sync, or messages can take longer than predicted to arrive. 
Servers need extra mechanisms to deal with receiving transactions that are supposed to be processed in the past. 
Second, failures such as crashes and network partitions make it challenging to design a safe protocol that is resilient to corner cases such as recovering a dangling timestamp while the original coordinator has an unknown status (could be either crashed or just slowed). In such cases, the leader is holding the transaction with its timestamp, but the leader becomes isolated due to the network partition. Meanwhile, a new leader is elected and decides a new timestamp for the transaction. As a result, when the old leader becomes reconnected to the system, it must abandon the old timestamp for the transaction. Third, linearizability is a \emph{local} property that is defined for a single shard but strict serializability is not~\cite{linearizability,adya1999weak}, which leads to the fact that a simple timestamp ordering solution is only serializable, losing the ``strict'' prefix which covers the external consistency or simply the linearizability part. This is also known as the timestamp inversion pitfall by recent work~\cite{osdi23-ncc}. 
Extra cross-shard coordination steps need to be carefully designed to achieve correctness.

\section{\sysname Design}
\label{sec:design}

\begin{figure}[!t]
    \centering
    \includegraphics[width=0.9\linewidth]{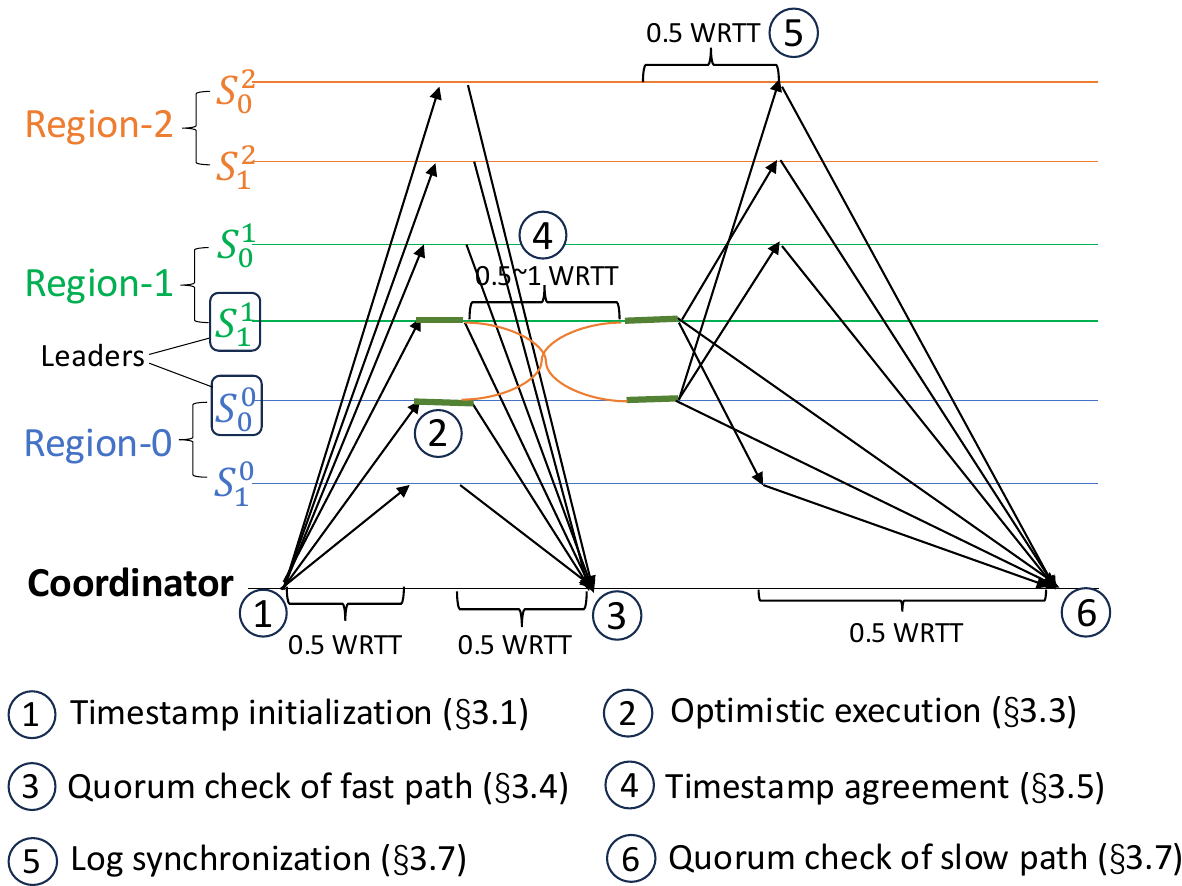}
    \caption{Workflow of \sysname. $S_{s}^r$ indicates the server's \emph{replica-id} is $r$ and \emph{shard-id} is $s$. The green solid bars \protect\circled{2} indicate that servers optimistically execute the transaction. However, servers can only know whether the execution is valid after timestamp agreement \protect\circled{4}. If the execution turns out to be invalid, servers will revoke the previous execution and re-execute the transaction (see Case-3 in \S\ref{sec:timestamp-agreement}).}
    \label{fig:tiga-workflow}    
\end{figure}

\begin{figure}[!t]
    \centering
    \begin{tcolorbox}[colback=white!10,
                      width=8cm,% Use 8cm total width,
                      arc=1mm, auto outer arc,
                      boxrule=0.5pt, 
                      % coltext=magenta
                     ] 
    \begin{itemize}[leftmargin=*,nosep]
        \item \emph{shard-id}--- shard identifier ($0,1,\cdots, m-1$).
        \item \emph{replica-id}--- replica identifier ($0,1,\cdots, 2f$). 
        \item \emph{g-view}--- the global view, indexed by an integer which is incremented after every view change.
        \item \emph{l-view}--- the local view, indexed by a integer which is incremented or remain unchanged after every view change.
        \item \emph{status}--- one of \textsc{normal}, \textsc{viewchange}, or \textsc{recovering}.
        \item \emph{pq}--- a priority queue used to hold incoming transactions, and release transactions according to their timestamp order.
        \item \emph{log}--- a list of transactions, which are appended in the order of their timestamps agreed upon by the participating shards.  
        \item \emph{sync-point}--- the log position up to which this server's \emph{log} is consistent with its leader (i.e., the leader that has the same \emph{shard-id} as this server).
        \item \emph{commit-point}--- the log position up to which the server has checkpointed the state. 
    \end{itemize}
    \end{tcolorbox}
    \caption{Local state of \sysname servers.}
    \label{fig:server-state-var}
\end{figure}

\captionsetup[algorithm]{font=footnotesize}

\begin{algorithm}[!t]
\footnotesize 
%\small, \footnotesize, \scriptsize, or \tiny
% \begin{multicols}{2}
\caption{Server action} 
\label{algo:server-action}
\begin{algorithmic}[1]
  \Event{$receiving$ $txn$, $T$} \label{algo-line:accept-start}
    \If{\textsc{conflict-detection}($T$)=OK}  \label{algo-line:conflict-detection}
         $pq.\textbf{insert}(T)$ \label{algo-line:accept-end}
    \ElsIf{\textsc{am-leader}()}  \Comment{Only leaders can update $T.t$}
    \State $T.t\leftarrow \textsc{clock-time}()$ \label{algo-line:timestamp-update}
    \State $pq.\textbf{insert}(T)$
    \EndIf
  \EndEvent

  \Event{$clock\ time$ $progressing$}  \Comment{Periodically check the clock time}
    \State $nowTime\leftarrow\textsc{clock-time}()$
    \State $releaseTxns\leftarrow []$
    \LeftComment{Enumerate txns based on timestamp order}
    \For{$T \in \emph{pq}$}
        \If{$T.t > nowTime$}
           \textbf{break}  \Comment{$T$ has not expired}
        \EndIf
        \If{$\not\exists T'\in pq: T'.t<T.t \textbf{ and } T' \text{ conflicts with } T $} 
        \State \emph{releaseTxns}.\textbf{append}($T$)
        \EndIf
    \EndFor  
    \For{$T\in \emph{releaseTxns}$}
        \State $\forall$ \emph{key} $\in$ \emph{T.readSet}, \  \emph{rMap[key]}$\leftarrow$$T.t$   \label{algo-line:timestamp-record1}
        \State $\forall$ \emph{key} $\in$ \emph{T.writeSet}, \  \emph{wMap[key]}$\leftarrow$$T.t$   \Comment{For conflict detection} \label{algo-line:timestamp-record2}
        \If{\textsc{am-leader}()} 
         \State $ret\leftarrow\textsc{execute}(T)$ \Comment{Only leaders execute $T$}
        \State $hash=\textsc{calculate-hash}(log)$
        \State \textsc{send-fast-reply($T$, $hash$, $ret$)}
        \LeftComment{$tSet$ contains $T$'s timestamps used by each leader}
        \State $tSet\leftarrow$\textsc{timestamp-agreement}($T$) 
        \If{$T.t=\max \{t:t\in tSet\}$} 
        \If{$tSet$.\textbf{size}()>1} \Comment{Some leaders used incorrect $T.t$}
        \LeftComment{After completing second round, leaders agree on $T.t$}
        \State \textsc{timestamp-agreement}($T$) 
        \EndIf        
        \State Append $T$ to $log$ and syncs $T.t$ with followers
        \State $\emph{pq}$.\textbf{erase}(T) \label{algo-line:release}
        \Else \Comment{This leader used smaller timestamp}
        \State $T.t\leftarrow \max \{t:t\in tSet\}$
        \State $\emph{pq}$.\textbf{reposition}($T$) \label{algo-line:reposition}
        \EndIf
        \Else \Comment{Follower sends fast-reply without execution result}
        \State \textsc{send-fast-reply($T$, $hash$, $null$)}
        \State $\emph{pq}$.\textbf{erase}(T)
        \EndIf
    \EndFor
  \EndEvent

   \Event{$follower's$ $receiving$ log-sync, $msg$} \label{algo-line:slow-path-begin}
    \State Update \emph{log} to keep consistent with leader's \emph{log}
    \State Advance follower's \emph{sync-point}
    \State \textsc{send-slow-reply}($T$)
  \EndEvent \label{algo-line:slow-path-end}

\end{algorithmic}
% \end{multicols}
\end{algorithm}

Figure~\ref{fig:tiga-workflow} illustrates the workflow of \sysname. We next follow the workflow to explain the protocol details. Figure~\ref{fig:server-state-var} summarizes the state variables and data structures maintained at each \sysname server, and Algorithm~\ref{algo:server-action} sketches the server's action in response to different events. We refer to these state variables and actions during our explanation. The full algorithm description of server action and coordination action is included in Appendix~\ref{appendix:tiga-algo}.

\subsection{Timestamp Initialization}
 \label{sec:timestamp-initilization}

The protocol starts with the coordinator multicasting transactions to servers, i.e., \circled{1} in Figure~\ref{fig:tiga-workflow}. When multicasting the transaction $T$, the coordinator needs to predict a future timestamp for $T$, so that $T$ can arrive at all involved servers just before the timestamp. The future timestamp $t$ is calculated by adding the headroom to the transaction's sending time $t_{send}$. We next describe how we estimate the headroom.

The headroom estimation is based on the measurement of the one-way delays (OWDs) between the coordinator and the servers. We use $C$ to represent the coordinator, and use $S_{s}^{r}$ to represent a server whose \emph{replica-id} is $r$ and \emph{shard-id} is $s$. We use $O(C,S_{s}^{r})$ to represent the OWD from $C$ to $S_{s}^{r}$. 
These OWDs can be easily measured since the clocks have been synchronized among coordinators and servers. Huygens achieves clock synchronization errors of only a few microseconds (\S\ref{sec:ablation-study}), which are negligible compared to WAN OWDs (tens to hundreds of milliseconds), thereby enabling accurate OWD measurement.

Assume that $T$ will be submitted to $m$ shards, and \emph{shard-id}s are $0,\cdots,m-1$. $C$ will assign a future timestamp for $T$, by adding the headroom to its sending time $t_{send}$. The size of the added headroom will decide how likely $T$ can be committed in the fast or slow path. 

To commit $T$ via the fast path, its future timestamp $t$ should be sufficiently large for $T$ to reach at least a super quorum ($1+f+\lceil f/2 \rceil$) of replicas in each shard.
\begin{equation*}
t = t_{send} + \max\limits_{0\leq s < m}  \max_{r\in SQ_s} O(C, S_s^{r})+\Delta
\end{equation*}
$SQ_s$ represents a super quorum of replicas from the shard (\emph{shard-id} is $s$) that are closest to $C$, i.e., the replicas with smaller OWDs to $C$ than the remaining replicas in the shard. We choose $\Delta=10 ms$ in our implementation so that the headroom added to $t_{send}$ is slightly larger than the OWDs of the super quorum. The necessity of a super quorum (rather than a simple quorum of $f+1$ servers) for fast path will be explained later in \S\ref{sec:tiga-qc-fast}.

\subsection{Conflict Detection and Timestamp Update}
 \label{sec:timestamp-update}

On each server, \sysname maintains a priority queue (denoted $pq$ in Algorithm~\ref{algo:server-action}) to buffer transactions and release them according to their timestamp order. Given a transaction $T$, the server performs \emph{conflict detection} (line~\ref{algo-line:conflict-detection} in Algorithm~\ref{algo:server-action}) to decide whether $T$ can be accepted into $pq$.

\Para{Conflict detection.} The server checks $T$'s timestamp and will not accept it into $pq$ if $pq$ has already released another transaction $T'$, which has a larger timestamp and has read-write or write-write conflict with $T$ on the same keys. Since transactions are written as (or can be decomposed as) one-shot stored procedures, the server knows their read/write sets before execution. Thus, conflict detection can be implemented very efficiently: The server maintains two maps ($rMap$ and $wMap$). Both maps associate every data item (key) of the key-value store with a timestamp. When $T'$ is released from $pq$ (the release conditions will be explained in \S\ref{sec:spec-exec}), the server uses $T'$'s timestamp to update the timestamp of every key that falls in $T'$'s read/write set (line~\ref{algo-line:timestamp-record1}--\ref{algo-line:timestamp-record2} in Algorithm~\ref{algo:server-action}). 
When $T$ arrives, the server directly compares $T$'s timestamp with the recorded timestamps of keys that $T$ will read/write. $T$ will be accepted into $pq$ if its timestamp is larger than the timestamps of all conflicting transactions that have been released from $pq$ (line~\ref{algo-line:accept-start}--\ref{algo-line:accept-end}).

Not every transaction can be accepted into $pq$ after conflict detection. When $T$ arrives late at a leader server due to network delay or packet loss, its timestamp may be too small to be accepted into $pq$. 
In such cases, the leader updates $T$'s timestamp to the local clock time (line~\ref{algo-line:timestamp-update}), after which $T$ can enter the leader's queue.

Followers, by contrast, do not perform timestamp updates. If $T.t$ is smaller than acceptable for the queue, the follower holds $T$ and waits for synchronization instructions from the leader in the slow path (\S\ref{sec:tiga-qc-slow}, line~\ref{algo-line:slow-path-begin}--\ref{algo-line:slow-path-end} in Algorithm~\ref{algo:server-action}).

\subsection{Optimistic Execution}
\label{sec:spec-exec}

For each transaction $T$ in the queue $pq$, followers refer to the local clock to determine when to release it. 
Once the local time surpasses $T.t$, the follower releases $T$ without executing it: $T$ is removed from the queue and then appended to the log list. 
After that, the follower sends a fast-reply to the coordinator to perform a quorum check (\S\ref{sec:tiga-qc-fast}).

Leaders, on the other hand, must execute transactions before releasing them. To minimize latency (1 WRTT), leaders optimistically execute transactions without coordination. Periodically, the leader refers to its local clock to identify \emph{expired} transactions (i.e., transactions whose timestamps have been passed by the current clock time) in its queue. 
It checks these transactions in timestamp order to decide whether each can be optimistically executed. When $T$ has reached the head of the queue without any conflicting transactions ahead, $T$ can be executed. However, $T$ will remain at the head of the queue after execution.

After executing $T$, the leader sends a fast-reply to the coordinator, including the execution results. $T$ stays at the head of the leader's queue to undergo timestamp agreement (see \S\ref{sec:timestamp-agreement}). After that, $T$ is either released (line~\ref{algo-line:release}) or repositioned in the queue with a larger timestamp (line~\ref{algo-line:reposition}). 

Before $T$ can be released, the leader/follower records $T$'s timestamp with its read set and write set (line~\ref{algo-line:timestamp-record1}-\ref{algo-line:timestamp-record2}) for subsequent conflict detection (\S\ref{sec:timestamp-update}, line~\ref{algo-line:conflict-detection} in Algorithm~\ref{algo:server-action}). The follow-up transaction is no longer acceptable into the queue if it conflicts with $T$ but has a smaller timestamp.

\subsection{Quorum Check of Fast Path}
\label{sec:tiga-qc-fast}
A server's fast-reply regarding transaction $T$ includes a hash value of the log list to represent its state before $T$. 
The hash of the log list is computed as the bitwise exclusive-or (XOR) of the hashes of all its entries. 
This allows the server to \emph{incrementally} compute the hash. 
When adding/deleting a log entry $e$, the new hash is computed as $H_{new} = XOR(H_{old}, hash(e))$. 
We use a 160-bit SHA-1 hash and assume hashes do not collide in practice. 
Note that applying XOR on hashes does not make them vulnerable to collisions~\cite{crypto97-incremental-hash,crpyto95-xormac}. 
It is a commonly applied technique in systems using incremental hash~\cite{incremental-hash-memory-checker,incremental-hash-memory,incremental-hash-sign,stoc95-incremental-hash-virus,incremental-hash-trend,vldb23-nezha}. Appendix~\ref{appendix:incremental-hash} provides more details on how \sysname uses incremental hash. In addition, the fast-reply includes $T$'s timestamp.

The coordinator receives $T$'s fast-replies from all servers. $T$ is considered \emph{fast-committed} on a shard if, from this shard, the coordinator receives a super quorum of fast-replies that have the same hash and timestamp for $T$. 
The super quorum must satisfy two conditions: 
(1) it contains the leader, and (2) its size is at least 
$1+f+\lceil f/2 \rceil$. 
If these are met, the coordinator uses the optimistic results in the leaders' replies as $T$'s execution results on this shard.

The reason that \sysname's fast path requires a super quorum ($1+f+\lceil f/2 \rceil$) instead of a simple quorum ($f+1$) is similar to Fast Paxos~\cite{fastpaxos}: Because the fast path omits leader–follower communication, a simple quorum lacks sufficient information for a new leader to distinguish committed from uncommitted transactions. Consider the leader and $f$ followers append $T_1$ and $T_3$ ($T_1\rightarrow T_3$) in their log lists whereas the other $f$ followers append $T_2$ and $T_3$ ($T_2\rightarrow T_3$) in the same positions of their log lists. Assuming the fast path only requires a simple quorum, then $T_1$ and $T_3$ will be considered committed. However, when the leader fails, both $T_1$ and $T_2$ exist among half of the remaining servers, thus the new leader cannot know whether $T_1\rightarrow T_3$ or $T_2\rightarrow T_3$ is previously committed. If the new leader mistakenly believes $T_2\rightarrow T_3$ is previously committed, then $T_3$ will have a different execution result compared to that before the crash.

If $T$ is fast-committed on all its involved shards, the coordinator additionally checks whether leaders have consistent timestamps for $T$ in their fast-replies, because some leaders may have updated $T$'s timestamp whereas the others have not. If all participating leaders have used the same timestamp to execute $T$, $T$ is committed in the fast path.

\subsection{Timestamp Agreement}
\label{sec:timestamp-agreement}

After $T$'s execution, the leaders need to verify whether the execution is valid; i.e., all participating leaders should execute $T$ in the same timestamp order; otherwise, the execution may violate strict serializability and should be revoked. To support revoking, \sysname maintains multiple versions for each data item (key). $T$'s optimistic execution creates new versions of data. Once the server detects the execution is invalid, it erases the corresponding data versions. Note that the revoking operation is internal to \sysname, and does not cause application-related rollback.

To check the validity of $T$'s execution, leaders start a round of message exchange. Each leader notifies the other participating leaders of its local timestamp $T.t$. Then, each leader collects the full set of $T$'s timestamps used by different leaders, and computes the maximum as the agreed timestamp, $T.t_{agreed}$. Since all leaders operate on the same timestamp set, they deterministically compute the same $T.t_{agreed}$. The subsequent actions depend on three possible cases:

\Para{Case-1:} All timestamps match. This is the ideal case, which takes only 0.5 WRTT (\protect\circled{4} in Figure~\ref{fig:tiga-workflow}) for leaders to notify each other. If every leader’s local timestamp equals $T.t_{\text{agreed}}$, the timestamp agreement succeeds immediately. Each leader releases $T$ and then appends $T$ to its log.

\Para{Case-2:} This leader used $T.t_{agreed}$, but others did not. In this case, the leader’s optimistic execution remains valid, but some other leaders used smaller timestamps. To avoid potential timestamp inversion (discussed in \S\ref{sec:fix-timetamp-inversion}), the leader cannot release $T$ immediately. Instead, it initiates a second round of timestamp exchange (another 0.5 WRTT) to confirm that all leaders have updated $T$'s timestamp to $T.t_{\text{agreed}}$. Once confirmed, the leader proceeds as in Case-1. In this case, the timestamp agreement \circled{4} takes 1 WRTT in total.

\Para{Case-3:} This leader used a timestamp smaller than $T.t_{agreed}$. This indicates that the leader’s optimistic execution is invalid, so it revokes $T$'s execution. Then, it updates $T$’s timestamp: $T.t \leftarrow T.t_{agreed}$. After that, the leader initiates the second round of timestamp exchange (another 0.5 WRTT) to notify the other leaders. Since $T$'s timestamp changes to a larger value, $T$ will be repositioned in the leader's queue. Eventually, $T$ will come to the head again, and then the leader will re-execute $T$ with the agreed timestamp.

\subsection{Avoiding Timestamp Inversion}
\label{sec:fix-timetamp-inversion}

\begin{figure}[!t]
\includegraphics[width=0.48\textwidth]{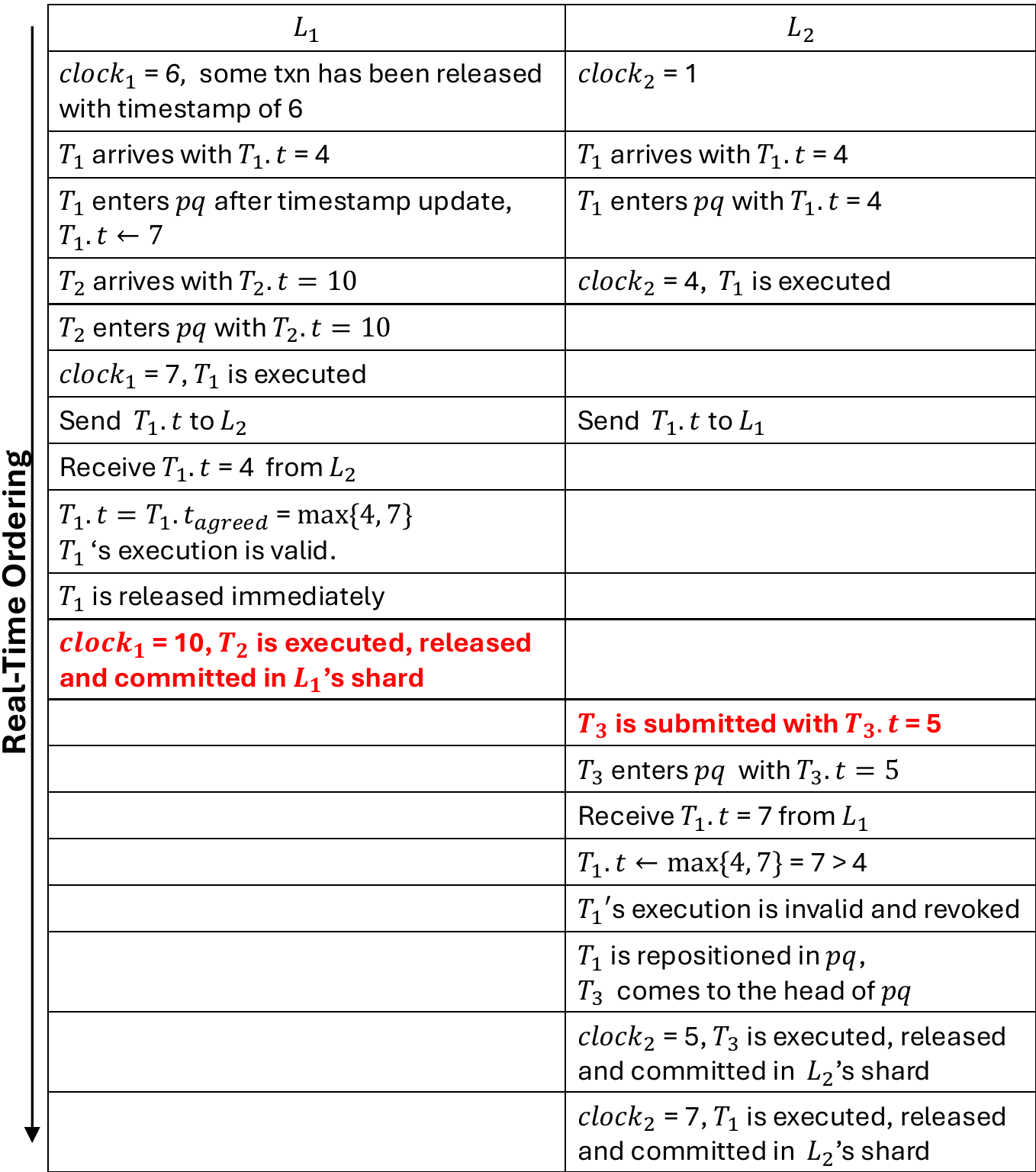}
\caption{Illustration of timestamp inversion. With only one round of message exchange between $L_1$ and $L_2$, $T_3$ may be submitted after $T_2$ is committed, leading to real-time ordering $T_2\rightarrow T_3$ that contradicts serializable order $T_3\rightarrow T_1\rightarrow T_2$.}
\label{fig:timestamp-inversion-demo}
\end{figure}

Readers may wonder why the leader in Case-2, denoted $L_1$, cannot immediately release $T$ as this leader already used the agreed timestamp for $T$. The reason is the timestamp inversion pitfall~\cite{osdi23-ncc}, which hurts correctness, in particular strict serializability. 
While $L_1$ has confirmed it used the proper timestamp $T.t_{agreed}$ for $T$, the other leader(s), denoted $L_2$, used a smaller timestamp $T.t<T.t_{agreed}$, and $L_1$ is uncertain whether $L_2$ has completed timestamp agreement and updated $T.t$ to $T.t_{agreed}$. As a result, if $L_1$ releases $T$ immediately, timestamp inversion may occur: After $L_1$ has committed some transactions with timestamps larger than $T.t_{agreed}$, $L_2$ could still commit other transactions with smaller timestamps than $T.t_{agreed}$.

Figure~\ref{fig:timestamp-inversion-demo} shows a concrete sequence of events illustrating how timestamp inversion occurs. $L_1$ and $L_2$ are the leaders of two shards, and they process three transactions $T_1$, $T_2$ and $T_3$. $T_1$ is a multi-shard transaction that involves both $L_1$'s and $L_2$'s shards. $T_2$ is only processed by $L_1$'s shard; $T_3$ is only processed by $L_2$'s shard. The leaders' clocks (e.g., $clock_1$ and $clock_2$) are badly synchronized. 

$L_1$'s event sequence indicates the dependency relation $T_1\rightarrow T_2$ and $L_2$'s event sequence indicates $T_3\rightarrow T_1$, so the only valid serializable schedule is $T_3\rightarrow T_1\rightarrow T_2$. However, in real time, $T_3$ starts after $T_2$ has been completed, indicating the real-time ordering relation $T_2\rightarrow T_3$, which contradicts the serializable order.

The fundamental reason behind timestamp inversion lies in the different guarantees of linearizability versus strict serializability. Linearizability is a \textbf{local} property within each shard—e.g., $L_1$ only needs to ensure its followers in the same shard use a consistent order between $T_1$ and $T_2$ with $L_1$ itself, and does not consider the ordering between $T_2$ and $T_3$, which are processed by different shards. The same holds for $L_2$. In contrast, strict serializability enforces a \textbf{global} order across shards. Although $T_2$ and $T_3$ do not directly conflict, they both access data involved in $T_1$, forming a dependency chain that induces a real-time order between them. This indirect dependency is not captured by linearizability, but is essential for preserving strict serializability.

To avoid timestamp inversion, when a leader notices it is holding a different timestamp from the other leaders for a transaction, it must ensure no other transactions with smaller timestamps (e.g., $T_3$) can be committed later. Specifically in Figure~\ref{fig:timestamp-inversion-demo}, $L_1$ should release $T_1$ after it confirms that $L_2$ has updated $T_1$ with the agreed timestamp, and $T_1$ has come to the head of the queue again (after repositioning). At this point, (1) $T_3$ has been executed on $L_2$ whereas $T_2$ remains in $L_1$'s queue, because $T_1$ is at the head of the queue and blocking $T_2$ from execution. (2) $L_2$ will no longer allow the other transactions (which conflict with $T_1$) to enter its queue with smaller timestamps than $T_1$'s agreed timestamp ($T_1.t=7$). Thus, the second round of timestamp agreement rules out any real-time ordering violations that would otherwise conflict with the serializable schedule. We include the proof in Appendix~\ref{appendix:correctness-proof}.

In contrast to the leaders, the followers do not engage in timestamp update and agreement, so they may have different timestamps from the leaders at this point. The potential leader-follower inconsistency will be detected in the fast path by comparing the hashes and $T$'s timestamp (\S\ref{sec:tiga-qc-fast}) and resolved in the slow path (\S\ref{sec:tiga-qc-slow}).

\subsection{Log Synchronization and Slow Path}
\label{sec:tiga-cross-region-sync}
\sysname does not guarantee that all transactions are committed in the fast path. 
If a leader updates a timestamp, it causes inconsistency between itself and its followers. 
Therefore, after appending the transaction to its log, the leader advances its \emph{sync-point} and also sends the followers a log synchronization message. 
In the synchronization messages, the leader includes the entry's position, unique identifier\footnote{The coordinator attaches a sequence number to the transaction at submission. The unique identifier for this transaction is to combine the coordinator-id and the sequence number.} and the timestamp agreed by leaders. When receiving the synchronization message, the followers update their logs to keep consistent with the leader's log: (1) If the follower's log contains some entry that does not exist in the leader's log, then the follower removes the entry; 
(2) If the leader's log contains some entry that does not exist in the follower's log, then the follower first tries to obtain the missing entry locally from its server. If the entry is not found, then the follower fetches it from the leader. 
(3) If some entry exists in both the leader's and the follower's logs but has different timestamps, then the follower updates the entry's timestamp to keep consistent with the leader.

\label{sec:tiga-qc-slow}

After the log update, the follower advances its \emph{sync-point} to indicate its log list has been synchronized with the leader's log list up to this point. Then, the follower sends slow-replies to the coordinators which have multicast those synchronized entries, notifying the coordinators that entries for these transactions have become consistent with the leader. Appendix~\ref{appendix:optim-slow-reply} describes an optimization that does not require the followers to send the slow reply for every entry.

Followers also periodically report their \emph{sync-point}s to the leader, so that the leader knows which log entries have been sufficiently replicated. After the leader confirms that the log entries have been surpassed by the \emph{sync-points} from $f+1$ servers of the same shard, the leader knows these entries are committed. The leader then advances its \emph{commit-point} and notifies its followers of the updated \emph{commit-point}. Followers can execute the log entries up to their \emph{commit-point}s and generate checkpoints to accelerate failure recovery (\S\ref{sec:tiga-failure-recovery}). 

A transaction is considered \emph{slow-committed} on a shard if the coordinator (1) receives the fast-reply from the leader and (2) receives slow-replies from at least $f$ followers. 
If the transaction is either fast- or slow-committed on every involved shard, it is considered committed.

\subsection{Optimization based on Leaders' Co-location}
\label{sec:tiga-colocation-optimization}
\begin{figure}[!t]
    \centering
    \includegraphics[width=0.9\linewidth]{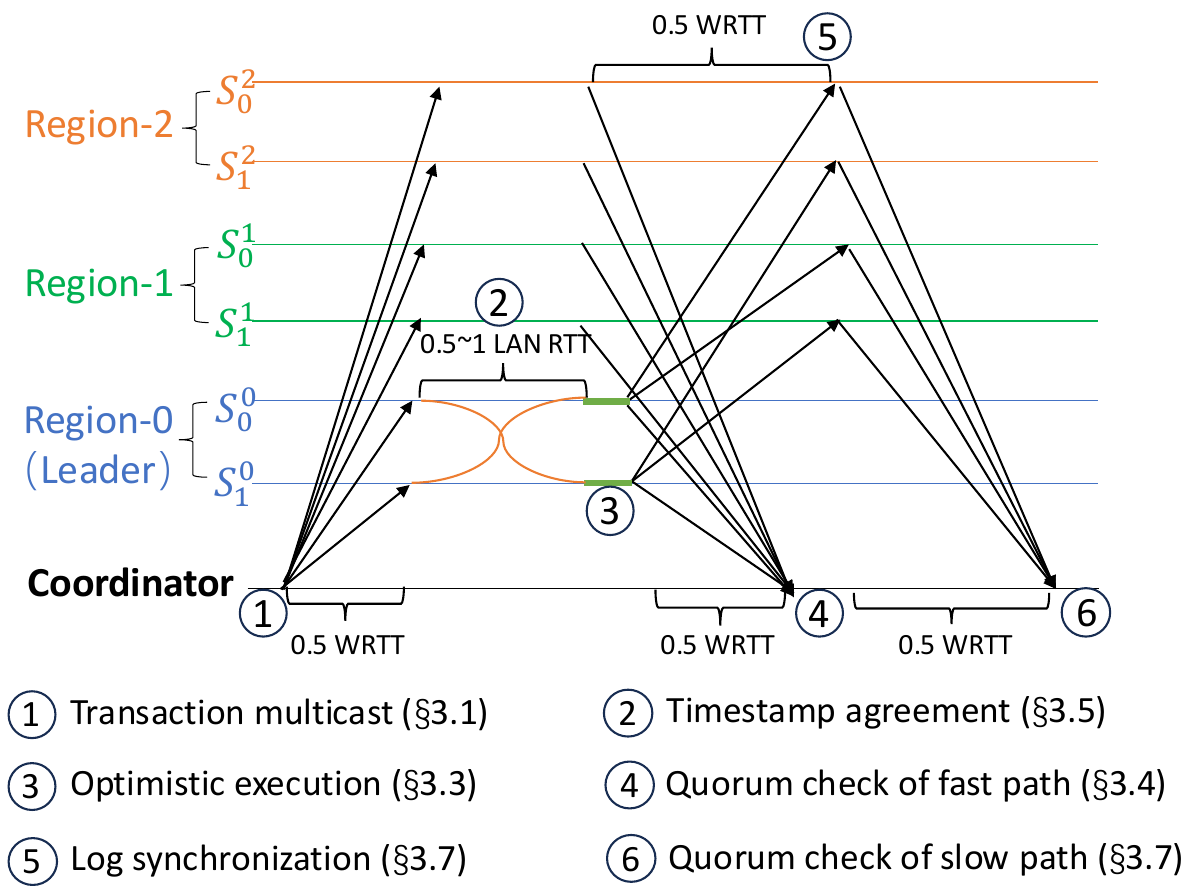}
    \caption{Workflow of \sysname (preventive approach). } 
    \label{fig:tiga-workflow-colo}    
\end{figure}

In \S\ref{sec:spec-exec}-\S\ref{sec:timestamp-agreement}, we let the leaders start optimistic execution without waiting for timestamp agreement. The purpose is to minimize the latency of the fast path, because timestamp agreement costs additional WAN latency when leaders are separated across regions. However, skipping timestamp agreement in the fast path introduces the risk of invalid execution, i.e., different shards execute transactions based on inconsistent timestamp orders, incurring expensive rollback. 

Alternatively, if timestamp agreement is \emph{cheap}, i.e., it only costs LAN latency, then prioritizing timestamp agreement over execution is more desirable: It only adds negligible overhead to the commit latency, but avoids the rollback of invalid execution, because all the leaders execute the transactions according to their agreed timestamp order. Fortunately, we realize that this approach is commonly feasible in practical deployment. In typical geo-distributed OLTP systems~\cite{sigmod15-helios, osdi16-janus,sosp11-cops,aws-shard, tikv-multi-region-deployment,azure-cosmos-partition,eurosys21-dast,vldb23-caerus}, each datacenter (region) usually contains a full copy of data, thus enabling co-location of all leaders within the same region. In addition, industry workloads also exhibit strong data locality. For example, the Yahoo! trace~\cite{vldb08-pnuts} reveals 85\% regional locality for user data accesses; the typical edge workload~\cite{eurosys21-dast} has 90\% of intra-region transactions. By leveraging the co-location property, we can schedule timestamp agreement ahead of execution, as shown in Figure~\ref{fig:tiga-workflow-colo}, in contrast to Figure~\ref{fig:tiga-workflow}.

\Para{Choose the approach of timestamp agreement.} Since there is no one-size-fits-all approach towards different deployments, \sysname incorporates both approaches into the protocol design, with the choice being configurable through its modified view change protocol (\S\ref{sec:tiga-failure-recovery}). Specifically, \sysname leverages Huygens' probe mesh to continuously monitor the OWDs between servers. Based on the measured OWDs, \sysname initializes a view change to designate the leader for every shard. \sysname tries to co-locate all leaders close to each other, so that it can schedule timestamp agreement before execution, which costs negligible LAN overhead but prevents invalid execution at its root. However, if co-location is infeasible, i.e., \sysname cannot find a group of leaders with OWDs below a predefined threshold (e.g., \SI{10}{\milli\second}), then the preventive approach becomes inefficient, prompting \sysname to adopt the detective approach (Figure~\ref{fig:tiga-workflow}). The view change message includes the planned approach (i.e., preventive or detective), so that all servers consistently adopt the planned approach after entering the new view.

\section{Failure Recovery}
\label{sec:tiga-failure-recovery}
Server failures in \sysname can be classified into two categories: leader failures and follower failures. 
Follower failures are relatively easier to deal with. 
A minority of follower failures in any shard do not interrupt service availability. 
\sysname can always use the slow path to commit transactions if the servers alive are insufficient for the super quorum in the fast path. When failed followers reboot, they catch up by synchronizing logs with the leader. Here, we mainly discuss leader failure handling. Further details and the correctness proof are included in Appendix~\ref{appendix:recovery} and Appendix~\ref{appendix:correctness-proof}.

\sysname uses a \emph{view-based}~\cite{viewstamp-original} protocol to facilitate leader failure recovery. 
A view records information on membership, including each member's role, i.e., as a leader or follower.

\sysname distinguishes between two views: a local view (\emph{l-view}) which stores information about a shard, and a global view (\emph{g-view}) about all shards. 
A global view includes all local views.  
Both the global views and the local views are indexed by unique and monotonically increasing integers. 
The views are managed by a \emph{view manager}. 
The view manager is a simple service implemented on a replicated state machine that is resilient to failures, e.g., it could be built with ZooKeeper~\cite{zookeeper}. 
It is off the critical path of transaction processing in the common cases, so its performance has no significant impact.

Every server stores both the global and its local view. 
When the system is stable (no failures), all servers have the same global view, which also implies that servers of the same shard share the same local view. 
A server attaches the global and local view-ids to every message it sends out. 
When receiving a message from other shards, a server always checks and rejects the message if the message has a different global view. 
If the message is from within the shard, the server also checks whether it has the same local view.

The view manager detects server failure(s) using heartbeats, 
and launches a view change if a leader fails. 
The view change proceeds in the following steps. 

\circled{1} The view manager creates a new view that has new leader(s) to replace the failed one(s). When selecting new leaders, the view manager prioritizes choices that can make most leaders co-located in the same region, so that inter-leader timestamp agreement only costs LAN overhead after the system resumes normal processing in the new view. Based on the latency cost of timestamp agreement, the view manager decides whether to use the preventive or detective approach (\S\ref{sec:tiga-colocation-optimization}) in the new view. The view manager also creates new view-ids by incrementing the old view-ids. 
This includes a new global view-id and new local view-ids for the shards whose leaders are changed.

\circled{2} The view manager broadcasts the new view to all servers in the system. 
When a server receives a newer view (i.e., higher \emph{<g-view,l-view>}), it will update its view, and switch its \emph{status} from \textsc{normal} to \textsc{viewchange}. 

At the start of the view change, the servers stop processing new transactions. Each server empties its queue and appends the transactions in the queue to its log list according to their timestamp order. The new leader is responsible for collecting the servers' log lists and rebuilding a new log list that contains all the previously committed transactions.

\circled{3} If a server is the new leader of a shard, it rebuilds a new log list based on the log lists from any $f+1$ servers that remain alive in this shard. The reconstruction of the log list includes two parts: (a) The leader finds the server that is holding the largest \emph{sync-point} among the $f+1$ servers, and copies its log list up to the \emph{sync-point}. (b) The leader continues to check the remaining entries. For any remaining entry with a larger timestamp than those recovered in (a), if it exists in the log lists of $\lceil f/2 \rceil +1$ participating servers, then this entry will also be appended to the leader's log list according to its timestamp order.

\circled{4} Because the leaders' timestamp agreement \emph{happens before} followers advance their \emph{sync-points}, (a)'s log entries have the agreed timestamps across shards. But (b)'s log entries may have inconsistent timestamps across shards. Therefore, after rebuilding the log lists, the leaders conduct timestamp agreement for (b)'s log entries: (1) If a recovered transaction involves both $shard_1$ and $shard_2$, but it is only recovered in $shard_1$, then $shard_2$'s leader will pick the transaction from $shard_1$ to add to its own log list. (2) If a recovered transaction has inconsistent timestamps across shards, the leaders pick the maximum one as the agreed timestamp. 

\circled{5} After timestamp agreement, each leader broadcasts its log list to its followers. Leaders execute the recovered logs and switch back to \textsc{normal}. Followers use leaders' log lists to replace their old ones, then switch back to \textsc{normal}. 

To complete the overall design, the coordinator(s) in \sysname also cache the global view from the view manager. 
It only accepts replies that have the same global view-id. 
In case of a view change, the coordinator retries the transaction. 

\Para{Coordinator failure.}
If a coordinator fails, the servers will detect it after a timeout and launch a recovery coordinator to commit the transaction following the same coordinator procedure (\S\ref{sec:timestamp-initilization}, \S\ref{sec:tiga-qc-fast} and \S\ref{sec:tiga-qc-slow}). The newly launched coordinator can always fetch the view information from the view manager, and itself is stateless. As a result, the coordinator failure does not trigger any view change.

\Para{Checkpoints to accelerate recovery.}
\sysname incorporates a periodic checkpoint mechanism, a common practice for accelerating the recovery of transaction processing systems~\cite{osdi16-nopaxos,nsdi15-specpaxos,nsdi23-hydra,vldb23-nezha}. Since each server maintains the \emph{commit-point}, the server can safely execute the log entries prior to its \emph{commit-point}, and checkpoint the system state. When servers fail, the new leader can restore the system state from the latest checkpoint rather than from scratch. The failed follower can first fetch the latest checkpoint from the leader and catch up, significantly speeding up recovery.

\section{Evaluation}
\label{sec:eval}

We build on the Janus codebase~\cite{janus-repo}, which provides a high-performance implementation of several baseline protocols, including 2PL+Paxos, OCC+Paxos, Tapir, Janus and NCC. Using the same RPC library and runtime environment, we implement \sysname along with additional baselines, such as Detock~\cite{sigmod23-detock} and an enhanced version of Calvin~\cite{sigmod12-calvin}, namely Calvin+. Calvin+ replaces Calvin’s Paxos-based consensus layer with Nezha~\cite{vldb23-nezha}, saving at least 1 WRTT in committing transactions. In total, we compare \sysname against 8 baselines across a range of workloads.

\subsection{Evaluation Setup}
\Para{Workloads.} We employ 2 benchmarks: a micro-benchmark (MicroBench) and the widely used TPC-C~\cite{tpcc}. 
MicroBench pre-populates each shard with 1 million key-value pairs. Each transaction performs 3 read-write operations across different shards by incrementing 3 key-value pairs. The key-value pairs are selected using a Zipfian distribution~\cite{sigmod94_zipfian}. 
We tune the skew factor of the Zipfian distribution to control the contention in MicroBench, where higher skew factors yield more contention. 
For TPC-C, we implement all 5 types of transactions according to the specification~\cite{tpcc}. Additionally, we follow NCC's approach~\cite{osdi23-ncc} and make 2 types (\texttt{Payment} and \texttt{Order-Status}) multi-shot (interactive) transactions.

\Para{Baselines and testbed.} We compare \sysname to the 8 baseline protocols. 2PL+Paxos utilizes the wound-wait mechanism~\cite{tocs78-2pl-woundwait} to prevent deadlocks. Detock performs only local replication at commit and performs geo-replication asynchronously~\cite{sigmod23-detock}. To tolerate region failures, we make Detock perform synchronous geo-replication during transaction commit. In Detock, we evenly distribute the home directories of data items across regions. NCC's implementation does not tolerate server failure, and suggests using Paxos to achieve fault tolerance, so we implement NCC+ by placing NCC atop a Paxos replication layer. 2PL/OCC+Paxos and NCC inherently support interactive transactions by design. For the other protocols (Calvin+, Janus, Detock and \sysname), we integrate the decomposition technique~\cite{vldb10-txn-determinisim} (explained in Appendix~\ref{appendix:decomposition}) to support interactive transactions. 

All experiments are conducted in Google Cloud. We use \texttt{n2-standard-16} VMs to run servers and coordinators (clients are co-located with coordinators on the same VMs). Data is replicated across 3 regions: South Carolina, Finland, and Brazil. In practical deployment, clients/coordinators can either be co-located or separated from the servers, so we consider both cases: (1) We deploy 2 coordinators in each of the 3 regions (local regions). (2) We also deploy 2 coordinators in the 4th region (remote region), Hong Kong, because some coordinators might not be allowed to co-locate with servers due to governmental regulations (e.g., GDPR~\cite{gdpr-personal-data}, DSL~\cite{npc-dsl}) or proprietary business reasons. The system is configured with 3 shards (9 servers in total) for MicroBench and 6 shards (18 servers in total) for TPC-C to be consistent with Janus' original setup.

\Para{Evaluation method.} We evaluate the performance of the protocols using an open-loop approach~\cite{nsdi21-epaxos}: Each coordinator submits transactions at a given rate. The coordinator maintains a cap on the outstanding transactions and stops submitting new transactions once this cap is reached. Each test is repeated 5 times, and we report the median of the 5 trials. 
Since each region contains a full copy of the data, \sysname adopts the preventive approach in all evaluations except in \S\ref{sec:leader-separate} and \S\ref{sec:headroom-tradeoff}. 
\S\ref{sec:leader-separate} compares the performance of \sysname's preventive and detective approaches. \S\ref{sec:headroom-tradeoff} evaluates the impact of headroom on \sysname's latency and rollback rate.

\begin{table}[!t]
\centering
\caption{Maximum throughput ($10^3$ txns/s).}
\label{tab:micro-tpcc-max-tp}
\small
\begin{adjustbox}{max width=0.5\textwidth}
    \begin{tabular}{ccccccccc}
    \hline
    \textbf{Benchmark} & \textbf{2PL+Paxos} & \textbf{OCC+Paxos}  & \textbf{Tapir} & \textbf{Janus} & \textbf{Calvin+} & \textbf{Detock}  & \textbf{NCC} & \textbf{Tiga} \\ \hline
    
    MicroBench & 22.9 & 21.8 & 44.2 & 77.8  & 119.6 & 34.5 &47.4  & 157.3 \\ 

    TPC-C & 2.1 & 0.9 & 1.1 & 10.8   & 6.1 & 13.3  & 0.86 & 21.6 \\ \hline
        
    \end{tabular}
\end{adjustbox}
\end{table}

\subsection{MicroBench}
\label{sec:microbench-eval}

\begin{figure*}
\centering
\includegraphics[width=0.9\textwidth]{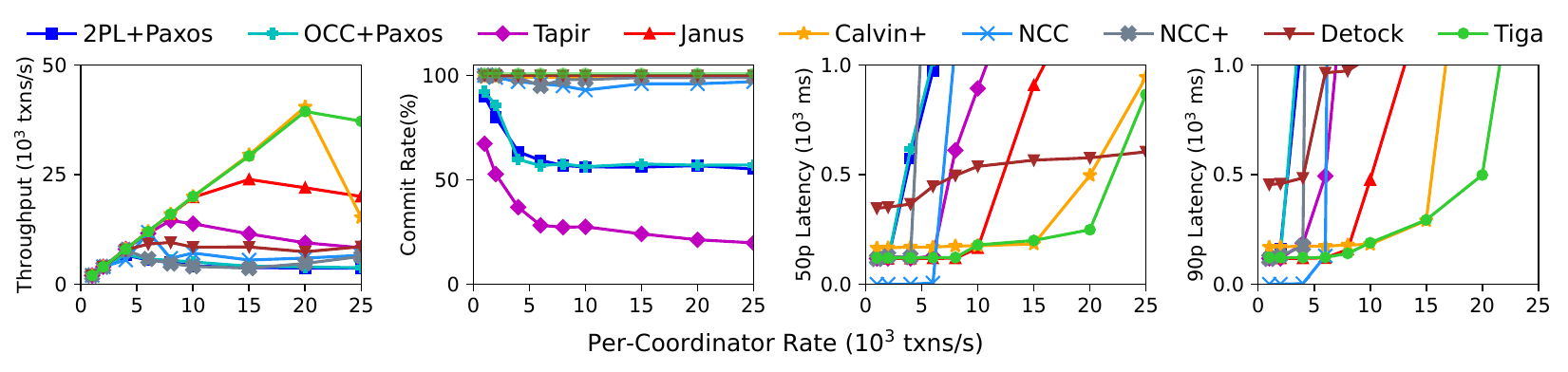}
\caption{MicroBench (skew factor=0.5) with varying rates in local region (South Carolina).}
\label{fig:Hybrid-Stats-Micro-Local-1--zipfian-50-1000000-50p}
\end{figure*}

\begin{figure*}
\centering
\includegraphics[width=0.9\textwidth]{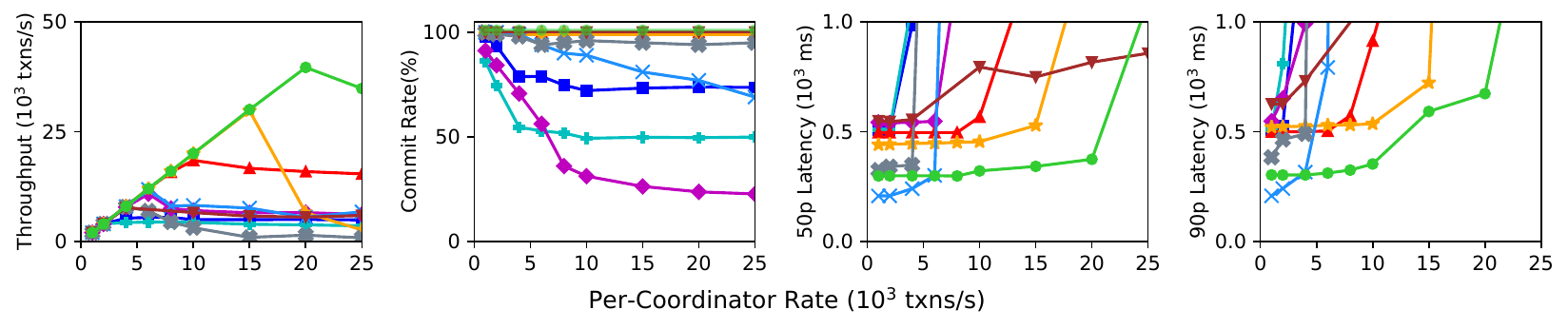}
\caption{MicroBench (skew factor=0.5) with varying rates in remote region (Hong Kong).}
\label{fig:Hybrid-Stats-Micro-Remote--zipfian-50-1000000-50p}
\end{figure*}

\begin{figure*}
\centering
\includegraphics[width=0.9\textwidth]{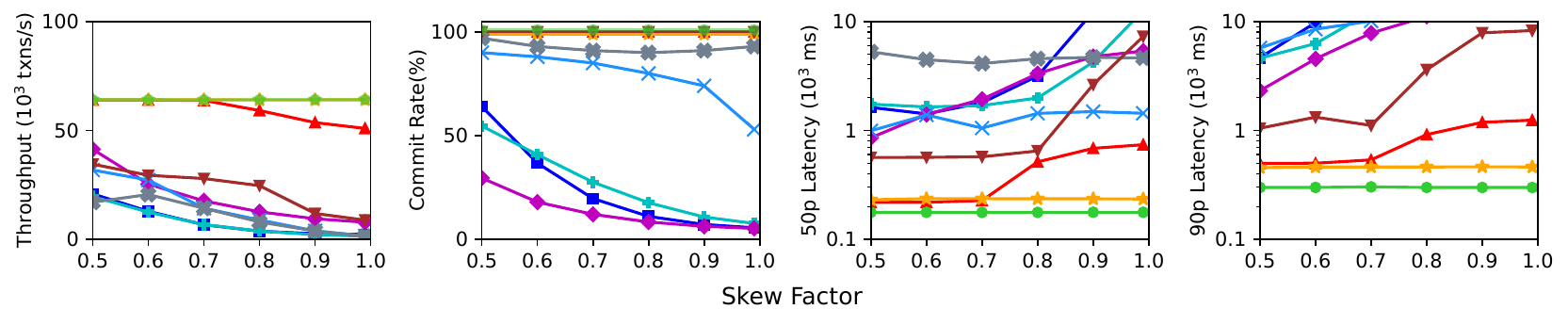}
\caption{MicroBench (per-coordinator rate=8K txns/s) with varying skew factors (all regions).}
\label{fig:micro-bench-vary-skew}
\end{figure*}

\begin{figure*}[!t]
\centering
\includegraphics[width=0.9\textwidth]{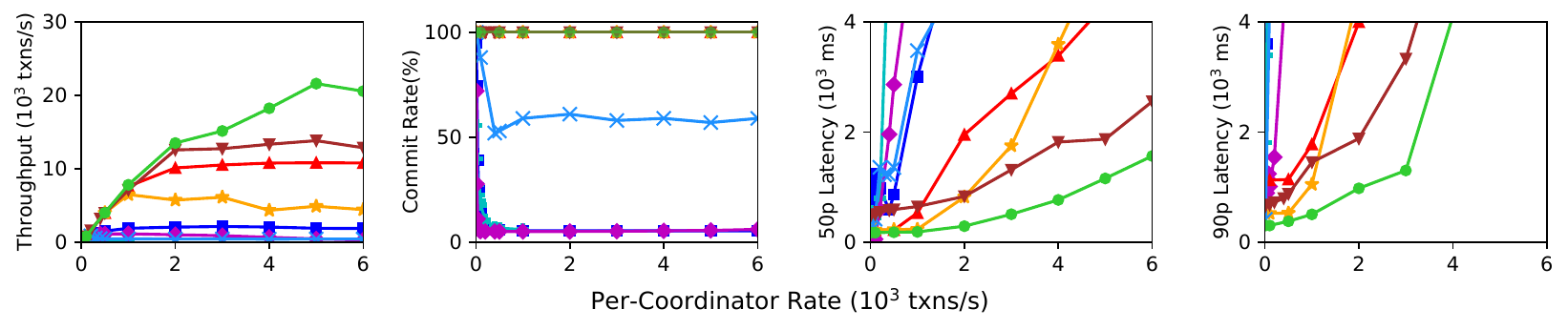}
\caption{TPC-C with varying rates (all regions).}
\label{fig:tpcc-region-all}
\end{figure*}

We first run MicroBench with a fixed skew factor of 0.5 and compare protocols' performance by increasing the submission rate of each coordinator (Table~\ref{tab:micro-tpcc-max-tp} and Figure~\ref{fig:Hybrid-Stats-Micro-Local-1--zipfian-50-1000000-50p}-\ref{fig:Hybrid-Stats-Micro-Remote--zipfian-50-1000000-50p}). 
Then we fix per-coordinator rate at 8K txns/s, and compare the protocols' performance by varying skew factor from 0.5 to 0.99 (Figure~\ref{fig:micro-bench-vary-skew}). We measure the coordinators' throughput, commit rate, 50th and 90th percentile latency in each region.

Our evaluation highlights \sysname's efficiency in achieving strict serializability and fault tolerance, outperforming state-of-the-art protocols across various metrics. Specifically: (1) 2PL/OCC+Paxos reach their throughput bottlenecks very early due to the Paxos consensus layer. Besides, the added WRTTs by the consensus layer inflate commit latency and extend the locking window, leading to more aborts. (2) Tapir's commit rate decreases rapidly as the load increases, because more concurrent transactions arrive at the servers in different orders, making Tapir abort more transactions to resolve ordering inconsistencies. (3) Janus and Detock run expensive graph algorithms to resolve inconsistencies between servers. When the submission rate grows and/or the contention (skew factor) increases, the graph computation becomes a bottleneck. Detock incurs even more WRTTs due to its layered design. In addition, since the home directories of different data items are distributed across regions, Detock pays extra WAN RTTs for dependency collection, further impacting performance.  (4) Calvin+ uses an epoch-based mechanism to predefine transactions' order, which is more robust to the various skew factors. However, it suffers from the straggler problem---when one shard is overloaded and slows down, the entire system is affected, reducing throughput and increasing latency. (5) NCC does not include fault tolerance for servers, and all servers are located in one region (South Carolina), so it costs only LAN latency in this region, and requires at least 1 WRTT in the other three regions. However, NCC uses Response Time Control (RTC) to guarantee strict serializability. RTC makes servers release a transaction only after the previous conflicting transaction sends back the commit notification. Thus, RTC artificially creates a 1-WRTT gap between these conflicting transactions. This leads to significant queueing delay. Under high load and contention, RTC limits NCC's throughput and causes latency to rise rapidly. Besides, after adding fault tolerance, NCC+ experiences further performance degradation.

Compared to the local region (South Carolina), \sysname's latency advantage becomes more pronounced in the remote region (Hong Kong). In the local region (Figure~\ref{fig:Hybrid-Stats-Micro-Local-1--zipfian-50-1000000-50p}), Janus/Tapir/Calvin+ can yield 1-WRTT latency at a low submission rate. However, in the remote region without co-located servers (Figure~\ref{fig:Hybrid-Stats-Micro-Remote--zipfian-50-1000000-50p}), they all require at least 2 WRTTs to commit. In contrast, \sysname consistently achieves 1-WRTT latency in both regions due to its efficient fast path design, delivering higher performance in more general deployment scenarios.

\subsection{TPC-C}
\label{sec:tpcc-eval}

Compared with MicroBench, TPC-C exhibits more complexity and higher contention: (1) Over 92\% of transactions are read-modify-write operations, with some requiring multiple shots to complete; (2) Since the data is stored in a column-based manner (as implemented by Janus), transactions can conflict frequently as long as they attempt to write the same column. (3) TPC-C transactions are more CPU-intensive than MicroBench, resulting in lower throughput for all protocols.

Table~\ref{tab:micro-tpcc-max-tp} and Figure~\ref{fig:tpcc-region-all} present the evaluation results, with three main takeaways. (1) Under such a high-contention workload, 2PL+Paxos, OCC+Paxos and Tapir all suffer from very low throughput due to frequent transaction aborts.  Among them, 2PL+Paxos performs slightly better because its wound-wait mechanism reduces many transaction aborts. (2) NCC only achieves hundreds of txns/s of throughput, and NCC+'s throughput is even lower (not shown in Figure~\ref{fig:tpcc-region-all}). NCC's poor performance stems not only from aborts but also from high queueing delays caused by its RTC mechanism. The queueing delay leads to a buildup of outstanding transactions, which can easily reach the cap during our open-loop tests, and prevent coordinators from issuing more transactions. (3) Janus, Calvin+ and Detock all benefit from being largely abort-free, as does \sysname. Under TPC-C, Calvin+ becomes less efficient than Janus and Detock, as more shards are involved and the straggler effect becomes more distinct. However, all baselines are less efficient than \sysname's approach based on synchronized clocks, enabling \sysname to achieve the highest throughput and lowest latency.

\subsection{Failure Recovery Evaluation}
\label{sec:failure-recovery-eval}

\begin{figure}[!t]
\centering
\begin{subfigure}{.22\textwidth}
    \centering
    \includegraphics[width=\linewidth]{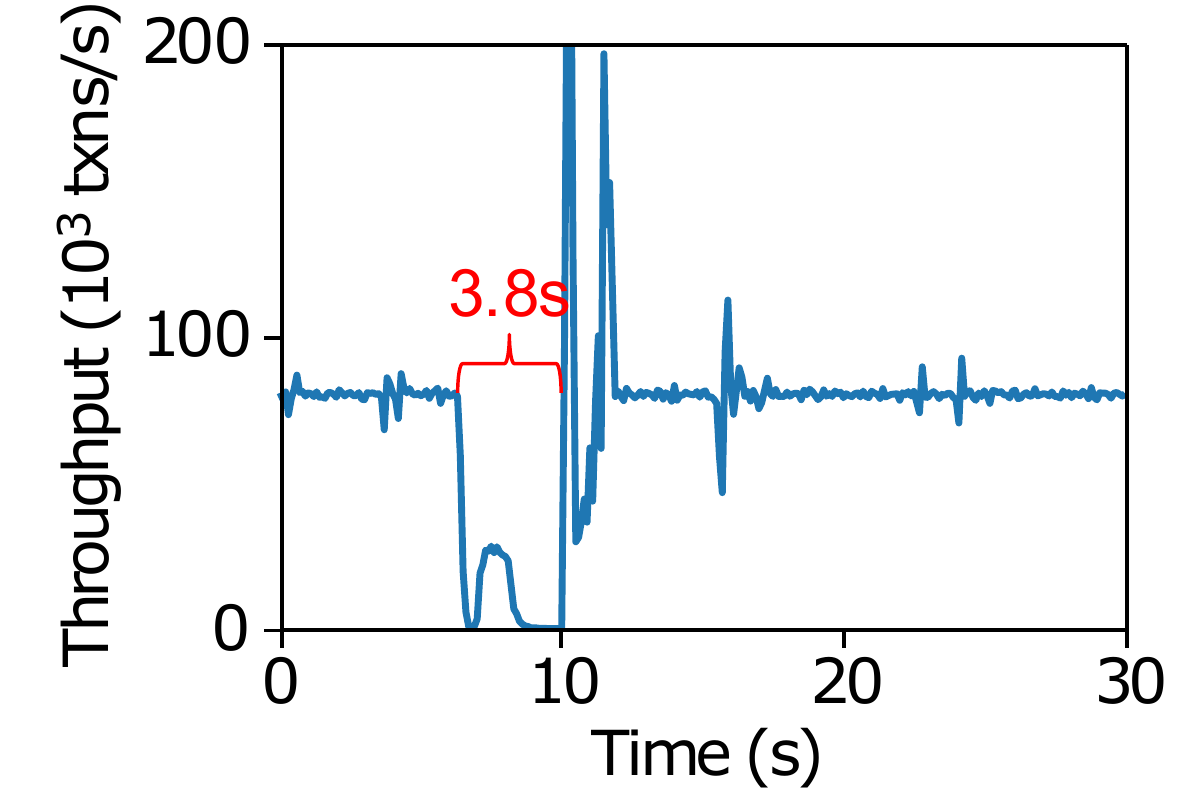} 
    \caption{Total throughput}
    \label{fig:failure-recovery-thpt} 
\end{subfigure}
\begin{subfigure}{.22\textwidth}
    \centering
    \includegraphics[width=\linewidth]{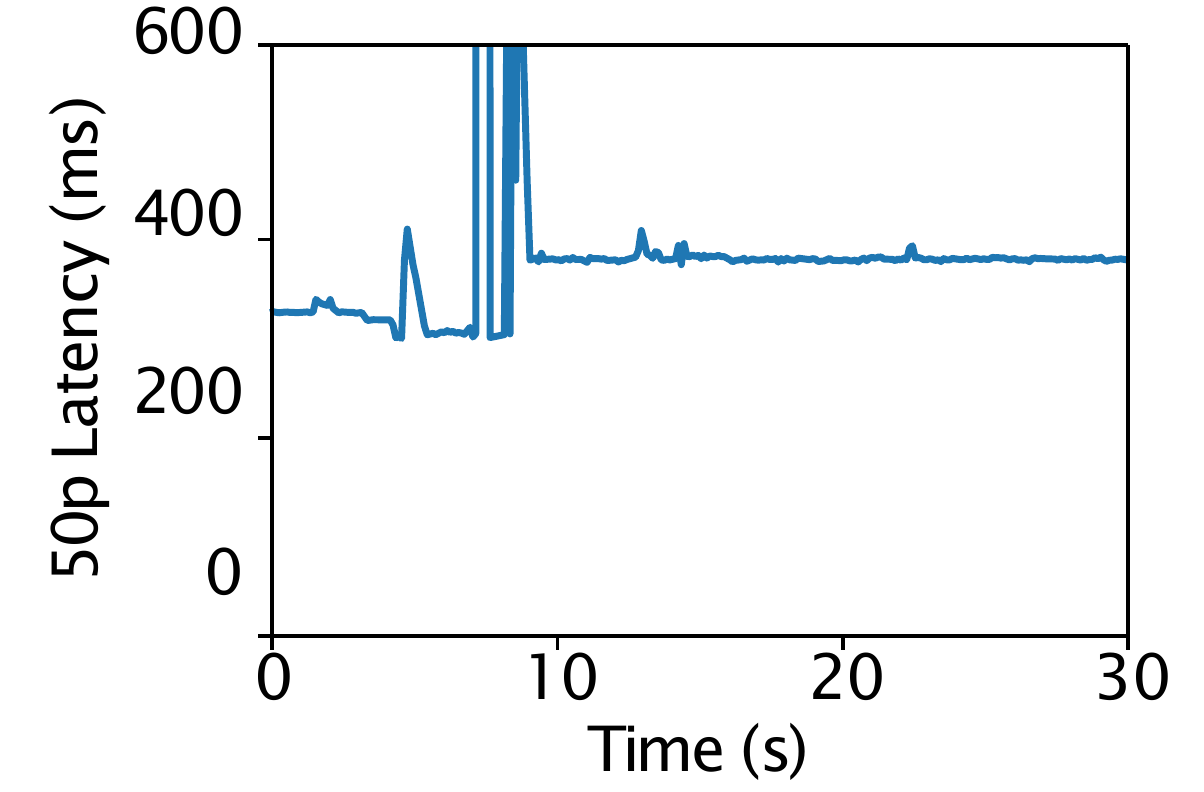}  
    \caption{Latency (Hong Kong)}
    \label{fig:failure-recovery-latency-region-4} 
\end{subfigure}
\caption{\sysname performance before/after leader failure.}
\label{fig:tiga-failure-recovery}
\end{figure}

We re-run MicroBench (skew factor=0.5), and each coordinator submits 10K txns/s (80K txns/s in total). We kill the leader in one shard and compare the performance (latency and throughput) before and after the leader failure. Figure~\ref{fig:failure-recovery-thpt} shows that \sysname takes only 3.8 seconds to complete the global view change and recover to the same level of throughput. After the recovery, the commit latency increases (Figure~\ref{fig:failure-recovery-latency-region-4}) because one of the shards only has $f+1=2$ remaining servers. When transactions involve the data from this shard, they can only be committed in the slow path. However, even in such cases, \sysname's coordinators in the remote region (Hong Kong) still yield lower latency than the other protocols under the same workload (Figure~\ref{fig:Hybrid-Stats-Micro-Remote--zipfian-50-1000000-50p}).

\subsection{Leaders Separation vs. Leaders Colocation}
\label{sec:leader-separate}

\begin{table}[!t]
\centering
\caption{Performance comparison after server rotation.}
\label{tab:micro-max-tp-separate-leaders}
\Large
\begin{adjustbox}{max width=0.5\textwidth}
    \begin{tabular}{c@{\hskip 3pt}ccccccc}
    \hline
   \  & \textbf{2PL+Paxos} & \textbf{OCC+Paxos}  & \textbf{Tapir} & \textbf{Janus} &\textbf{Calvin+} & \textbf{NCC}  & \textbf{Tiga} \\ 
   \hline
   \textbf{Thpt} &  18.6 & 18.0 & 44.7 & 71.9  & 120.0 & 40.7  & 141.9 \\
   \textbf{$+/-\%$} &  -18.8\% & -17.4\% & +1.1\% & -7.5\%  & +0.3\% & -16.5\%  & -9.7\% \\\hline
   \textbf{Latency} & 1.09   & 1.11  & 0.44 & 0.46   & 0.67 & 0.73 & 0.30  
   \\
   \textbf{$+/-\%$} &  +47.2\% & +38.9\% & +83.3\% & +39.3\%  & +162\%  & +72.5\% & +34.0\% \\
   % \textbf{Latency} & 0.74 (4000)  & 0.799 (4000)  & 0.24 (6000) & 0.303 (10000)   & 0.255 (18000)  & 0.198  \\
        \hline
    \end{tabular}
\end{adjustbox}
\caption*{\small Since Detock already distributes the home directories of data items \emph{across regions}, server rotation does not affect its performance. }
\end{table}

\begin{figure}[!t]
\centering
\includegraphics[width=0.85\linewidth]{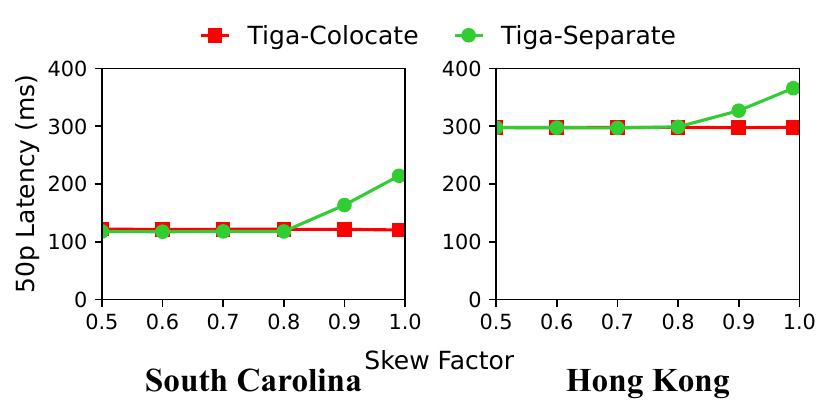}
\caption{MicroBench latency performance with varying skew factors (per-coordinator rate=8K txns/s).}
\label{fig:tiga-cmp-colo-separate}
\end{figure}
When leaders cannot be located in the same region, \sysname prioritizes optimistic execution without waiting for timestamp agreement, to achieve 1-WRTT latency. To evaluate \sysname in this setting, we rotate the \emph{shard-id}s and \emph{replica-id}s for each server so that servers with the same \emph{shard-id} are located in different regions. We continue to run MicroBench (skew factor=0.5). Table~\ref{tab:micro-max-tp-separate-leaders} summarizes the maximum throughput and the 50th percentile latency at this throughput, as well as the relative difference ($+/-\%$) compared to the previous setting (Table~\ref{tab:micro-tpcc-max-tp}) where leaders are co-located.

Table~\ref{tab:micro-max-tp-separate-leaders} indicates that \sysname's throughput decreases by 9.7\%, but it still outperforms the other protocols in both throughput and latency. Calvin+ achieves the highest throughput among baselines, but its latency increases significantly (+162\%) after server rotation because each server needs to collect the epoch messages across regions, costing additional WAN overhead and exacerbating the straggler problem. 

Figure~\ref{fig:tiga-cmp-colo-separate} compares \sysname's performance in the two settings with varying skew factors, represented as \sysname-Separate and \sysname-Colocate. \sysname-Separate incurs higher latency than \sysname-Colocate, as the skew factor (i.e., contention) increases. This is because \sysname-Separate involves more complexity to manage transactions; some transactions also require an additional WRTT to roll back when the execution results prove to be non-serializable. Even so, \sysname-Separate still achieves much lower latency than the other protocols.

\subsection{Sensitivity Analysis of Headroom}
\label{sec:headroom-tradeoff}
\begin{figure}[!t]
\centering
\includegraphics[width=\linewidth]{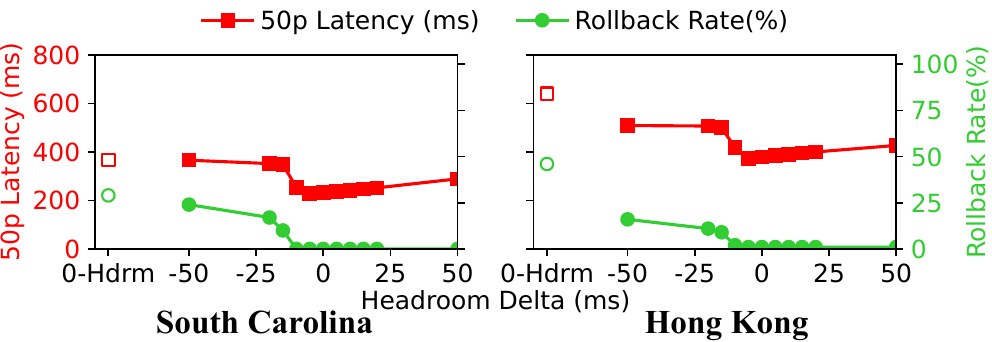}
\caption{\sysname performance with varying headroom (per-coordinator rate=8K txns/s), 0-Hdrm (i.e., headroom=\SI{0}{\milli\second}) directly uses sending time for proactive ordering.}
\label{fig:tiga-headroom}
\end{figure}

To evaluate the impact of headroom on \sysname's performance, we run MicroBench (skew factor=0.99) with leaders separated in different regions. \sysname continues to use the approach in \S\ref{sec:timestamp-initilization} to estimate the headroom for transactions, but we further adjust the headroom by adding different offsets (Headroom Delta in Figure~\ref{fig:tiga-headroom}), ranging from \SI{-50}{\milli\second} to \SI{50}{\milli\second}. We study \sysname's latency and rollback rate. As shown in Figure~\ref{fig:tiga-headroom}, \sysname's estimation approach (Headroom Delta=\SI{0}{\milli\second}) yields a headroom that is close to optimal: Reducing headroom incurs more rollback and worse latency; increasing headroom eliminates rollback but still prolongs latency because transactions are held unnecessarily long at servers. We also evaluate a baseline that uses the sending time directly (0-Hdrm in Figure~\ref{fig:tiga-headroom}). This approach yields the worst latency and rollback rate, as it cannot tolerate network message reordering (illustrated in Figure~\ref{fig:janus-reorder}), thereby highlighting the effectiveness of \sysname's headroom estimation based on synchronized clocks.

\subsection{\sysname with Different Clocks}
\label{sec:ablation-study}

\begin{table}[!t]
\centering
\caption{Throughput and clock synchronization errors with different clocks.}
\label{tab:micro-max-tp-cmp-clocks}
\footnotesize
\begin{adjustbox}{max width=0.5\textwidth}
    \begin{tabular}{ccccc}
    \hline
    \textbf{\ \ }  &  \textbf{Ntpd} & \textbf{Chrony}  & \textbf{Huygens}  & 
    \textbf{Bad-Clock} \\ \hline
    \textbf{Thpt ($10^3$ txns/s)} & 156.8 & 157.1   & 158.1 & 154.7 \\
        \hline
        
    \textbf{Clock errors (ms)} & 16.45 & 4.54 & 0.012 & 62.55\\
        \hline
        
    \end{tabular}
\end{adjustbox}
\begin{tablenotes}
  \small
  \item The stats of clock synchronization errors are collected by using Huygens' real-time monitor functionality~\cite{huygens-latency-sensei}. 
\end{tablenotes}
\end{table}

\begin{figure}[!t]
\centering
\includegraphics[width=0.9\linewidth]{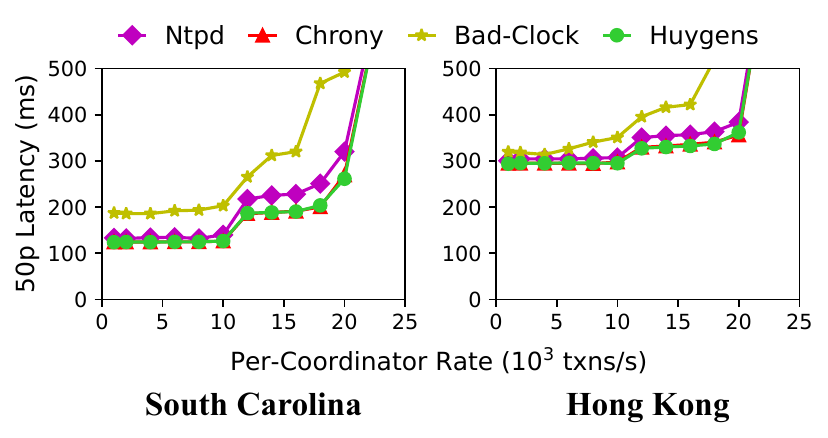}
\caption{\sysname latency using different clocks.}
\label{fig:cmp-clocks}
\end{figure}

To understand the impact of synchronized clocks on \sysname's performance, we conduct an ablation study to compare \sysname's performance with different clocks. We use different synchronization algorithms and design the following variants.

(1) \sysname-Ntpd. We use \texttt{ntpd}, which manages time synchronization in most older Linux distributions in Google Cloud~\cite{gcp-ntp}. We configure \texttt{ntpd} to only use Google's internal NTP server as the reference clock. 

(2) \sysname-Chrony. We use \texttt{chrony}, which is a newer implementation of NTP~\cite{chrony} as well as the current default NTP service in Google Cloud~\cite{gcp-chrony}. We configure \texttt{chrony} to only use Google's internal NTP server as the reference clock. 

(3) \sysname-Huygens. We use the Huygens algorithm to synchronize the clocks for coordinators and servers. 

(4) \sysname-Bad-Clock. We simulate the situation when the NTP service becomes unstable (e.g., due to network congestion and partition) by running a local NTP server as the reference clock. We keep periodically restarting and shutting down the NTP server. In this case, the clock synchronization becomes much worse than the previous variants (Table~\ref{tab:micro-max-tp-cmp-clocks}).

We run MicroBench (skew factor=0.99) to compare the performance of the four \sysname variants (Table~\ref{tab:micro-max-tp-cmp-clocks} and Figure~\ref{fig:cmp-clocks}).

While \texttt{chrony} and Huygens yield different levels of synchronization errors (Table~\ref{tab:micro-max-tp-cmp-clocks}), \sysname's latency remains similar when equipped with either of them. This is because cross-region delays range from \SI{60}{\milli\second} to \SI{150}{\milli\second}; the synchronization error of \texttt{chrony}, though not as good as Huygens, is still negligible compared to the cross-region delay.  As a result, both \texttt{chrony} and Huygens enable \sysname to accurately measure the one-way delay between coordinators and servers and decide a proper timestamp for the transaction at submission. By contrast, \texttt{ntpd}'s synchronization error is larger and causes extra holding time at servers due to the inaccurate measurement of one-way delay. In the worst case, when clocks are poorly synchronized (as in \sysname-Bad-Clock) and the error approaches the one-way delay between regions, \sysname’s latency inflates substantially.

\section{Discussion}
\label{sec:discussion}
\Para{Timestamp initialization.} \sysname initializes transaction timestamps based on the maximum latency from the coordinator to a super quorum of servers in each shard (\S\ref{sec:timestamp-initilization}). This approach aims to increase the likelihood of fast-path commits. However, in certain deployments, the fast path may actually incur higher latency than the slow path. This situation arises when each shard has a simple quorum located close to the coordinator, while the remaining servers are geographically distant. In such scenarios, committing through the slow path may be more efficient. To accommodate this, the coordinator can estimate the latencies for both paths and then choose whether to use a super quorum or a simple quorum, based on which option can yield better performance.

\Para{Dynamic sharding.} Dynamic sharding~\cite{eurosys21-zeus,vldb19-slog,osdi18-ushard,osdi16-auto-sharding} allows OLTP systems to distribute heavy-hitter keys and co-locate frequently accessed data. We believe it could further enhance \sysname's performance, and we plan to support it in future versions. Because single-shard transactions do not require timestamp agreement, dynamic sharding can convert multi-shard transactions into single-shard transactions, thereby reducing the overhead of timestamp agreement and rollback.

\begin{figure}[!t]
\centering
\includegraphics[width=0.95\linewidth]{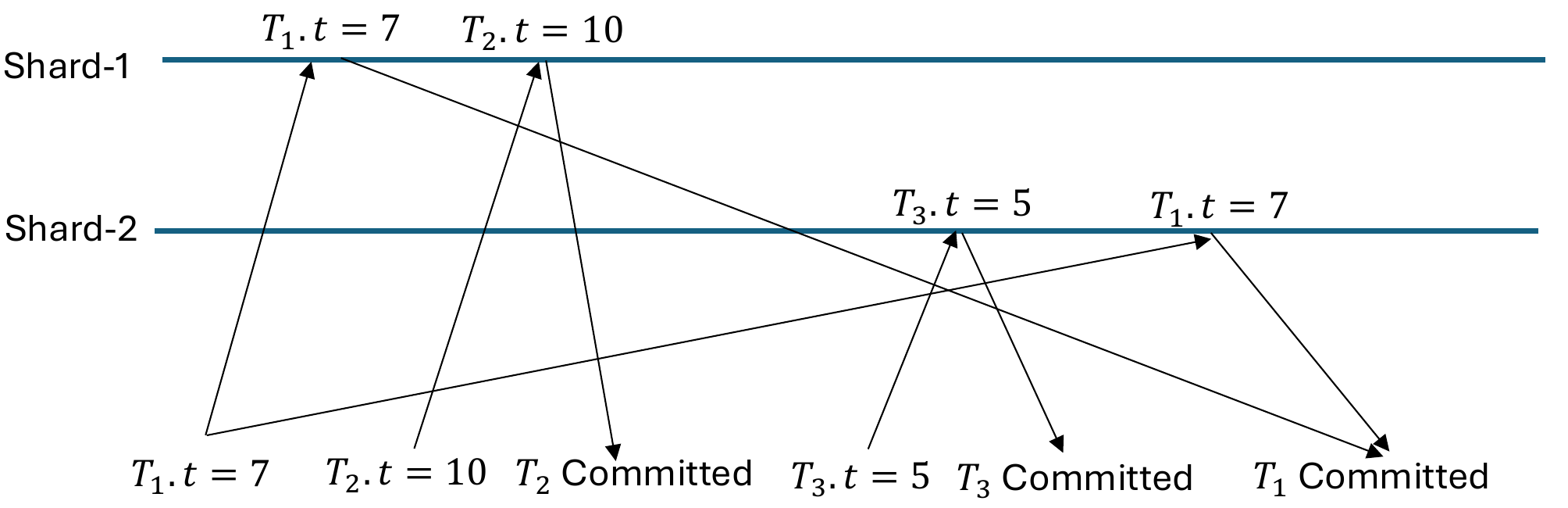}
\caption{Undetectable timestamp inversion without inter-shard (leader) coordination.}
\label{fig:timestamp-inversion-undetect}
\vspace{-0.5cm}
\end{figure}

\Para{Clock accuracy and timestamp inversion.} To prevent timestamp inversion and ensure strict serializability, \sysname introduces timestamp agreement (\S\ref{sec:timestamp-agreement}), which requires leaders of different shards to coordinate and confirm that their transactions respect real-time ordering. Without this inter-leader coordination, a shard cannot detect timestamp inversion when it occurs. Figure~\ref{fig:timestamp-inversion-undetect} illustrates such a case: Shard-1’s servers run with faster clocks than Shard-2’s. As a result, Shard-1 commits a single-shard transaction $T_2$ with a larger timestamp ($T_2.t=10$), while Shard-2 later commits another single-shard transaction $T_3$ with a smaller timestamp ($T_3.t=5$). Although $T_2$ and $T_3$ are processed independently, both conflict with the multi-shard transaction $T_1$. This yields a serializable schedule $T_2 \rightarrow T_1 \rightarrow T_3$, which contradicts the real-time order $T_2 \rightarrow T_3$. Since all transactions arrive at their shards before their assigned timestamps, both shards treat them as valid, leaving the timestamp inversion undetected. To avoid such violations of strict serializability, the shards (leaders) must coordinate.

However, such inter-leader coordination incurs 0.5--1 RTT of blocking latency for subsequent transactions: if the transaction at the head of the priority queue has not completed timestamp agreement, any conflicting transactions behind it cannot be executed or released. This blocking latency can be costly when leaders are distributed across regions and workloads exhibit high contention. This raises a natural question: \emph{Can we avoid coordination by leveraging synchronized clocks?}

In fact, if we could assume synchronized clocks with a bounded error $\epsilon$, \sysname can eliminate inter-leader coordination while still avoiding timestamp inversion. The coordination-free approach works as follows:

(1) Each leader updates an incoming transaction’s timestamp to its local clock time if the initial timestamp is smaller.

(2) Each leader defers the release of the transaction $T$ until its local clock exceeds $T.t + \epsilon$, ensuring that all leaders’ clocks have passed $T.t$ before $T$ is released.

Then, we revisit the example in Figure~\ref{fig:timestamp-inversion-undetect}. Suppose the local clock time of Shard-1’s leader is $clock_1$. Then the local clock time of Shard-2’s leader $clock_2 \in [clock_1 - \epsilon, clock_1 + \epsilon]$. When Shard-1 receives $T_2$, it defers release until $clock_1 > T_2.t + \epsilon$, ensuring that every shard’s clock has already passed $T_2.t$. Meanwhile, when Shard-2 receives $T_3$ and $T_1$, it updates their timestamps if they are smaller than $clock_2$. Two outcomes follow: (1) if $\epsilon \to 0$, Shard-2 updates the timestamps for both $T_3$ and $T_1$ to values greater than $T_2.t=10$, yielding the order $T_2 \rightarrow T_3 \rightarrow T_1$; (2) if $\epsilon \to \infty$, Shard-1 defers $T_2$’s release until after Shard-2 releases $T_3$ and $T_1$, yielding the order $T_3 \rightarrow T_1 \rightarrow T_2$. In both cases, the serializable order remains consistent with the real-time order. Thus, the prior knowledge of $\epsilon$ provides a straightforward way to prevent timestamp inversion without inter-leader coordination, allowing more transactions to commit in 1 RTT.

We do not assume a deterministic error bound in \sysname's design due to the probabilistic nature of Huygens. Nonetheless, several clock synchronization systems provide deterministic guarantees. For example, Spanner~\cite{spanner} achieves millisecond-level error bounds, and Sundial~\cite{osdi20-sundial} further reduces them to $\sim$\SI{100}{\nano\second}. While such synchronization requires specialized hardware, we expect these solutions to become increasingly deployable in the future, offering promising opportunities for \sysname to preserve strict serializability more efficiently.

\section{Related Work}
\label{sec:related-work}

Table~\ref{tab:cmp-protocols} compares \sysname with state-of-the-art protocols. While several existing protocols can achieve 1-WRTT commit latency, this optimal performance typically holds only under narrow conditions---such as co-locating servers and/or coordinators. Moreover, they often sacrifice correctness guarantees or incur costly aborts. In contrast, \sysname achieves 1-WRTT latency in more general deployments, and ensures strict serializability with few or no transaction aborts.\footnote{\sysname is abort-free for one-shot transactions when leaders are co-located.}

\Para{Ordering guarantees in multicast.} Several network primitives have been proposed to accelerate distributed protocols. Ordered Unreliable Multicast (OUM)~\cite{osdi16-nopaxos} and Multi-Sequencing Groupcast (MSG)~\cite{sosp17-eris} both leverage a single sequencer to establish ordering, which can incur centralized bottlenecks with a software-based sequencer in cloud settings. Hydra~\cite{nsdi23-hydra} extends OUM and MSG with multiple sequencers. However, it requires all sequencers to continually send flush messages to receivers. The slowdown of any sequencer can impede the progress of all receivers. \sysname's design is inspired by the deadline-ordered multicast (DOM) primitive of Nezha~\cite{vldb23-nezha}, but DOM does not consider inter-shard timestamp agreement of transactions. Moreover, Nezha, as a pure consensus protocol, cannot be easily extended to work in the multi-shard setting that \sysname targets.

\Para{\sysname vs. Mako.} The recent protocol, Mako~\cite{osdi25-warbler}, advocates for decoupling consensus and concurrency control to improve throughput. In contrast, \sysname prioritizes latency optimization. Accordingly, we argue that a consolidated design is better suited for minimizing transaction latency. In geo-distributed settings, Mako needs multiple WRTTs to commit transactions when they are issued from followers, or when leaders are not co-located. In contrast, \sysname can consistently commit transactions in the 1-WRTT fast path.

\newcommand{\mycell}[2]{\multirow{2}{*}{\parbox{#1}{\centering #2} } }
\newcommand{\red}[1]{\textcolor{red}{#1}}
\definecolor{darkgreen}{rgb}{0.0, 0.8, 0.0}
\newcommand{\green}[1]{\textcolor{darkgreen}{#1}}
\newcommand{\orange}[1]{\textcolor{orange}{#1}}

\begin{table}[!t]
\centering
\caption{Summary of protocol comparison.}
\label{tab:cmp-protocols}
\footnotesize
\begin{adjustbox}{max width=\textwidth}
    \begin{tabular}{
    c@{}cc@{}>{\centering\arraybackslash}p{0.6cm}>{\centering\arraybackslash}p{0.6cm}
    >{\centering\arraybackslash}p{2cm}}
    \hline
\mycell{1cm}{System}    
& \mycell{1cm}{Consistency}  
& \mycell{0.8cm}{Aborts}   
& \multicolumn{2}{>{\centering\arraybackslash}p{1.5cm}}{WRTTs\tnote{a}} 
& \mycell{3cm}{Require co-locating \\ leaders for best latency?}  \\
\cline{4-5}
& &  & Best & Worst &  \\
\hline  
% ROW-1 Complete

Spanner~\cite{spanner} &
\green{Strict Ser.} &
\red{High} &
\red{3}  &
\red{$\geq$4}  &
\red{Required}  
\\ 
% ROW-2 Complete
AOCC~\cite{sigmod95-clocc} &
\green{Strict Ser.}&
\red{High} &
\red{2} &
\red{$\geq$3}  &
\green{Not Required} 
\\ 
% ROW-3 Complete

MVTO~\cite{tocs-mvto} &
\red{Ser.}&
\orange{Med} &
\red{2}  &
\red{$\geq$3}  &
\green{Not Required}
\\ 
% ROW-3 Complete

MDCC~\cite{eurosys13-mdcc} &
\red{Ser.} &
\red{High} &
\red{2} &
\red{$\geq$3} &
\red{Required} 
\\ 
% ROW-3 Complete

Calvin~\cite{sigmod12-calvin} &
\green{Strict Ser.} &
\green{None} &
\red{2} &
\orange{2.5} &
\red{Required} 
\\ 
% ROW-4 Complete

Tapir~\cite{sosp15-tapir} &
% \red{Ser. (-~\cite{vldb25-verlso})} &
\red{Ser.} &
\red{High} &
\green{1} &
\red{$\geq$2} &
\green{Not Required} 
\\ 
% ROW-4 Complete

Janus~\cite{osdi16-janus} &
\green{Strict Ser.} &
\green{None} &
\red{2} &
\red{3} &
\red{Required} 
\\ 
% ROW-5 Complete

OceanVista~\cite{vldb19-ocean-vista} &
\green{Strict Ser.} &
\green{None} &
\red{2} &
\orange{2.5} &
\red{Required} 
\\ 
% ROW-6 Complete

Natto~\cite{sigmod22-natto} &
\green{Strict Ser.} &
\orange{Med} &
\red{2} &
\red{$\geq$3} &
\green{Not Required} 
\\ 
% ROW-6 Complete

Detock~\cite{sigmod23-detock} &
\green{Strict Ser.} &
\green{None} &
\red{2} &
\orange{2.5} &
\red{Required} 
\\ 
% ROW-6 Complete

NCC~\cite{osdi23-ncc} &
\green{Strict Ser.} &
\orange{Med} &
\red{2} &
\red{$\geq$3} &
\red{Required} 
\\ 
% ROW-6 Complete

Mako~\cite{osdi25-warbler} &
\green{Strict Ser.} &
\orange{Med} &
\red{2} &
\red{$\geq$5} &
\red{Required} 
\\ 
% ROW-6 Complete

\sysname &
\green{Strict Ser.} &
\green{None} &
\green{1} &
\orange{2} &
\green{Not Required} 
\\ \hline
% ROW-6 Complete

\end{tabular}
\end{adjustbox}

\begin{tablenotes}
  \footnotesize
  % \item 1. The two columns of Min WRTTs represent the minimum number of WRTTs required by the system commit one transaction, under its best deployment (i.e., co-locating servers/coordinators) and its worst deployment, respectively.
  \item We discuss the commit latency for each system assuming no co-location requirement between coordinators and servers. For AOCC, MVTO and NCC, we assume they achieve geo-distributed fault tolerance via replication. Some systems incur $\geq x$ WRTTs because of aborts and retries.  
\end{tablenotes}
\end{table}

\section{Conclusion}
\label{sec:conclusion}

The rapid advancement of accurate clock synchronization enables new protocols that exploit timestamp ordering to accelerate geo-distributed transaction processing. In this paper, we have presented the design, implementation, and evaluation of \sysname, a consolidated protocol that uses synchronized clocks to proactively order transactions at predesignated timestamps and efficiently resolve inconsistencies among servers. Compared with conventional layered designs (e.g., 2PL/OCC+Paxos, Calvin+, Detock, NCC) and state-of-the-art consolidated designs (e.g., Tapir and Janus), \sysname can achieve significantly higher throughput and lower latency.

% We have presented the design, implementation, and evaluation of \sysname, a new consolidated protocol for geo-distributed transactional systems. 
% \sysname leverages synchronized clocks to proactively order transactions at pre-designated timestamps and efficiently resolve inconsistencies among servers. 
% Compared with the conventional layered designs (e.g., 2PL/OCC+Paxos, Calvin+, Detock, NCC) and the state-of-the-art consolidated designs (e.g., Tapir and Janus), \sysname can achieve significantly higher throughput and lower latency.

This work does not raise any ethical issues.

\section*{Acknowledgments}
We thank the anonymous reviewers for suggestions that improved our work. This project was funded in part by NSF
awards CNS-2321725, CNS-2238768, CNS-2130590 and NSF CAREER award 2340748. We also appreciate the support from Google Cloud Research Credits program.

\balance

\bibliographystyle{ACM-Reference-Format}
\bibliography{tiga-ref}


\begin{thebibliography}{97}

%%% the Web address.

\ifx \showCODEN    \undefined \def \showCODEN     #1{\unskip}     \fi
\ifx \showDOI      \undefined \def \showDOI       #1{#1}\fi
\ifx \showISBNx    \undefined \def \showISBNx     #1{\unskip}     \fi
\ifx \showISBNxiii \undefined \def \showISBNxiii  #1{\unskip}     \fi
\ifx \showISSN     \undefined \def \showISSN      #1{\unskip}     \fi
\ifx \showLCCN     \undefined \def \showLCCN      #1{\unskip}     \fi
\ifx \shownote     \undefined \def \shownote      #1{#1}          \fi
\ifx \showarticletitle \undefined \def \showarticletitle #1{#1}   \fi
\ifx \showURL      \undefined \def \showURL       {\relax}        \fi
% invisible to TeX
\providecommand\bibfield[2]{#2}
\providecommand\bibinfo[2]{#2}
\providecommand\natexlab[1]{#1}
\providecommand\showeprint[2][]{arXiv:#2}

\bibitem[\protect\citeauthoryear{Adya}{Adya}{1999}]%
        {adya1999weak}
\bibfield{author}{\bibinfo{person}{Atul Adya}.} \bibinfo{year}{1999}\natexlab{}.
\newblock \bibinfo{booktitle}{{\em Weak Consistency: A Generalized Theory and Optimistic Implementations for Distributed Transactions}}.
\newblock \bibinfo{type}{{T}echnical {R}eport}. \bibinfo{address}{USA}.
\newblock
\showURL{%
\url{https://hdl.handle.net/1721.1/149899}}


\bibitem[\protect\citeauthoryear{Adya, Gruber, Liskov, and Maheshwari}{Adya et~al\mbox{.}}{1995}]%
        {sigmod95-clocc}
\bibfield{author}{\bibinfo{person}{Atul Adya}, \bibinfo{person}{Robert Gruber}, \bibinfo{person}{Barbara Liskov}, {and} \bibinfo{person}{Umesh Maheshwari}.} \bibinfo{year}{1995}\natexlab{}.
\newblock \showarticletitle{Efficient Optimistic Concurrency Control Using Loosely Synchronized Clocks}.
\newblock \bibinfo{journal}{{\em SIGMOD Record\/}} \bibinfo{volume}{24}, \bibinfo{number}{2} (\bibinfo{date}{May} \bibinfo{year}{1995}), \bibinfo{pages}{23--34}.
\newblock
\showDOI{%
\url{https://doi.org/10.1145/568271.223787}}


\bibitem[\protect\citeauthoryear{Adya, Myers, Howell, Elson, Meek, Khemani, Fulger, Gu, Bhuvanagiri, Hunter, Peon, Kai, Shraer, Merchant, and Lev-Ari}{Adya et~al\mbox{.}}{2016}]%
        {osdi16-auto-sharding}
\bibfield{author}{\bibinfo{person}{Atul Adya}, \bibinfo{person}{Daniel Myers}, \bibinfo{person}{Jon Howell}, \bibinfo{person}{Jeremy Elson}, \bibinfo{person}{Colin Meek}, \bibinfo{person}{Vishesh Khemani}, \bibinfo{person}{Stefan Fulger}, \bibinfo{person}{Pan Gu}, \bibinfo{person}{Lakshminath Bhuvanagiri}, \bibinfo{person}{Jason Hunter}, \bibinfo{person}{Roberto Peon}, \bibinfo{person}{Larry Kai}, \bibinfo{person}{Alexander Shraer}, \bibinfo{person}{Arif Merchant}, {and} \bibinfo{person}{Kfir Lev-Ari}.} \bibinfo{year}{2016}\natexlab{}.
\newblock \showarticletitle{Slicer: Auto-Sharding for Datacenter Applications}. In \bibinfo{booktitle}{{\em Proceedings of the 12th USENIX Symposium on Operating Systems Design and Implementation (OSDI 2016)}}. \bibinfo{publisher}{USENIX Association}, \bibinfo{address}{Savannah, GA}, \bibinfo{pages}{739--753}.
\newblock
\showISBNx{978-1-931971-33-1}
\showURL{%
\url{https://www.usenix.org/conference/osdi16/technical-sessions/presentation/adya}}


\bibitem[\protect\citeauthoryear{Al-Kuwari, Davenport, and Bradford}{Al-Kuwari et~al\mbox{.}}{2011}]%
        {incremental-hash-trend}
\bibfield{author}{\bibinfo{person}{Saif Al-Kuwari}, \bibinfo{person}{James~H. Davenport}, {and} \bibinfo{person}{Russell~J. Bradford}.} \bibinfo{year}{2011}\natexlab{}.
\newblock \bibinfo{title}{Cryptographic Hash Functions: Recent Design Trends and Security Notions}.
\newblock   (\bibinfo{year}{2011}).
\newblock
\showURL{%
\url{https://eprint.iacr.org/2011/565.pdf}}


\bibitem[\protect\citeauthoryear{Alomari, Cahill, Fekete, and Rohm}{Alomari et~al\mbox{.}}{2008}]%
        {smallbank}
\bibfield{author}{\bibinfo{person}{Mohammad Alomari}, \bibinfo{person}{Michael Cahill}, \bibinfo{person}{Alan Fekete}, {and} \bibinfo{person}{Uwe Rohm}.} \bibinfo{year}{2008}\natexlab{}.
\newblock \showarticletitle{The Cost of Serializability on Platforms That Use Snapshot Isolation}. In \bibinfo{booktitle}{{\em 2008 IEEE 24th International Conference on Data Engineering}}. \bibinfo{pages}{576--585}.
\newblock
\showDOI{%
\url{https://doi.org/10.1109/ICDE.2008.4497466}}


\bibitem[\protect\citeauthoryear{{Andy Pavlo and et al.}}{{Andy Pavlo and et al.}}{[n. d.]a}]%
        {auctionMark-bench}
\bibfield{author}{\bibinfo{person}{{Andy Pavlo and et al.}}} \bibinfo{year}{[n. d.]}\natexlab{a}.
\newblock \bibinfo{title}{{AuctionMark Benchmark}}.
\newblock \bibinfo{howpublished}{\url{https://hstore.cs.brown.edu/projects/auctionmark}}.   (\bibinfo{year}{[n. d.]}).
\newblock


\bibitem[\protect\citeauthoryear{{Andy Pavlo and et al.}}{{Andy Pavlo and et al.}}{[n. d.]b}]%
        {Seats}
\bibfield{author}{\bibinfo{person}{{Andy Pavlo and et al.}}} \bibinfo{year}{[n. d.]}\natexlab{b}.
\newblock \bibinfo{title}{{SEATS Benchmark}}.
\newblock \bibinfo{howpublished}{\url{https://github.com/apavlo/h-store/tree/master/src/benchmarks/edu/brown/benchmark/seats/}}.   (\bibinfo{year}{[n. d.]}).
\newblock


\bibitem[\protect\citeauthoryear{Annamalai, Ravichandran, Srinivas, Zinkovsky, Pan, Savor, Nagle, and Stumm}{Annamalai et~al\mbox{.}}{2018}]%
        {osdi18-ushard}
\bibfield{author}{\bibinfo{person}{Muthukaruppan Annamalai}, \bibinfo{person}{Kaushik Ravichandran}, \bibinfo{person}{Harish Srinivas}, \bibinfo{person}{Igor Zinkovsky}, \bibinfo{person}{Luning Pan}, \bibinfo{person}{Tony Savor}, \bibinfo{person}{David Nagle}, {and} \bibinfo{person}{Michael Stumm}.} \bibinfo{year}{2018}\natexlab{}.
\newblock \showarticletitle{Sharding the Shards: Managing Datastore Locality at Scale with Akkio}. In \bibinfo{booktitle}{{\em Proceedings of the 13th USENIX Symposium on Operating Systems Design and Implementation (OSDI 2018)}}. \bibinfo{publisher}{USENIX Association}, \bibinfo{address}{Carlsbad, CA}, \bibinfo{pages}{445--460}.
\newblock
\showISBNx{978-1-939133-08-3}
\showURL{%
\url{https://www.usenix.org/conference/osdi18/presentation/annamalai}}


\bibitem[\protect\citeauthoryear{{Apache Software Foundation}}{{Apache Software Foundation}}{2021}]%
        {zookeeper}
\bibfield{author}{\bibinfo{person}{{Apache Software Foundation}}.} \bibinfo{year}{2021}\natexlab{}.
\newblock \bibinfo{title}{ZooKeeper}.
\newblock \bibinfo{howpublished}{\url{https://zookeeper.apache.org}}.   (\bibinfo{year}{2021}).
\newblock
\newblock
\shownote{Accessed: 2025-08-31.}


\bibitem[\protect\citeauthoryear{Armstrong, Ponnekanti, Borthakur, and Callaghan}{Armstrong et~al\mbox{.}}{2013}]%
        {sigmod13-linkbench}
\bibfield{author}{\bibinfo{person}{Timothy~G. Armstrong}, \bibinfo{person}{Vamsi Ponnekanti}, \bibinfo{person}{Dhruba Borthakur}, {and} \bibinfo{person}{Mark Callaghan}.} \bibinfo{year}{2013}\natexlab{}.
\newblock \showarticletitle{LinkBench: A Database Benchmark Based on the Facebook Social Graph}. In \bibinfo{booktitle}{{\em Proceedings of the 2013 ACM SIGMOD International Conference on Management of Data}} {\em (\bibinfo{series}{SIGMOD '13})}. \bibinfo{publisher}{Association for Computing Machinery}, \bibinfo{address}{New York, NY, USA}, \bibinfo{pages}{1185–1196}.
\newblock
\showISBNx{9781450320375}
\showDOI{%
\url{https://doi.org/10.1145/2463676.2465296}}


\bibitem[\protect\citeauthoryear{{AWS}}{{AWS}}{2019}]%
        {aws-shard}
\bibfield{author}{\bibinfo{person}{{AWS}}.} \bibinfo{year}{2019}\natexlab{}.
\newblock \bibinfo{title}{Global Tables: Multi-Region Replication for DynamoDB}.
\newblock \bibinfo{howpublished}{\url{https://docs.aws.amazon.com/amazondynamodb/latest/developerguide/GlobalTables.html}}.   (\bibinfo{year}{2019}).
\newblock
\newblock
\shownote{Accessed: 2025-08-31.}


\bibitem[\protect\citeauthoryear{{AWS}}{{AWS}}{2024}]%
        {aws-clock-sync}
\bibfield{author}{\bibinfo{person}{{AWS}}.} \bibinfo{year}{2024}\natexlab{}.
\newblock \bibinfo{title}{{Amazon Time Sync Service expands Microsecond-Accurate time to 87 additonal EC2 instance types}}.
\newblock \bibinfo{howpublished}{\url{https://aws.amazon.com/about-aws/whats-new/2024/04/amazon-time-sync-service-microsecond-accurate-time-additonal-ec2-instance-types/}}.   (\bibinfo{year}{2024}).
\newblock
\newblock
\shownote{Accessed: 08/31/2024.}


\bibitem[\protect\citeauthoryear{{Baidu}}{{Baidu}}{2022}]%
        {server-push}
\bibfield{author}{\bibinfo{person}{{Baidu}}.} \bibinfo{year}{2022}\natexlab{}.
\newblock \bibinfo{title}{{Server Push}}.
\newblock \bibinfo{howpublished}{\url{https://brpc.apache.org/docs/server/server-push/}}.   (\bibinfo{year}{2022}).
\newblock
\newblock
\shownote{Accessed: 2025-08-31.}


\bibitem[\protect\citeauthoryear{Bellare, Goldreich, and Goldwasser}{Bellare et~al\mbox{.}}{1994}]%
        {incremental-hash-sign}
\bibfield{author}{\bibinfo{person}{Mihir Bellare}, \bibinfo{person}{Oded Goldreich}, {and} \bibinfo{person}{Shafi Goldwasser}.} \bibinfo{year}{1994}\natexlab{}.
\newblock \showarticletitle{Incremental Cryptography: The Case of Hashing and Signing}. In \bibinfo{booktitle}{{\em Proceedings of the 14th Annual International Cryptology Conference on Advances in Cryptology}} {\em (\bibinfo{series}{CRYPTO '94})}. \bibinfo{publisher}{Springer-Verlag}, \bibinfo{address}{Berlin, Heidelberg}, \bibinfo{pages}{216–233}.
\newblock
\showISBNx{3540583335}


\bibitem[\protect\citeauthoryear{Bellare, Goldreich, and Goldwasser}{Bellare et~al\mbox{.}}{1995a}]%
        {stoc95-incremental-hash-virus}
\bibfield{author}{\bibinfo{person}{Mihir Bellare}, \bibinfo{person}{Oded Goldreich}, {and} \bibinfo{person}{Shafi Goldwasser}.} \bibinfo{year}{1995}\natexlab{a}.
\newblock \showarticletitle{Incremental Cryptography and Application to Virus Protection}. In \bibinfo{booktitle}{{\em Proceedings of the 27th Annual ACM Symposium on Theory of Computing (STOC 1995)}}.
\newblock


\bibitem[\protect\citeauthoryear{Bellare, Guérin, and Rogaway}{Bellare et~al\mbox{.}}{1995b}]%
        {crpyto95-xormac}
\bibfield{author}{\bibinfo{person}{Mihir Bellare}, \bibinfo{person}{Roch Guérin}, {and} \bibinfo{person}{Phillip Rogaway}.} \bibinfo{year}{1995}\natexlab{b}.
\newblock \showarticletitle{XOR MACs: New Methods for Message Authentication Using Finite Pseudorandom Functions}. In \bibinfo{booktitle}{{\em Proceedings of the Annual International Cryptology Conference (CRYPTO 1995)}}.
\newblock


\bibitem[\protect\citeauthoryear{Bellare and Rogaway}{Bellare and Rogaway}{1997}]%
        {crypto97-incremental-hash}
\bibfield{author}{\bibinfo{person}{Mihir Bellare} {and} \bibinfo{person}{Phillip Rogaway}.} \bibinfo{year}{1997}\natexlab{}.
\newblock \showarticletitle{Collision-Resistant Hashing: Towards Making UOWHFs Practical}. In \bibinfo{booktitle}{{\em Proceedings of the Annual International Cryptology Conference (CRYPTO 1997)}}.
\newblock


\bibitem[\protect\citeauthoryear{Bernstein, Hadzilacos, and Goodman}{Bernstein et~al\mbox{.}}{1987}]%
        {bernstein-book}
\bibfield{author}{\bibinfo{person}{Philip~A. Bernstein}, \bibinfo{person}{Vassos Hadzilacos}, {and} \bibinfo{person}{Nathan Goodman}.} \bibinfo{year}{1987}\natexlab{}.
\newblock \bibinfo{booktitle}{{\em Concurrency Control and Recovery in Database Systems}}.
\newblock \bibinfo{publisher}{Addison-Wesley}, \bibinfo{address}{Reading, MA, USA}.
\newblock


\bibitem[\protect\citeauthoryear{Bernstein and Newcomer}{Bernstein and Newcomer}{2009}]%
        {berstein-chapter}
\bibfield{author}{\bibinfo{person}{Philip~A. Bernstein} {and} \bibinfo{person}{Eric Newcomer}.} \bibinfo{year}{2009}\natexlab{}.
\newblock \showarticletitle{Chapter 6: Locking}.
\newblock In \bibinfo{booktitle}{{\em Principles of Transaction Processing (Second Edition)}}. \bibinfo{publisher}{Morgan Kaufmann}, \bibinfo{address}{Burlington, MA, USA}.
\newblock


\bibitem[\protect\citeauthoryear{Cahill, R\"{o}hm, and Fekete}{Cahill et~al\mbox{.}}{2009}]%
        {tods-sibench}
\bibfield{author}{\bibinfo{person}{Michael~J. Cahill}, \bibinfo{person}{Uwe R\"{o}hm}, {and} \bibinfo{person}{Alan~D. Fekete}.} \bibinfo{year}{2009}\natexlab{}.
\newblock \showarticletitle{Serializable Isolation for Snapshot Databases}.
\newblock \bibinfo{journal}{{\em ACM Trans. Database Syst.\/}} \bibinfo{volume}{34}, \bibinfo{number}{4}, Article \bibinfo{articleno}{20} (\bibinfo{date}{dec} \bibinfo{year}{2009}), \bibinfo{numpages}{42}~pages.
\newblock
\showISSN{0362-5915}
\showDOI{%
\url{https://doi.org/10.1145/1620585.1620587}}


\bibitem[\protect\citeauthoryear{Chen, Song, Jiang, Ruan, Li, Wang, Zhang, Cheng, and Cui}{Chen et~al\mbox{.}}{2021}]%
        {eurosys21-dast}
\bibfield{author}{\bibinfo{person}{Xusheng Chen}, \bibinfo{person}{Haoze Song}, \bibinfo{person}{Jianyu Jiang}, \bibinfo{person}{Chaoyi Ruan}, \bibinfo{person}{Cheng Li}, \bibinfo{person}{Sen Wang}, \bibinfo{person}{Gong Zhang}, \bibinfo{person}{Reynold Cheng}, {and} \bibinfo{person}{Heming Cui}.} \bibinfo{year}{2021}\natexlab{}.
\newblock \showarticletitle{Achieving Low Tail-Latency and High Scalability for Serializable Transactions in Edge Computing}. In \bibinfo{booktitle}{{\em Proceedings of the 16th European Conference on Computer Systems (EuroSys 2021)}}. \bibinfo{pages}{1--16}.
\newblock
\showDOI{%
\url{https://doi.org/10.1145/3447786.3456238}}


\bibitem[\protect\citeauthoryear{Cheng, Shi, Kabcenell, Lawande, Qadeer, Chan, Tin, Zhao, Bailis, Balakrishnan, Bronson, Crooks, and Stoica}{Cheng et~al\mbox{.}}{2022}]%
        {vldb22-taobench}
\bibfield{author}{\bibinfo{person}{Audrey Cheng}, \bibinfo{person}{Xiao Shi}, \bibinfo{person}{Aaron Kabcenell}, \bibinfo{person}{Shilpa Lawande}, \bibinfo{person}{Hamza Qadeer}, \bibinfo{person}{Jason Chan}, \bibinfo{person}{Harrison Tin}, \bibinfo{person}{Ryan Zhao}, \bibinfo{person}{Peter Bailis}, \bibinfo{person}{Mahesh Balakrishnan}, \bibinfo{person}{Nathan Bronson}, \bibinfo{person}{Natacha Crooks}, {and} \bibinfo{person}{Ion Stoica}.} \bibinfo{year}{2022}\natexlab{}.
\newblock \showarticletitle{TAOBench: An End-to-End Benchmark for Social Network Workloads}.
\newblock \bibinfo{journal}{{\em Proc. VLDB Endow.\/}} \bibinfo{volume}{15}, \bibinfo{number}{9} (\bibinfo{date}{jul} \bibinfo{year}{2022}), \bibinfo{pages}{1965–1977}.
\newblock
\showISSN{2150-8097}
\showDOI{%
\url{https://doi.org/10.14778/3538598.3538616}}


\bibitem[\protect\citeauthoryear{Choi, Michael, Li, Ports, and Li}{Choi et~al\mbox{.}}{2023}]%
        {nsdi23-hydra}
\bibfield{author}{\bibinfo{person}{Inho Choi}, \bibinfo{person}{Ellis Michael}, \bibinfo{person}{Yunfan Li}, \bibinfo{person}{Dan Ports}, {and} \bibinfo{person}{Jialin Li}.} \bibinfo{year}{2023}\natexlab{}.
\newblock \showarticletitle{Hydra: Serialization-Free Network Ordering for Strongly Consistent Distributed Applications}. In \bibinfo{booktitle}{{\em Proceedings of the 20th USENIX Conference on Networked Systems Design and Implementation (NSDI 2023)}}. \bibinfo{pages}{1--16}.
\newblock
\showURL{%
\url{https://www.usenix.org/conference/nsdi23/presentation/choi}}


\bibitem[\protect\citeauthoryear{{Chrony Team}}{{Chrony Team}}{2024}]%
        {chrony}
\bibfield{author}{\bibinfo{person}{{Chrony Team}}.} \bibinfo{year}{2024}\natexlab{}.
\newblock \bibinfo{title}{Chrony}.
\newblock \bibinfo{howpublished}{\url{https://chrony-project.org/index.html}}.   (\bibinfo{year}{2024}).
\newblock
\newblock
\shownote{Accessed: 09/11/2024.}


\bibitem[\protect\citeauthoryear{Clarke, Devadas, van Dijk, Gassend, and Suh}{Clarke et~al\mbox{.}}{2003}]%
        {incremental-hash-memory}
\bibfield{author}{\bibinfo{person}{Dwaine Clarke}, \bibinfo{person}{Srinivas Devadas}, \bibinfo{person}{Marten van Dijk}, \bibinfo{person}{Blaise Gassend}, {and} \bibinfo{person}{G.~Edward Suh}.} \bibinfo{year}{2003}\natexlab{}.
\newblock \showarticletitle{Incremental Multiset Hash Functions and Their Application to Memory Integrity Checking}. In \bibinfo{booktitle}{{\em Advances in Cryptology -- Proceedings of CRYPTO 2003}}. \bibinfo{pages}{1--18}.
\newblock


\bibitem[\protect\citeauthoryear{Clockwork.io}{Clockwork.io}{2022}]%
        {clockwork-clock-cmp}
\bibfield{author}{\bibinfo{person}{Clockwork.io}.} \bibinfo{year}{2022}\natexlab{}.
\newblock \bibinfo{title}{Cloud Clocksync Showdown: Ntpd vs Chrony vs Clockwork}.
\newblock \bibinfo{howpublished}{\url{https://www.clockwork.io/cloud-clocksync-showdown-ntpd-vs-chrony-vs-clockwork/}}.   (\bibinfo{year}{2022}).
\newblock


\bibitem[\protect\citeauthoryear{Clockwork.io}{Clockwork.io}{2024a}]%
        {huygens-latency-sensei}
\bibfield{author}{\bibinfo{person}{Clockwork.io}.} \bibinfo{year}{2024}\natexlab{a}.
\newblock \bibinfo{title}{Clockwork Latency Sensei}.
\newblock \bibinfo{howpublished}{\url{https://www.clockwork.io/latency-sensei/}}.   (\bibinfo{year}{2024}).
\newblock
\newblock
\shownote{Accessed: 09/11/2024.}


\bibitem[\protect\citeauthoryear{Clockwork.io}{Clockwork.io}{2024b}]%
        {clockwork-accurate-owd}
\bibfield{author}{\bibinfo{person}{Clockwork.io}.} \bibinfo{year}{2024}\natexlab{b}.
\newblock \bibinfo{title}{Why One-Way Latency Measures Are Critical for Distributed Databases, Microservices, and AI Workloads}.
\newblock \bibinfo{howpublished}{\url{https://www.clockwork.io/why-one-way-latency-measures-are-critical-for-distributed-databases-microservices-and-ai-workloads/}}.   (\bibinfo{year}{2024}).
\newblock
\newblock
\shownote{Accessed: 08/31/2025.}


\bibitem[\protect\citeauthoryear{Cooper, Ramakrishnan, Srivastava, Silberstein, Bohannon, Jacobsen, Puz, Weaver, and Yerneni}{Cooper et~al\mbox{.}}{2008}]%
        {vldb08-pnuts}
\bibfield{author}{\bibinfo{person}{Brian~F. Cooper}, \bibinfo{person}{Raghu Ramakrishnan}, \bibinfo{person}{Utkarsh Srivastava}, \bibinfo{person}{Adam Silberstein}, \bibinfo{person}{Philip Bohannon}, \bibinfo{person}{Hans-Arno Jacobsen}, \bibinfo{person}{Nick Puz}, \bibinfo{person}{Daniel Weaver}, {and} \bibinfo{person}{Ramana Yerneni}.} \bibinfo{year}{2008}\natexlab{}.
\newblock \showarticletitle{PNUTS: Yahoo!'s Hosted Data Serving Platform}.
\newblock \bibinfo{journal}{{\em Proceedings of the VLDB Endowment\/}} \bibinfo{volume}{1}, \bibinfo{number}{2} (\bibinfo{date}{August} \bibinfo{year}{2008}), \bibinfo{pages}{1277--1288}.
\newblock
\showDOI{%
\url{https://doi.org/10.14778/1454159.1454167}}


\bibitem[\protect\citeauthoryear{Corbett, Dean, Epstein, Fikes, Frost, Furman, Ghemawat, Gubarev, and et~al.}{Corbett et~al\mbox{.}}{2012}]%
        {spanner}
\bibfield{author}{\bibinfo{person}{James~C. Corbett}, \bibinfo{person}{Jeffrey Dean}, \bibinfo{person}{Michael Epstein}, \bibinfo{person}{Andrew Fikes}, \bibinfo{person}{Christopher Frost}, \bibinfo{person}{JJ Furman}, \bibinfo{person}{Sanjay Ghemawat}, \bibinfo{person}{Andrey Gubarev}, {and} \bibinfo{person}{et al.}} \bibinfo{year}{2012}\natexlab{}.
\newblock \showarticletitle{Spanner: Google's Globally-Distributed Database}. In \bibinfo{booktitle}{{\em 10th USENIX Symposium on Operating Systems Design and Implementation (OSDI 2012)}}. \bibinfo{pages}{251--264}.
\newblock
\showURL{%
\url{https://www.usenix.org/conference/osdi12/technical-sessions/presentation/corbett}}


\bibitem[\protect\citeauthoryear{Council}{Council}{2022}]%
        {tpcc}
\bibfield{author}{\bibinfo{person}{Transaction Processing~Performance Council}.} \bibinfo{year}{2022}\natexlab{}.
\newblock \bibinfo{title}{TPC-C}.
\newblock \bibinfo{howpublished}{\url{https://www.tpc.org/tpcc/}}.   (\bibinfo{year}{2022}).
\newblock
\newblock
\shownote{Accessed: 08/31/2025.}


\bibitem[\protect\citeauthoryear{Data}{Data}{2025}]%
        {voltdb}
\bibfield{author}{\bibinfo{person}{Volt~Active Data}.} \bibinfo{year}{2025}\natexlab{}.
\newblock \bibinfo{title}{How VoltDB Works}.
\newblock \bibinfo{howpublished}{\url{https://docs.voltdb.com/UsingVoltDB/IntroHowVoltDBWorks.php}}.   (\bibinfo{year}{2025}).
\newblock
\newblock
\shownote{Accessed: 08/31/2025.}


\bibitem[\protect\citeauthoryear{Difallah, Pavlo, Curino, and Cudre-Mauroux}{Difallah et~al\mbox{.}}{2013}]%
        {vldb13-oltpbench}
\bibfield{author}{\bibinfo{person}{Djellel~Eddine Difallah}, \bibinfo{person}{Andrew Pavlo}, \bibinfo{person}{Carlo Curino}, {and} \bibinfo{person}{Philippe Cudre-Mauroux}.} \bibinfo{year}{2013}\natexlab{}.
\newblock \showarticletitle{OLTP-Bench: An Extensible Testbed for Benchmarking Relational Databases}.
\newblock \bibinfo{journal}{{\em Proc. VLDB Endow.\/}} \bibinfo{volume}{7}, \bibinfo{number}{4} (\bibinfo{date}{dec} \bibinfo{year}{2013}), \bibinfo{pages}{277–288}.
\newblock
\showISSN{2150-8097}
\showDOI{%
\url{https://doi.org/10.14778/2732240.2732246}}


\bibitem[\protect\citeauthoryear{Eswaran, Gray, Lorie, and Traiger}{Eswaran et~al\mbox{.}}{1976}]%
        {acm76-2pl}
\bibfield{author}{\bibinfo{person}{K.~P. Eswaran}, \bibinfo{person}{J.~N. Gray}, \bibinfo{person}{R.~A. Lorie}, {and} \bibinfo{person}{I.~L. Traiger}.} \bibinfo{year}{1976}\natexlab{}.
\newblock \showarticletitle{The Notions of Consistency and Predicate Locks in a Database System}.
\newblock \bibinfo{journal}{{\it Commun. ACM}} \bibinfo{volume}{19}, \bibinfo{number}{11} (\bibinfo{year}{1976}), \bibinfo{pages}{624--633}.
\newblock
\showDOI{%
\url{https://doi.org/10.1145/360363.360369}}


\bibitem[\protect\citeauthoryear{{European Union}}{{European Union}}{2018}]%
        {gdpr-personal-data}
\bibfield{author}{\bibinfo{person}{{European Union}}.} \bibinfo{year}{2018}\natexlab{}.
\newblock \bibinfo{title}{GDPR Personal Data – What Information Does This Cover?}
\newblock \bibinfo{howpublished}{\url{https://www.gdpreu.org/the-regulation/key-concepts/personal-data/}}.   (\bibinfo{year}{2018}).
\newblock
\newblock
\shownote{Accessed: 08/31/2025.}


\bibitem[\protect\citeauthoryear{Fan and Golab}{Fan and Golab}{2019}]%
        {vldb19-ocean-vista}
\bibfield{author}{\bibinfo{person}{Hua Fan} {and} \bibinfo{person}{Wojciech Golab}.} \bibinfo{year}{2019}\natexlab{}.
\newblock \showarticletitle{Ocean Vista: Gossip-Based Visibility Control for Speedy Geo-Distributed Transactions}.
\newblock \bibinfo{journal}{{\em Proceedings of the VLDB Endowment\/}} \bibinfo{volume}{12}, \bibinfo{number}{6} (\bibinfo{year}{2019}), \bibinfo{pages}{1471–1484}.
\newblock
\showDOI{%
\url{https://doi.org/10.14778/3342263.3342627}}


\bibitem[\protect\citeauthoryear{Fischlin}{Fischlin}{1997}]%
        {incremental-hash-memory-checker}
\bibfield{author}{\bibinfo{person}{Marc Fischlin}.} \bibinfo{year}{1997}\natexlab{}.
\newblock \showarticletitle{Incremental Cryptography and Memory Checkers}. In \bibinfo{booktitle}{{\em Proceedings of the International Conference on the Theory and Application of Cryptographic Techniques (EUROCRYPT 1997)}}. \bibinfo{pages}{275--291}.
\newblock


\bibitem[\protect\citeauthoryear{Ganesan, Alagappan, Arpaci-Dusseau, and Arpaci-Dusseau}{Ganesan et~al\mbox{.}}{2020}]%
        {fast20-cad}
\bibfield{author}{\bibinfo{person}{Aishwarya Ganesan}, \bibinfo{person}{Ramnatthan Alagappan}, \bibinfo{person}{Andrea Arpaci-Dusseau}, {and} \bibinfo{person}{Remzi Arpaci-Dusseau}.} \bibinfo{year}{2020}\natexlab{}.
\newblock \showarticletitle{Strong and Efficient Consistency with Consistency-Aware Durability}. In \bibinfo{booktitle}{{\em Proceedings of the 18th USENIX Conference on File and Storage Technologies (FAST 2020)}}. \bibinfo{pages}{1--16}.
\newblock
\showDOI{%
\url{https://doi.org/10.1145/3423138}}


\bibitem[\protect\citeauthoryear{Geng}{Geng}{2025}]%
        {tiga-tla}
\bibfield{author}{\bibinfo{person}{Jinkun Geng}.} \bibinfo{year}{2025}\natexlab{}.
\newblock \bibinfo{title}{TLA+ Specification of Tiga}.
\newblock \bibinfo{howpublished}{\url{https://github.com/New-Consensus-Concurrency-Control/Tiga-TLA-plus}}.   (\bibinfo{year}{2025}).
\newblock


\bibitem[\protect\citeauthoryear{Geng, Sivaraman, Prabhakar, and Rosenblum}{Geng et~al\mbox{.}}{2022}]%
        {nezha-tech-report}
\bibfield{author}{\bibinfo{person}{Jinkun Geng}, \bibinfo{person}{Anirudh Sivaraman}, \bibinfo{person}{Balaji Prabhakar}, {and} \bibinfo{person}{Mendel Rosenblum}.} \bibinfo{year}{2022}\natexlab{}.
\newblock \bibinfo{title}{Nezha: Deployable and High-Performance Consensus Using Synchronized Clocks [Technical Report]}.
\newblock   (\bibinfo{year}{2022}).
\newblock
\showURL{%
\url{https://arxiv.org/abs/2206.03285}}


\bibitem[\protect\citeauthoryear{Geng, Sivaraman, Prabhakar, and Rosenblum}{Geng et~al\mbox{.}}{2023}]%
        {vldb23-nezha}
\bibfield{author}{\bibinfo{person}{Jinkun Geng}, \bibinfo{person}{Anirudh Sivaraman}, \bibinfo{person}{Balaji Prabhakar}, {and} \bibinfo{person}{Mendel Rosenblum}.} \bibinfo{year}{2023}\natexlab{}.
\newblock \showarticletitle{Nezha: Deployable and High-Performance Consensus Using Synchronized Clocks}.
\newblock \bibinfo{journal}{{\em Proceedings of the VLDB Endowment\/}}  \bibinfo{volume}{16} (\bibinfo{year}{2023}), \bibinfo{pages}{629–642}.
\newblock
\showDOI{%
\url{https://doi.org/10.14778/3574245.3574250}}


\bibitem[\protect\citeauthoryear{Geng, Liu, Yin, Naik, Prabhakar, Rosenblum, and Vahdat}{Geng et~al\mbox{.}}{2018}]%
        {huygens}
\bibfield{author}{\bibinfo{person}{Yilong Geng}, \bibinfo{person}{Shiyu Liu}, \bibinfo{person}{Zi Yin}, \bibinfo{person}{Ashish Naik}, \bibinfo{person}{Balaji Prabhakar}, \bibinfo{person}{Mendel Rosenblum}, {and} \bibinfo{person}{Amin Vahdat}.} \bibinfo{year}{2018}\natexlab{}.
\newblock \showarticletitle{Exploiting a Natural Network Effect for Scalable, Fine-grained Clock Synchronization}. In \bibinfo{booktitle}{{\em Proceedings of the 15th USENIX Symposium on Networked Systems Design and Implementation (NSDI 18)}}. \bibinfo{publisher}{USENIX Association}, \bibinfo{address}{Renton, WA}, \bibinfo{pages}{81--94}.
\newblock
\showISBNx{978-1-939133-01-4}
\showURL{%
\url{https://www.usenix.org/conference/nsdi18/presentation/geng}}


\bibitem[\protect\citeauthoryear{Google}{Google}{2025a}]%
        {gcp-ntp}
\bibfield{author}{\bibinfo{person}{Google}.} \bibinfo{year}{2025}\natexlab{a}.
\newblock \bibinfo{title}{Configure NTP on a VM}.
\newblock \bibinfo{howpublished}{\url{https://cloud.google.com/compute/docs/instances/configure-ntp\#linux-ntpd}}.   (\bibinfo{year}{2025}).
\newblock
\newblock
\shownote{Accessed: 2025-08-31.}


\bibitem[\protect\citeauthoryear{Google}{Google}{2025b}]%
        {gcp-chrony}
\bibfield{author}{\bibinfo{person}{Google}.} \bibinfo{year}{2025}\natexlab{b}.
\newblock \bibinfo{title}{Configure NTP on a VM (Chrony)}.
\newblock \bibinfo{howpublished}{\url{https://cloud.google.com/compute/docs/instances/configure-ntp\#linux-chrony}}.   (\bibinfo{year}{2025}).
\newblock
\newblock
\shownote{Accessed: 2025-08-31.}


\bibitem[\protect\citeauthoryear{Gray and Lamport}{Gray and Lamport}{2006}]%
        {tods06-consensus-commit}
\bibfield{author}{\bibinfo{person}{Jim Gray} {and} \bibinfo{person}{Leslie Lamport}.} \bibinfo{year}{2006}\natexlab{}.
\newblock \showarticletitle{Consensus on Transaction Commit}.
\newblock \bibinfo{journal}{{\em ACM Transactions on Database Systems\/}} \bibinfo{volume}{31}, \bibinfo{number}{1} (\bibinfo{date}{March} \bibinfo{year}{2006}), \bibinfo{pages}{133--160}.
\newblock
\showDOI{%
\url{https://doi.org/10.1145/1132863.1132867}}


\bibitem[\protect\citeauthoryear{Gray, Sundaresan, Englert, Baclawski, and Weinberger}{Gray et~al\mbox{.}}{1994}]%
        {sigmod94_zipfian}
\bibfield{author}{\bibinfo{person}{Jim Gray}, \bibinfo{person}{Prakash Sundaresan}, \bibinfo{person}{Susanne Englert}, \bibinfo{person}{Ken Baclawski}, {and} \bibinfo{person}{Peter~J. Weinberger}.} \bibinfo{year}{1994}\natexlab{}.
\newblock \showarticletitle{Quickly Generating Billion-Record Synthetic Databases}.
\newblock \bibinfo{journal}{{\em Proceedings of the International Conference on Management of Data (SIGMOD 1994)\/}} (\bibinfo{year}{1994}), \bibinfo{pages}{243--252}.
\newblock
\showDOI{%
\url{https://doi.org/10.1145/191839.191886}}


\bibitem[\protect\citeauthoryear{Herlihy and Wing}{Herlihy and Wing}{1990}]%
        {linearizability}
\bibfield{author}{\bibinfo{person}{Maurice~P. Herlihy} {and} \bibinfo{person}{Jeannette~M. Wing}.} \bibinfo{year}{1990}\natexlab{}.
\newblock \showarticletitle{Linearizability: A Correctness Condition for Concurrent Objects}.
\newblock \bibinfo{journal}{{\em ACM Transactions on Programming Languages and Systems\/}} \bibinfo{volume}{12}, \bibinfo{number}{3} (\bibinfo{year}{1990}), \bibinfo{pages}{463--492}.
\newblock
\showDOI{%
\url{https://doi.org/10.1145/78969.78972}}


\bibitem[\protect\citeauthoryear{Hildred, Abebe, and Daudjee}{Hildred et~al\mbox{.}}{2023}]%
        {vldb23-caerus}
\bibfield{author}{\bibinfo{person}{Joshua Hildred}, \bibinfo{person}{Michael Abebe}, {and} \bibinfo{person}{Khuzaima Daudjee}.} \bibinfo{year}{2023}\natexlab{}.
\newblock \showarticletitle{Caerus: Low-Latency Distributed Transactions for Geo-Replicated Systems}.
\newblock \bibinfo{journal}{{\em Proceedings of the VLDB Endowment\/}} \bibinfo{volume}{17}, \bibinfo{number}{3} (\bibinfo{date}{November} \bibinfo{year}{2023}), \bibinfo{pages}{469--482}.
\newblock
\showDOI{%
\url{https://doi.org/10.14778/3632093.3632109}}


\bibitem[\protect\citeauthoryear{Huang, Liu, Cui, Fang, Ma, Xu, Shen, Tang, Zhou, Huang, Wei, Liu, Zhang, Li, Wu, Song, Sun, Yu, Zhao, Cameron, Pei, and Tang}{Huang et~al\mbox{.}}{2020}]%
        {vldb20-tidb}
\bibfield{author}{\bibinfo{person}{Dongxu Huang}, \bibinfo{person}{Qi Liu}, \bibinfo{person}{Qiu Cui}, \bibinfo{person}{Zhuhe Fang}, \bibinfo{person}{Xiaoyu Ma}, \bibinfo{person}{Fei Xu}, \bibinfo{person}{Li Shen}, \bibinfo{person}{Liu Tang}, \bibinfo{person}{Yuxing Zhou}, \bibinfo{person}{Menglong Huang}, \bibinfo{person}{Wan Wei}, \bibinfo{person}{Cong Liu}, \bibinfo{person}{Jian Zhang}, \bibinfo{person}{Jianjun Li}, \bibinfo{person}{Xuelian Wu}, \bibinfo{person}{Lingyu Song}, \bibinfo{person}{Ruoxi Sun}, \bibinfo{person}{Shuaipeng Yu}, \bibinfo{person}{Lei Zhao}, \bibinfo{person}{Nicholas Cameron}, \bibinfo{person}{Liquan Pei}, {and} \bibinfo{person}{Xin Tang}.} \bibinfo{year}{2020}\natexlab{}.
\newblock \showarticletitle{TiDB: A Raft-Based HTAP Database}.
\newblock \bibinfo{journal}{{\em Proceedings of the VLDB Endowment\/}} \bibinfo{volume}{13}, \bibinfo{number}{12} (\bibinfo{year}{2020}), \bibinfo{pages}{3072--3084}.
\newblock
\showDOI{%
\url{https://doi.org/10.14778/3415478.3415535}}


\bibitem[\protect\citeauthoryear{{IBM Software Group}}{{IBM Software Group}}{[n. d.]}]%
        {tatp-bench}
\bibfield{author}{\bibinfo{person}{{IBM Software Group}}.} \bibinfo{year}{[n. d.]}\natexlab{}.
\newblock \bibinfo{title}{{Telecommunication Application Transaction Processing (TATP) Benchmark Description}}.
\newblock \bibinfo{howpublished}{\url{https://tatpbenchmark.sourceforge.net/TATP_Description.pdf}}.   (\bibinfo{year}{[n. d.]}).
\newblock


\bibitem[\protect\citeauthoryear{Kallman, Kimura, Natkins, Pavlo, Rasin, Zdonik, Jones, Madden, Stonebraker, and Zhang}{Kallman et~al\mbox{.}}{2008}]%
        {vldb08-hstore}
\bibfield{author}{\bibinfo{person}{Robert Kallman}, \bibinfo{person}{Hideaki Kimura}, \bibinfo{person}{Jonathan Natkins}, \bibinfo{person}{Andrew Pavlo}, \bibinfo{person}{Alexander Rasin}, \bibinfo{person}{Stanley Zdonik}, \bibinfo{person}{Evan P.~C. Jones}, \bibinfo{person}{Samuel Madden}, \bibinfo{person}{Michael Stonebraker}, {and} \bibinfo{person}{Yang Zhang}.} \bibinfo{year}{2008}\natexlab{}.
\newblock \showarticletitle{H-Store: A High-Performance, Distributed Main Memory Transaction Processing System}.
\newblock \bibinfo{journal}{{\em Proceedings of the VLDB Endowment\/}} \bibinfo{volume}{1}, \bibinfo{number}{2} (\bibinfo{year}{2008}), \bibinfo{pages}{1496--1499}.
\newblock
\showDOI{%
\url{https://doi.org/10.14778/1454159.1454211}}


\bibitem[\protect\citeauthoryear{Katsarakis, Ma, Tan, Bainbridge, Balkwill, Dragojevic, Grot, Radunovic, and Zhang}{Katsarakis et~al\mbox{.}}{2021}]%
        {eurosys21-zeus}
\bibfield{author}{\bibinfo{person}{Antonios Katsarakis}, \bibinfo{person}{Yijun Ma}, \bibinfo{person}{Zhaowei Tan}, \bibinfo{person}{Andrew Bainbridge}, \bibinfo{person}{Matthew Balkwill}, \bibinfo{person}{Aleksandar Dragojevic}, \bibinfo{person}{Boris Grot}, \bibinfo{person}{Bozidar Radunovic}, {and} \bibinfo{person}{Yongguang Zhang}.} \bibinfo{year}{2021}\natexlab{}.
\newblock \showarticletitle{Zeus: Locality-Aware Distributed Transactions}. In \bibinfo{booktitle}{{\em Proceedings of the Sixteenth European Conference on Computer Systems (EuroSys 2021)}} {\em (\bibinfo{series}{EuroSys '21})}. \bibinfo{publisher}{Association for Computing Machinery}, \bibinfo{address}{New York, NY, USA}, \bibinfo{pages}{145--161}.
\newblock
\showISBNx{9781450383349}
\showDOI{%
\url{https://doi.org/10.1145/3447786.3456234}}


\bibitem[\protect\citeauthoryear{Kończak, Wojciechowski, Santos, Żurkowski, and Schiper}{Kończak et~al\mbox{.}}{2021}]%
        {tdsn21-recovery}
\bibfield{author}{\bibinfo{person}{Jan Kończak}, \bibinfo{person}{Paweł~T. Wojciechowski}, \bibinfo{person}{Nuno Santos}, \bibinfo{person}{Tomasz Żurkowski}, {and} \bibinfo{person}{André Schiper}.} \bibinfo{year}{2021}\natexlab{}.
\newblock \showarticletitle{Recovery Algorithms for Paxos-Based State Machine Replication}.
\newblock \bibinfo{journal}{{\em IEEE Transactions on Dependable and Secure Computing\/}} \bibinfo{volume}{18}, \bibinfo{number}{4} (\bibinfo{date}{July--August} \bibinfo{year}{2021}), \bibinfo{pages}{1234--1247}.
\newblock
\showISSN{1545-5971}
\showDOI{%
\url{https://doi.org/10.1109/TDSC.2021.3051234}}


\bibitem[\protect\citeauthoryear{Kraska, Pang, Franklin, Madden, and Fekete}{Kraska et~al\mbox{.}}{2013}]%
        {eurosys13-mdcc}
\bibfield{author}{\bibinfo{person}{Tim Kraska}, \bibinfo{person}{Gene Pang}, \bibinfo{person}{Michael~J. Franklin}, \bibinfo{person}{Samuel Madden}, {and} \bibinfo{person}{Alan Fekete}.} \bibinfo{year}{2013}\natexlab{}.
\newblock \showarticletitle{MDCC: Multi-Data Center Consistency}. In \bibinfo{booktitle}{{\em Proceedings of the 8th ACM European Conference on Computer Systems (EuroSys 2013)}} {\em (\bibinfo{series}{EuroSys '13})}. \bibinfo{publisher}{Association for Computing Machinery}, \bibinfo{address}{New York, NY, USA}, \bibinfo{pages}{113--126}.
\newblock
\showISBNx{9781450319942}
\showDOI{%
\url{https://doi.org/10.1145/2465351.2465363}}


\bibitem[\protect\citeauthoryear{Lamport}{Lamport}{2001}]%
        {paxos}
\bibfield{author}{\bibinfo{person}{Leslie Lamport}.} \bibinfo{year}{2001}\natexlab{}.
\newblock \showarticletitle{Paxos Made Simple}.
\newblock \bibinfo{journal}{{\em ACM SIGACT News\/}} \bibinfo{volume}{32}, \bibinfo{number}{4} (\bibinfo{year}{2001}), \bibinfo{pages}{51--58}.
\newblock
\showDOI{%
\url{https://doi.org/10.1145/568425.568433}}


\bibitem[\protect\citeauthoryear{Lamport}{Lamport}{2006}]%
        {fastpaxos}
\bibfield{author}{\bibinfo{person}{Leslie Lamport}.} \bibinfo{year}{2006}\natexlab{}.
\newblock \showarticletitle{Fast Paxos}.
\newblock \bibinfo{journal}{{\em Distributed Computing\/}} \bibinfo{volume}{19}, \bibinfo{number}{2} (\bibinfo{date}{October} \bibinfo{year}{2006}), \bibinfo{pages}{79--103}.
\newblock
\showDOI{%
\url{https://doi.org/10.1007/s00446-006-0016-x}}


\bibitem[\protect\citeauthoryear{Li, Michael, and Ports}{Li et~al\mbox{.}}{2017}]%
        {sosp17-eris}
\bibfield{author}{\bibinfo{person}{Jialin Li}, \bibinfo{person}{Ellis Michael}, {and} \bibinfo{person}{Dan R.~K. Ports}.} \bibinfo{year}{2017}\natexlab{}.
\newblock \showarticletitle{Eris: Coordination-Free Consistent Transactions Using In-Network Concurrency Control}. In \bibinfo{booktitle}{{\em Proceedings of the 26th ACM Symposium on Operating Systems Principles (SOSP 2017)}}. \bibinfo{publisher}{ACM}, \bibinfo{address}{Shanghai, China}, \bibinfo{pages}{17}.
\newblock
\showDOI{%
\url{https://doi.org/10.1145/3132747.3132751}}


\bibitem[\protect\citeauthoryear{Li, Michael, Sharma, Szekeres, and Ports}{Li et~al\mbox{.}}{2016}]%
        {osdi16-nopaxos}
\bibfield{author}{\bibinfo{person}{Jialin Li}, \bibinfo{person}{Ellis Michael}, \bibinfo{person}{Naveen~Kr. Sharma}, \bibinfo{person}{Adriana Szekeres}, {and} \bibinfo{person}{Dan R.~K. Ports}.} \bibinfo{year}{2016}\natexlab{}.
\newblock \showarticletitle{Just Say No to Paxos Overhead: Replacing Consensus with Network Ordering}. In \bibinfo{booktitle}{{\em Proceedings of the 12th USENIX Symposium on Operating Systems Design and Implementation (OSDI 2016)}}. \bibinfo{publisher}{USENIX Association}, \bibinfo{address}{Savannah, GA}, \bibinfo{pages}{395--410}.
\newblock
\showURL{%
\url{https://www.usenix.org/conference/osdi16/technical-sessions/presentation/li}}


\bibitem[\protect\citeauthoryear{Li, Kumar, Hariharan, Wassel, Hochschild, Platt, Sabato, Yu, Dukkipati, Chandra, and Vahdat}{Li et~al\mbox{.}}{2020}]%
        {osdi20-sundial}
\bibfield{author}{\bibinfo{person}{Yuliang Li}, \bibinfo{person}{Gautam Kumar}, \bibinfo{person}{Hema Hariharan}, \bibinfo{person}{Hassan Wassel}, \bibinfo{person}{Peter Hochschild}, \bibinfo{person}{Dave Platt}, \bibinfo{person}{Simon Sabato}, \bibinfo{person}{Minlan Yu}, \bibinfo{person}{Nandita Dukkipati}, \bibinfo{person}{Prashant Chandra}, {and} \bibinfo{person}{Amin Vahdat}.} \bibinfo{year}{2020}\natexlab{}.
\newblock \showarticletitle{Sundial: Fault-Tolerant Clock Synchronization for Datacenters}. In \bibinfo{booktitle}{{\em Proceedings of the 14th USENIX Symposium on Operating Systems Design and Implementation (OSDI 2020)}}. \bibinfo{publisher}{USENIX Association}, \bibinfo{address}{Santa Clara, CA}, \bibinfo{pages}{611--630}.
\newblock
\showURL{%
\url{https://www.usenix.org/conference/osdi20/presentation/li}}


\bibitem[\protect\citeauthoryear{Liskov}{Liskov}{1991}]%
        {liskov_clock_sync}
\bibfield{author}{\bibinfo{person}{Barbara Liskov}.} \bibinfo{year}{1991}\natexlab{}.
\newblock \showarticletitle{{Practical Uses of Synchronized Clocks in Distributed Systems}}. In \bibinfo{booktitle}{{\em Proceedings of the Tenth Annual ACM Symposium on Principles of Distributed Computing}}.
\newblock
\showDOI{%
\url{https://doi.org/10.1145/112600.112601}}


\bibitem[\protect\citeauthoryear{Liskov and Cowling}{Liskov and Cowling}{2012}]%
        {viewstamp}
\bibfield{author}{\bibinfo{person}{Barbara Liskov} {and} \bibinfo{person}{James Cowling}.} \bibinfo{year}{2012}\natexlab{}.
\newblock \showarticletitle{Viewstamped replication revisited}.
\newblock  (\bibinfo{year}{2012}).
\newblock


\bibitem[\protect\citeauthoryear{Lloyd, Freedman, Kaminsky, and Andersen}{Lloyd et~al\mbox{.}}{2011}]%
        {sosp11-cops}
\bibfield{author}{\bibinfo{person}{Wyatt Lloyd}, \bibinfo{person}{Michael~J. Freedman}, \bibinfo{person}{Michael Kaminsky}, {and} \bibinfo{person}{David~G. Andersen}.} \bibinfo{year}{2011}\natexlab{}.
\newblock \showarticletitle{Don't Settle for Eventual: Scalable Causal Consistency for Wide-Area Storage with COPS}. In \bibinfo{booktitle}{{\em Proceedings of the 23rd ACM Symposium on Operating Systems Principles (SOSP '11)}}. \bibinfo{publisher}{ACM}, \bibinfo{address}{Cascais, Portugal}, \bibinfo{pages}{401--416}.
\newblock
\showDOI{%
\url{https://doi.org/10.1145/2043556.2043593}}


\bibitem[\protect\citeauthoryear{Lu, Mu, Sen, and Lloyd}{Lu et~al\mbox{.}}{2023}]%
        {osdi23-ncc}
\bibfield{author}{\bibinfo{person}{Haonan Lu}, \bibinfo{person}{Shuai Mu}, \bibinfo{person}{Siddhartha Sen}, {and} \bibinfo{person}{Wyatt Lloyd}.} \bibinfo{year}{2023}\natexlab{}.
\newblock \showarticletitle{NCC: Natural Concurrency Control for Strictly Serializable Datastores by Avoiding the Timestamp-Inversion Pitfall}. In \bibinfo{booktitle}{{\em Proceedings of the 17th USENIX Symposium on Operating Systems Design and Implementation (OSDI 2023)}}. \bibinfo{publisher}{USENIX Association}, \bibinfo{address}{Santa Clara, CA}, \bibinfo{pages}{821--839}.
\newblock
\showURL{%
\url{https://www.usenix.org/conference/osdi23/presentation/lu}}


\bibitem[\protect\citeauthoryear{{Meta}}{{Meta}}{2021}]%
        {zippydb}
\bibfield{author}{\bibinfo{person}{{Meta}}.} \bibinfo{year}{2021}\natexlab{}.
\newblock \bibinfo{title}{ZippyDB: Facebook's Key-Value Store}.
\newblock \bibinfo{howpublished}{\url{https://engineering.fb.com/2021/08/06/core-infra/zippydb/}}.   (\bibinfo{year}{2021}).
\newblock
\newblock
\shownote{Accessed: 2025-08-31.}


\bibitem[\protect\citeauthoryear{Michael, Ports, Sharma, and Szekeres}{Michael et~al\mbox{.}}{2017}]%
        {disc17-dcr-techreport}
\bibfield{author}{\bibinfo{person}{Ellis Michael}, \bibinfo{person}{Dan R.~K. Ports}, \bibinfo{person}{Naveen~Kr. Sharma}, {and} \bibinfo{person}{Adriana Szekeres}.} \bibinfo{year}{2017}\natexlab{}.
\newblock \bibinfo{booktitle}{{\em Recovering Shared Objects Without Stable Storage [Extended Version]}}.
\newblock \bibinfo{type}{Technical Report}. \bibinfo{institution}{University of Washington}.
\newblock
\showURL{%
\url{https://www.microsoft.com/en-us/research/publication/recovering-shared-objects-without-stable-storage-extended-version/}}
\newblock
\shownote{Accessed: 2025-08-31.}


\bibitem[\protect\citeauthoryear{Microsoft}{Microsoft}{2022}]%
        {cosmosdb}
\bibfield{author}{\bibinfo{person}{Microsoft}.} \bibinfo{year}{2022}\natexlab{}.
\newblock \bibinfo{title}{Global Data Distribution with Azure Cosmos DB — Under the Hood}.
\newblock \bibinfo{howpublished}{\url{https://docs.microsoft.com/en-us/azure/cosmos-db/global-dist-under-the-hood}}.   (\bibinfo{year}{2022}).
\newblock
\newblock
\shownote{Accessed: 2025-08-31.}


\bibitem[\protect\citeauthoryear{{Microsoft}}{{Microsoft}}{2025}]%
        {azure-cosmos-partition}
\bibfield{author}{\bibinfo{person}{{Microsoft}}.} \bibinfo{year}{2025}\natexlab{}.
\newblock \bibinfo{title}{Partitioning and Horizontal Scaling in Azure Cosmos DB}.
\newblock \bibinfo{howpublished}{\url{https://learn.microsoft.com/en-us/azure/cosmos-db/partitioning-overview}}.   (\bibinfo{year}{2025}).
\newblock
\newblock
\shownote{Accessed: 2025-08-31.}


\bibitem[\protect\citeauthoryear{Mills}{Mills}{1991}]%
        {ntp}
\bibfield{author}{\bibinfo{person}{D.~L. Mills}.} \bibinfo{year}{1991}\natexlab{}.
\newblock \showarticletitle{Internet Time Synchronization: The Network Time Protocol}.
\newblock \bibinfo{journal}{{\em IEEE Transactions on Communications\/}} \bibinfo{volume}{39}, \bibinfo{number}{10} (\bibinfo{year}{1991}), \bibinfo{pages}{1482--1493}.
\newblock
\showDOI{%
\url{https://doi.org/10.1109/26.103043}}


\bibitem[\protect\citeauthoryear{Mu and et~al.}{Mu and et~al.}{2016}]%
        {janus-repo}
\bibfield{author}{\bibinfo{person}{Shuai Mu} {and} \bibinfo{person}{et al.}} \bibinfo{year}{2016}\natexlab{}.
\newblock \bibinfo{title}{Janus Repo}.
\newblock \bibinfo{howpublished}{\url{https://github.com/NYU-NEWS/janus}}.   (\bibinfo{year}{2016}).
\newblock
\newblock
\shownote{Accessed: 2025-08-31.}


\bibitem[\protect\citeauthoryear{Mu, Nelson, Lloyd, and Li}{Mu et~al\mbox{.}}{2016}]%
        {osdi16-janus}
\bibfield{author}{\bibinfo{person}{Shuai Mu}, \bibinfo{person}{Lamont Nelson}, \bibinfo{person}{Wyatt Lloyd}, {and} \bibinfo{person}{Jinyang Li}.} \bibinfo{year}{2016}\natexlab{}.
\newblock \showarticletitle{Consolidating Concurrency Control and Consensus for Commits under Conflicts}. In \bibinfo{booktitle}{{\em Proceedings of the 12th USENIX Symposium on Operating Systems Design and Implementation (OSDI 2016)}}. \bibinfo{publisher}{USENIX Association}, \bibinfo{address}{Savannah, GA}, \bibinfo{pages}{409--425}.
\newblock
\showURL{%
\url{https://www.usenix.org/conference/osdi16/technical-sessions/presentation/mu}}


\bibitem[\protect\citeauthoryear{Najafi and Wei}{Najafi and Wei}{2022}]%
        {nsdi22-graham}
\bibfield{author}{\bibinfo{person}{Ali Najafi} {and} \bibinfo{person}{Michael Wei}.} \bibinfo{year}{2022}\natexlab{}.
\newblock \showarticletitle{Graham: Synchronizing Clocks by Leveraging Local Clock Properties}. In \bibinfo{booktitle}{{\em Proceedings of the 19th USENIX Symposium on Networked Systems Design and Implementation (NSDI 2022)}}. \bibinfo{publisher}{USENIX Association}, \bibinfo{address}{Renton, WA, USA}, \bibinfo{pages}{453--466}.
\newblock
\showURL{%
\url{https://www.usenix.org/conference/nsdi22/presentation/najafi}}


\bibitem[\protect\citeauthoryear{{National Congress of the People's Republic of China}}{{National Congress of the People's Republic of China}}{2021}]%
        {npc-dsl}
\bibfield{author}{\bibinfo{person}{{National Congress of the People's Republic of China}}.} \bibinfo{year}{2021}\natexlab{}.
\newblock \bibinfo{title}{Data Security Law of the People's Republic of China}.
\newblock \bibinfo{howpublished}{\url{https://digichina.stanford.edu/work/translation-data-security-law-of-the-peoples-republic-of-china/}}.   (\bibinfo{year}{2021}).
\newblock
\newblock
\shownote{Accessed: 2025-08-31.}


\bibitem[\protect\citeauthoryear{Nawab, Arora, Agrawal, and Abbadi}{Nawab et~al\mbox{.}}{2015}]%
        {sigmod15-helios}
\bibfield{author}{\bibinfo{person}{Faisal Nawab}, \bibinfo{person}{Vaibhav Arora}, \bibinfo{person}{Divyakant Agrawal}, {and} \bibinfo{person}{Amr~El Abbadi}.} \bibinfo{year}{2015}\natexlab{}.
\newblock \showarticletitle{Minimizing Commit Latency of Transactions in Geo-Replicated Data Stores}. In \bibinfo{booktitle}{{\em Proceedings of the 33rd ACM SIGMOD International Conference on Management of Data (SIGMOD '15)}}. \bibinfo{publisher}{ACM}, \bibinfo{pages}{1279--1294}.
\newblock
\showDOI{%
\url{https://doi.org/10.1145/2723372.2723729}}


\bibitem[\protect\citeauthoryear{Nguyen, Miller, and Abadi}{Nguyen et~al\mbox{.}}{2023}]%
        {sigmod23-detock}
\bibfield{author}{\bibinfo{person}{Cuong D.~T. Nguyen}, \bibinfo{person}{Johann~K. Miller}, {and} \bibinfo{person}{Daniel~J. Abadi}.} \bibinfo{year}{2023}\natexlab{}.
\newblock \showarticletitle{Detock: High Performance Multi-Region Transactions at Scale}.
\newblock \bibinfo{journal}{{\em Proc. ACM Manag. Data\/}} \bibinfo{volume}{1}, \bibinfo{number}{2}, Article \bibinfo{articleno}{148} (\bibinfo{date}{June} \bibinfo{year}{2023}), \bibinfo{numpages}{27}~pages.
\newblock
\showDOI{%
\url{https://doi.org/10.1145/3589293}}


\bibitem[\protect\citeauthoryear{{ObjectDB Software Ltd.}}{{ObjectDB Software Ltd.}}{[n. d.]}]%
        {jpab-bench}
\bibfield{author}{\bibinfo{person}{{ObjectDB Software Ltd.}}} \bibinfo{year}{[n. d.]}\natexlab{}.
\newblock \bibinfo{title}{{JPA Performance Benchmark (JPAB)}}.
\newblock \bibinfo{howpublished}{\url{https://www.jpab.org/Benchmark_FAQ.html}}.   (\bibinfo{year}{[n. d.]}).
\newblock


\bibitem[\protect\citeauthoryear{Oki and Liskov}{Oki and Liskov}{1988}]%
        {viewstamp-original}
\bibfield{author}{\bibinfo{person}{Brian~M. Oki} {and} \bibinfo{person}{Barbara~H. Liskov}.} \bibinfo{year}{1988}\natexlab{}.
\newblock \showarticletitle{Viewstamped Replication: A New Primary Copy Method to Support Highly-Available Distributed Systems}. In \bibinfo{booktitle}{{\em Proceedings of the Seventh Annual ACM Symposium on Principles of Distributed Computing (PODC)}}. \bibinfo{publisher}{ACM}, \bibinfo{address}{New York, NY, USA}, \bibinfo{pages}{8--17}.
\newblock
\showDOI{%
\url{https://doi.org/10.1145/62546.62549}}


\bibitem[\protect\citeauthoryear{Ongaro and Ousterhout}{Ongaro and Ousterhout}{2014}]%
        {atc14-raft}
\bibfield{author}{\bibinfo{person}{Diego Ongaro} {and} \bibinfo{person}{John Ousterhout}.} \bibinfo{year}{2014}\natexlab{}.
\newblock \showarticletitle{In Search of an Understandable Consensus Algorithm}. In \bibinfo{booktitle}{{\em Proceedings of the 2014 USENIX Annual Technical Conference (USENIX ATC '14)}}. \bibinfo{publisher}{USENIX Association}, \bibinfo{address}{Philadelphia, PA}, \bibinfo{pages}{305--319}.
\newblock
\showURL{%
\url{https://www.usenix.org/conference/atc14/technical-sessions/presentation/ongaro}}


\bibitem[\protect\citeauthoryear{Park and Ousterhout}{Park and Ousterhout}{2019}]%
        {nsdi19-curp}
\bibfield{author}{\bibinfo{person}{Seo~Jin Park} {and} \bibinfo{person}{John Ousterhout}.} \bibinfo{year}{2019}\natexlab{}.
\newblock \showarticletitle{Exploiting Commutativity for Practical Fast Replication}. In \bibinfo{booktitle}{{\em Proceedings of the 16th USENIX Symposium on Networked Systems Design and Implementation (NSDI 2019)}}. \bibinfo{publisher}{USENIX Association}, \bibinfo{address}{Boston, MA, USA}, \bibinfo{pages}{183--198}.
\newblock
\showURL{%
\url{https://www.usenix.org/conference/nsdi19/presentation/park}}


\bibitem[\protect\citeauthoryear{{PingCap}}{{PingCap}}{2024}]%
        {tikv-multi-region-deployment}
\bibfield{author}{\bibinfo{person}{{PingCap}}.} \bibinfo{year}{2024}\natexlab{}.
\newblock \bibinfo{title}{{Three Availability Zones in Two Regions Deployment}}.
\newblock \bibinfo{howpublished}{\url{https://docs.pingcap.com/tidb/stable/multi-data-centers-in-one-city-deployment}}.   (\bibinfo{year}{2024}).
\newblock
\newblock
\shownote{Accessed: 2025-08-31.}


\bibitem[\protect\citeauthoryear{Ports, Li, Liu, Sharma, and Krishnamurthy}{Ports et~al\mbox{.}}{2015}]%
        {nsdi15-specpaxos}
\bibfield{author}{\bibinfo{person}{Dan R.~K. Ports}, \bibinfo{person}{Jialin Li}, \bibinfo{person}{Vincent Liu}, \bibinfo{person}{Naveen~Kr. Sharma}, {and} \bibinfo{person}{Arvind Krishnamurthy}.} \bibinfo{year}{2015}\natexlab{}.
\newblock \showarticletitle{Designing Distributed Systems Using Approximate Synchrony in Data Center Networks}. In \bibinfo{booktitle}{{\em Proceedings of the 12th USENIX Symposium on Networked Systems Design and Implementation (NSDI 2015)}} {\em (\bibinfo{series}{NSDI '15})}. \bibinfo{publisher}{USENIX Association}, \bibinfo{address}{Oakland, CA, USA}, \bibinfo{pages}{43--57}.
\newblock
\showISBNx{978-1-931971-21-8}
\showURL{%
\url{https://www.usenix.org/conference/nsdi15/technical-sessions/presentation/ports}}


\bibitem[\protect\citeauthoryear{Reed}{Reed}{1983}]%
        {tocs-mvto}
\bibfield{author}{\bibinfo{person}{David~P. Reed}.} \bibinfo{year}{1983}\natexlab{}.
\newblock \showarticletitle{Implementing Atomic Actions on Decentralized Data}.
\newblock \bibinfo{journal}{{\em ACM Transactions on Computer Systems\/}} \bibinfo{volume}{1}, \bibinfo{number}{1} (\bibinfo{date}{February} \bibinfo{year}{1983}), \bibinfo{pages}{3--23}.
\newblock
\showISSN{0734-2071}
\showDOI{%
\url{https://doi.org/10.1145/357353.357355}}


\bibitem[\protect\citeauthoryear{Ren, Li, and Abadi}{Ren et~al\mbox{.}}{2019}]%
        {vldb19-slog}
\bibfield{author}{\bibinfo{person}{Kun Ren}, \bibinfo{person}{Dennis Li}, {and} \bibinfo{person}{Daniel~J. Abadi}.} \bibinfo{year}{2019}\natexlab{}.
\newblock \showarticletitle{SLOG: Serializable, Low-Latency, Geo-Replicated Transactions}.
\newblock \bibinfo{journal}{{\em Proc. VLDB Endow.\/}} \bibinfo{volume}{12}, \bibinfo{number}{11} (\bibinfo{date}{July} \bibinfo{year}{2019}), \bibinfo{pages}{1747--1761}.
\newblock
\showISSN{2150-8097}
\showDOI{%
\url{https://doi.org/10.14778/3342263.3342647}}


\bibitem[\protect\citeauthoryear{Rosenkrantz, Stearns, and Lewis}{Rosenkrantz et~al\mbox{.}}{1978}]%
        {tocs78-2pl-woundwait}
\bibfield{author}{\bibinfo{person}{Daniel~J. Rosenkrantz}, \bibinfo{person}{Richard~E. Stearns}, {and} \bibinfo{person}{Philip~M. Lewis}.} \bibinfo{year}{1978}\natexlab{}.
\newblock \showarticletitle{System-Level Concurrency Control for Distributed Database Systems}.
\newblock \bibinfo{journal}{{\em ACM Transactions on Database Systems\/}} \bibinfo{volume}{3}, \bibinfo{number}{2} (\bibinfo{year}{1978}), \bibinfo{pages}{178--198}.
\newblock
\showDOI{%
\url{https://doi.org/10.1145/320080.320083}}


\bibitem[\protect\citeauthoryear{Shen, Cui, Sen, Angel, and Mu}{Shen et~al\mbox{.}}{2025}]%
        {osdi25-warbler}
\bibfield{author}{\bibinfo{person}{Weihai Shen}, \bibinfo{person}{Yang Cui}, \bibinfo{person}{Siddhartha Sen}, \bibinfo{person}{Sebastian Angel}, {and} \bibinfo{person}{Shuai Mu}.} \bibinfo{year}{2025}\natexlab{}.
\newblock \showarticletitle{Mako: Speculative Distributed Transactions with Geo-Replication}. In \bibinfo{booktitle}{{\em Proceedings of the 19th USENIX Symposium on Operating Systems Design and Implementation (OSDI 2025)}}. \bibinfo{publisher}{USENIX Association}, \bibinfo{address}{Santa Clara, CA, USA}, \bibinfo{pages}{1--16}.
\newblock
\showURL{%
\url{https://www.usenix.org/conference/osdi25/presentation/shen-weihai}}


\bibitem[\protect\citeauthoryear{Taft, Sharif, Matei, VanBenschoten, Lewis, Grieger, Niemi, Woods, Birzin, Poss, Bardea, Ranade, Darnell, Gruneir, Jaffray, Zhang, and Mattis}{Taft et~al\mbox{.}}{2020}]%
        {sigmod20-cockroachdb}
\bibfield{author}{\bibinfo{person}{Rebecca Taft}, \bibinfo{person}{Irfan Sharif}, \bibinfo{person}{Andrei Matei}, \bibinfo{person}{Nathan VanBenschoten}, \bibinfo{person}{Jordan Lewis}, \bibinfo{person}{Tobias Grieger}, \bibinfo{person}{Kai Niemi}, \bibinfo{person}{Andy Woods}, \bibinfo{person}{Anne Birzin}, \bibinfo{person}{Raphael Poss}, \bibinfo{person}{Paul Bardea}, \bibinfo{person}{Amruta Ranade}, \bibinfo{person}{Ben Darnell}, \bibinfo{person}{Bram Gruneir}, \bibinfo{person}{Justin Jaffray}, \bibinfo{person}{Lucy Zhang}, {and} \bibinfo{person}{Peter Mattis}.} \bibinfo{year}{2020}\natexlab{}.
\newblock \showarticletitle{CockroachDB: The Resilient Geo-Distributed SQL Database}. In \bibinfo{booktitle}{{\em Proceedings of the 2020 ACM SIGMOD International Conference on Management of Data}} {\em (\bibinfo{series}{SIGMOD '20})}. \bibinfo{publisher}{Association for Computing Machinery}, \bibinfo{address}{New York, NY, USA}, \bibinfo{pages}{1493–1509}.
\newblock
\showISBNx{9781450367356}
\showDOI{%
\url{https://doi.org/10.1145/3318464.3386134}}


\bibitem[\protect\citeauthoryear{Thomas}{Thomas}{1979}]%
        {tods79-timestamp-ordering}
\bibfield{author}{\bibinfo{person}{Robert~H. Thomas}.} \bibinfo{year}{1979}\natexlab{}.
\newblock \showarticletitle{A Majority Consensus Approach to Concurrency Control for Multiple Copy Databases}.
\newblock \bibinfo{journal}{{\em ACM Transactions on Database Systems\/}} \bibinfo{volume}{4}, \bibinfo{number}{2} (\bibinfo{date}{June} \bibinfo{year}{1979}), \bibinfo{pages}{180--209}.
\newblock
\showISSN{0362-5915}
\showDOI{%
\url{https://doi.org/10.1145/320071.320076}}


\bibitem[\protect\citeauthoryear{Thomson and Abadi}{Thomson and Abadi}{2010}]%
        {vldb10-txn-determinisim}
\bibfield{author}{\bibinfo{person}{Alexander Thomson} {and} \bibinfo{person}{Daniel~J. Abadi}.} \bibinfo{year}{2010}\natexlab{}.
\newblock \showarticletitle{The case for determinism in database systems}.
\newblock \bibinfo{journal}{{\em Proc. VLDB Endow.\/}} \bibinfo{volume}{3}, \bibinfo{number}{1–2} (\bibinfo{date}{Sept.} \bibinfo{year}{2010}), \bibinfo{pages}{70–80}.
\newblock
\showISSN{2150-8097}
\showDOI{%
\url{https://doi.org/10.14778/1920841.1920855}}


\bibitem[\protect\citeauthoryear{Thomson, Diamond, Weng, Ren, Shao, and Abadi}{Thomson et~al\mbox{.}}{2012}]%
        {sigmod12-calvin}
\bibfield{author}{\bibinfo{person}{Alexander Thomson}, \bibinfo{person}{Thaddeus Diamond}, \bibinfo{person}{Shu-Chun Weng}, \bibinfo{person}{Kun Ren}, \bibinfo{person}{Philip Shao}, {and} \bibinfo{person}{Daniel~J. Abadi}.} \bibinfo{year}{2012}\natexlab{}.
\newblock \showarticletitle{Calvin: Fast Distributed Transactions for Partitioned Database Systems}. In \bibinfo{booktitle}{{\em Proceedings of the 2012 ACM SIGMOD International Conference on Management of Data}} {\em (\bibinfo{series}{SIGMOD '12})}. \bibinfo{publisher}{Association for Computing Machinery}, \bibinfo{address}{New York, NY, USA}, \bibinfo{pages}{1–12}.
\newblock
\showISBNx{9781450312479}
\showDOI{%
\url{https://doi.org/10.1145/2213836.2213838}}


\bibitem[\protect\citeauthoryear{Tollman, Park, and Ousterhout}{Tollman et~al\mbox{.}}{2021}]%
        {nsdi21-epaxos}
\bibfield{author}{\bibinfo{person}{Sarah Tollman}, \bibinfo{person}{Seo~Jin Park}, {and} \bibinfo{person}{John Ousterhout}.} \bibinfo{year}{2021}\natexlab{}.
\newblock \showarticletitle{EPaxos Revisited}. In \bibinfo{booktitle}{{\em Proceedings of the 18th USENIX Symposium on Networked Systems Design and Implementation (NSDI 2021)}}.
\newblock
\showURL{%
\url{https://www.usenix.org/conference/nsdi21/presentation/tollman}}


\bibitem[\protect\citeauthoryear{{VoltDB}}{{VoltDB}}{[n. d.]}]%
        {voter}
\bibfield{author}{\bibinfo{person}{{VoltDB}}.} \bibinfo{year}{[n. d.]}\natexlab{}.
\newblock \bibinfo{title}{{Voter Benchmark}}.
\newblock \bibinfo{howpublished}{\url{https://github.com/VoltDB/voltdb/tree/master/examples/voter}}.   (\bibinfo{year}{[n. d.]}).
\newblock


\bibitem[\protect\citeauthoryear{Yahoo!}{Yahoo!}{[n. d.]}]%
        {ycsb-workload}
\bibfield{author}{\bibinfo{person}{Yahoo!}} \bibinfo{year}{[n. d.]}\natexlab{}.
\newblock \bibinfo{title}{{YCSB Workload}}.
\newblock \bibinfo{howpublished}{\url{https://github.com/brianfrankcooper/YCSB/tree/master/workloads}}.   (\bibinfo{year}{[n. d.]}).
\newblock


\bibitem[\protect\citeauthoryear{Yan, Yang, Zhang, Lin, Wong, Salem, and Brecht}{Yan et~al\mbox{.}}{2018}]%
        {sigmod18-carousel}
\bibfield{author}{\bibinfo{person}{Xinan Yan}, \bibinfo{person}{Linguan Yang}, \bibinfo{person}{Hongbo Zhang}, \bibinfo{person}{Xiayue~Charles Lin}, \bibinfo{person}{Bernard Wong}, \bibinfo{person}{Kenneth Salem}, {and} \bibinfo{person}{Tim Brecht}.} \bibinfo{year}{2018}\natexlab{}.
\newblock \showarticletitle{Carousel: Low-Latency Transaction Processing for Globally-Distributed Data}. In \bibinfo{booktitle}{{\em Proceedings of the 2018 ACM SIGMOD International Conference on Management of Data}}.
\newblock
\showURL{%
\url{https://dl.acm.org/doi/10.1145/3183713.3196924}}


\bibitem[\protect\citeauthoryear{Yang, Yan, and Wong}{Yang et~al\mbox{.}}{2022}]%
        {sigmod22-natto}
\bibfield{author}{\bibinfo{person}{Linguan Yang}, \bibinfo{person}{Xinan Yan}, {and} \bibinfo{person}{Bernard Wong}.} \bibinfo{year}{2022}\natexlab{}.
\newblock \showarticletitle{Natto: Providing Distributed Transaction Prioritization for High-Contention Workloads}. In \bibinfo{booktitle}{{\em Proceedings of the 2022 ACM SIGMOD International Conference on Management of Data}}.
\newblock
\showDOI{%
\url{https://doi.org/10.1145/3514221.3526161}}


\bibitem[\protect\citeauthoryear{YugabyteDB}{YugabyteDB}{2025}]%
        {yugabytedb}
\bibfield{author}{\bibinfo{person}{YugabyteDB}.} \bibinfo{year}{2025}\natexlab{}.
\newblock \bibinfo{title}{Isolation Levels}.
\newblock   (\bibinfo{year}{2025}).
\newblock
\showURL{%
\url{https://docs.yugabyte.com/preview/explore/transactions/isolation-levels/}}
\newblock
\shownote{Accessed: 2025-08-31.}


\bibitem[\protect\citeauthoryear{Zhang, Sharma, Szekeres, Krishnamurthy, and Ports}{Zhang et~al\mbox{.}}{2015}]%
        {sosp15-tapir}
\bibfield{author}{\bibinfo{person}{Irene Zhang}, \bibinfo{person}{Naveen~Kr. Sharma}, \bibinfo{person}{Adriana Szekeres}, \bibinfo{person}{Arvind Krishnamurthy}, {and} \bibinfo{person}{Dan R.~K. Ports}.} \bibinfo{year}{2015}\natexlab{}.
\newblock \showarticletitle{Building Consistent Transactions with Inconsistent Replication}. In \bibinfo{booktitle}{{\em Proceedings of the 25th ACM Symposium on Operating Systems Principles (SOSP 2015)}}.
\newblock
\showDOI{%
\url{https://doi.org/10.1145/2815400.2815404}}


\bibitem[\protect\citeauthoryear{Zheng, Tu, Kohler, and Liskov}{Zheng et~al\mbox{.}}{2014}]%
        {osdi14-silo}
\bibfield{author}{\bibinfo{person}{Wenting Zheng}, \bibinfo{person}{Stephen Tu}, \bibinfo{person}{Eddie Kohler}, {and} \bibinfo{person}{Barbara Liskov}.} \bibinfo{year}{2014}\natexlab{}.
\newblock \showarticletitle{Fast Databases with Fast Durability and Recovery Through Multicore Parallelism}. In \bibinfo{booktitle}{{\em Proceedings of the 11th USENIX Symposium on Operating Systems Design and Implementation (OSDI 2014)}}.
\newblock
\showURL{%
\url{https://www.usenix.org/conference/osdi14/technical-sessions/presentation/zheng_wenting}}


\bibitem[\protect\citeauthoryear{Zhou, Xu, Shraer, Namasivayam, Miller, Tschannen, Atherton, Beamon, et~al\mbox{.}}{Zhou et~al\mbox{.}}{2021}]%
        {sigmod21-foundationdb}
\bibfield{author}{\bibinfo{person}{Jingyu Zhou}, \bibinfo{person}{Meng Xu}, \bibinfo{person}{Alexander Shraer}, \bibinfo{person}{Bala Namasivayam}, \bibinfo{person}{Alex Miller}, \bibinfo{person}{Evan Tschannen}, \bibinfo{person}{Steve Atherton}, \bibinfo{person}{Andrew~J. Beamon}, {et~al\mbox{.}}} \bibinfo{year}{2021}\natexlab{}.
\newblock \showarticletitle{FoundationDB: A Distributed Unbundled Transactional Key-Value Store}. In \bibinfo{booktitle}{{\em Proceedings of the 2021 ACM SIGMOD International Conference on Management of Data}}.
\newblock
\showDOI{%
\url{https://doi.org/10.1145/3448016.3457559}}


\end{thebibliography}

\newpage
\appendix
\section*{Appendices}
\setlist[itemize]{leftmargin=10pt}
In this appendix, we include the following:
\begin{itemize}
    \item Appendix~\ref{appendix:tiga-algo} explains more details of \sysname's protocol in normal transaction processing and includes a more comprehensive version of pseudo-code. 
    % To Shrin (handle different failures)
    \item Appendix~\ref{appendix:recovery} discusses how \sysname handles different failures. Besides, it describes the detailed procedures (in pseudo-code) to replace leaders, recover logs and let failed servers rejoin the system.  
    \item  Appendix~\ref{appendix:correctness-proof} presents the proof of \sysname's correctness properties (per-shard linearizability and strict serializability).

    \item Appendix~\ref{appendix:incremental-hash} explains the details about how \sysname uses incremental hash to facilitate the quorum check.

    \item Appendix~\ref{appendix:optim-slow-reply} describes \sysname's optimization of slow path. 
    
    \item Appendix~\ref{appendix:decomposition} illustrates how \sysname decomposes dependent transactions into one-shot transactions, so that dependent transactions can be processed by \sysname.
    
    \item The TLA+ specification of \sysname is available at \href{https://github.com/New-Consensus-Concurrency-Control/Tiga-TLA-plus}{\color{blue}https://github.com/New-Consensus-Concurrency-Control/Tiga-TLA-plus}. 
\end{itemize}

\section{Normal Processing in \sysname}
\label{appendix:tiga-algo}
In \S\ref{sec:design}, we have included a simplified version of the pseudo-code (Algorithm~\ref{algo:server-action}) to describe the actions at servers to process transactions. Here, we continue to explain more details of \sysname's protocol design and implementation. 
Algorithm~\ref{algo:server-action-complete} and Algorithm~\ref{algo:coord-action-complete} describe the actions at servers and coordinators in more detailed pseudo-code. In the algorithm description, we omit the details of implementing message retransmission and reliable delivery (e.g., \textsc{send-message}).

In general, \sysname extends the view-based approach~\cite{viewstamp-original,viewstamp}. In each shard, every server maintains a local view identified by \emph{l-view}, and a global view identified by \emph{g-view}. 
In \sysname's implementation, \emph{g-view} is associated with an $m$-length vector \emph{g-vec}, recording the \emph{l-view} of each shard under this global view. More specifically, $shard_i$'s \emph{l-view} equals \emph{g-vec}[$i$].

\Para{View manager.} \sysname uses a stand-alone view manager to manage all the view information for \sysname's servers and coordinators. The view manager maintains a simple replicated state machine (i.e., \emph{<g-view, g-vec>} ) that is resilient to failures. The view manager can be built with ZooKeeper~\cite{zookeeper}, or backed by canonical consensus protocols like Paxos~\cite{paxos}, Raft~\cite{atc14-raft}, and Viewstamped Replication~\cite{viewstamp-original,viewstamp}. In our implementation, we implement the view manager by using Viewstamped Replication protocol. Besides, we also equip Viewstamped Replication with the crash vector technique~\cite{disc17-dcr-techreport}, so that Viewstamped Replication can maintain its state machine in the memory, without writing disks every time it updates \emph{<g-view, g-vec>}. The view manager initiates global view change for \sysname's servers in case of server failure or reconfiguration. After every global view change, each server's \emph{g-view} is incremented by 1; their \emph{l-view}s may remain unchanged or be incremented by at least 1. Note that the view manager is off the critical path of transaction processing in the common
cases so its performance has no significant impact.

\Para{Server action.} During normal processing, all servers share the same \emph{g-view}, as well as the same \emph{g-vec} because \emph{g-vec} is associated with \emph{g-view}. Besides, servers in the same shard also share the same \emph{l-view}, which decides the leader's \emph{replica-id} in this shard ($\emph{replica-id}=\emph{l-view}\%(2f+1)$),  but servers in different shards may have different \emph{l-view}s. (1) Inside each shard, servers communicate with the others under the same local view (e.g., the leader calls \textsc{send-log-sync} in Algorithm~\ref{algo:server-action-complete} to communicate with its followers); (2) Across shards, only leaders communicate with each other under the same global view (e.g., the leader calls \textsc{send-timestamp-notification}  in Algorithm~\ref{algo:server-action-complete} to communicate with the other leaders).

% Specifically, given two shards $shard_i$ and $shard_j$ under the same global view, their leaders share the same <\emph{g-view}, \emph{g-vec}>, but may have different local views, i.e., $\emph{l-view}_i$ and $\emph{l-view}_j$. $shard_i$'s leader looks up \emph{g-vec} to know which server is $shard_j$'s leader in this global view, i.e., $shard_j$'s leader has  \emph{replica-id}=$\emph{g-vec}[j]\% (2f+1)$. $shard_i$'s leader attaches its <\emph{g-view}, $\emph{l-view}_i$,  \emph{shard-id}, \emph{replica-id}> (\emph{shard-id=i}) in each message sent to $SG_j$'s leader. $SG_j$'s leader looks up its own <\emph{g-view}, \emph{g-vec}>, and it only processes the message if (1) the message's \emph{g-view} matches its \emph{g-view}; (2) the message's $\emph{l-view}_i$ matches its \emph{g-vec}[\emph{i}]; (3) the message's $\emph{replica-id}=\emph{l-view}\%(2f+1)$, i.e., the message comes from a leader. 

\Para{Coordinator action.} The coordinator needs to collect fast-replies/slow-replies from servers to conduct quorum check for its submitted transactions. Initially, the coordinator needs to contact the view manager to acquire the view information <\emph{g-view}, \emph{g-vec}>. Then, the coordinator only accepts the reply messages with the matching \emph{g-view} and \emph{l-view} (line~\ref{algoline:view-check-start}-\ref{algoline:view-check-end} in Algorithm~\ref{algo:coord-action-complete}). 

<\emph{g-view}, \emph{l-view}> is the key to guarantee both serializability across multiple shards and the linearizability in each shard. To commit a transaction, the coordinator must receive sufficient replies from every participating shard, and (1) all replies should have the same \emph{g-view}; (2) replies belonging to the same shard should also have the same \emph{l-view}. 
% %We explain details in \S\ref{sec:tiga-qc-fast} and \S\ref{sec:tiga-qc-slow}.

\captionsetup[algorithm]{font=small}
\begin{algorithm*}[!t]
\small 
%\small, \footnotesize, \scriptsize, or \tiny
\begin{multicols}{2}
\caption{Server action}
\label{algo:server-action-complete}
\textbf{Local State}: \Comment{} \\
\emph{s}, \Comment{the server's \emph{shard-id}} \\ 
\emph{r}, \Comment{the server's \emph{replica-id}, } 
\newline\hspace*{\algorithmicindent}\Comment{\emph{<s,r>} uniquely identify the server}
\\ 
\emph{l-view}, \Comment{the local view of this server} \\ 
\emph{g-view}, \Comment{the global view of this server} \\ 
\emph{pq}, \Comment{the server's priority queue} \\ 
\emph{g-vec}, \Comment{\emph{g-vec} records the local view of the each shard} \\ 
\emph{sync-point}, \Comment{the point to which the server's log is consistent  } \newline\hspace*{\algorithmicindent}\Comment{with the leader's log} \\ 

\begin{algorithmic}[1]
  \Event{$receiving$ $txn$ $T$}
    \State \textsc{$T'$}$\leftarrow$ last released txn that conflicts with \textsc{$T$}
    \If{\textsc{conflict-detection}($T$)} 
         $pq.\textbf{insert}(T)$
    \ElsIf{\textsc{am-leader}()}  \Comment{Only leaders can update $T.t$}
    \State $T.t\leftarrow \textsc{clock-time}()$
    \State $pq.\textbf{insert}(T)$
    \EndIf
  \EndEvent

\LeftComment{Periodically check the clock time}
  \Event{$clock\ time$ $progressing$}
    \State $nowTime\leftarrow\textsc{clock-time}()$
    \State $releaseTxns\leftarrow []$
    \LeftComment{Enumerate txns based on timestamp order}
    \For{$T \in \emph{pq}$}
        \If{$T.t > nowTime$}
           \textbf{break}  \Comment{$T$ has not expired}
        \EndIf
         \If{$\not\exists T'\in pq: T'.t<T.t \textbf{ and } T' \text{ conflicts with } T $} 
        \State \emph{releaseTxns}.\textbf{append}($T$)
        \EndIf
    \EndFor

    \For{$T\in \emph{releaseTxns}$}
    \State $\forall$ \emph{key}$\in$\emph{T.readSet}, \emph{rMap}[\emph{key}]$\leftarrow$\emph{T.t}

    \State $\forall$ \emph{key}$\in$\emph{T.writeSet}, \emph{wMap}[\emph{key}]$\leftarrow$\emph{T.t} \Comment{ Conflict detection}

        \If{\textsc{am-leader}()}    
         \State $ret\leftarrow\textsc{execute}(T)$ \Comment{Only leaders execute $T$}
        \State $hash=\textsc{calculate-hash}(log)$
        \State \textsc{send-fast-reply($T$, $hash$, $ret$)}
        \LeftComment{$tSet$ contains $T$'s timestamps used by each leader}
        \State $tSet\leftarrow$\textsc{timestamp-agreement}($T$) 
        \If{$T.t=\max \{t:t\in tSet\}$} 
        \LeftComment{This leader used correct (max) timestamp}
        \If{$tSet$.\textbf{size}()>1} \LeftComment{Some leaders used incorrect $T.t$}
        \LeftComment{After completing second round, leaders agree \newline\hspace*{2cm} on $T$'s timestamp, i.e., $T.t$}
        \State \textsc{timestamp-agreement}($T$) 
        \EndIf        
        \State Append $T$ to $log$ and syncs $T.t$ with followers
        \State $\emph{pq}$.\textbf{erase}($T$)
        \Else \Comment{This leader used smaller timestamp}
        \State $T.t\leftarrow \max \{t:t\in tSet\}$
        \State $\emph{pq}$.\textbf{reposition}($T$)
        \EndIf
        \Else \Comment{Follower sends fast-reply without execution}
        \State \textsc{send-fast-reply($T$, $hash$, $null$)}
        \State $\emph{pq}$.\textbf{erase}($T$)
        \EndIf
    \EndFor
  \EndEvent

    \LeftComment{Only leaders send/receive \textsc{timestamp-notification} to/from each other}
   \Event{$leader's\  receiving$ timestamp-notification, $msg$}
    \If{$m$.$\emph{g-view}\neq\emph{g-view}$}
        \State \Return \Comment{$msg$ does not come from the same global view}
    \EndIf
    \If{$m.r\neq\emph{g-vec}$[$msg.s$]$\%(2f+1)$}
        \State \Return \Comment{$msg$ does not come from the leader of $shard_s$}
    \EndIf
    \State Choose $T\in \emph{pq}:$ $T.\emph{txn-id}=msg.\emph{txn-id}$
    \LeftComment{Maintain a quorum set for each txn}
    \State $T.timestampQ\leftarrow T.timestampQ \cup \{msg\}$
    \If{\emph{T.timestampQ}.\textbf{size}()=\emph{T.shards}.\textbf{size}()}
        \LeftComment{All involved leaders' timestamps have been collected}
        \State $T.t\leftarrow\max\{mm.t:mm\in T.t \}$ 
        \LeftComment{The updated timestamp is commonly agreed}
    \EndIf
  \EndEvent

   \LeftComment{Only followers receive \textsc{log-sync} }
   \Event{\emph{follower's} \emph{receiving} log-sync, $msg$}
    \State Modify \emph{log} to keep consistent with leader's \emph{log}
    \State Update \emph{sync-point}
    \State \textsc{send-slow-reply}($T$)
  \EndEvent

\Function{timestamp-agreement}{$T$}
\If{$T.timestampQ$.\textbf{size}()<$T.shards$.\textbf{size}()}
\LeftComment{This is the first round of timestamp agreement}
    \State \textsc{broadcast-timestamp-notification}($T$)
    \State Wait until $T.timestampQ$.\textbf{size}()=$T.shards$.\textbf{size}()
    \State $tSet\leftarrow\{mm.t| mm\in T.timestampQ\}$
    \State \Return $tSet$
\Else \Comment{We have previously collected $T$'s timestamps}
\LeftComment{This is the second round of timestamp agreement,\newline\hspace*{0.5cm}because the first round detects inconsistent timestamps}
    \State $t_\emph{agreed}\leftarrow \max\{mm.t: mm\in T.timestampQ\}$
    \State $T.timestampQ\leftarrow\{mm: mm\in T.timestampQ \newline\hspace*{3.5cm} \textbf{and } mm.t=t_{agreed} \}$
    \State \textsc{broadcast-timestamp-notification}($T$)
    \State Wait until $T.timestampQ$.\textbf{size}()=$T.shards$.\textbf{size}()
    \State \Return $tSet$
\EndIf
\EndFunction

    \algstore{server-action-part-1}
\end{algorithmic}
\end{multicols}
\end{algorithm*}

\begin{algorithm*}[!t]
\small 
\begin{multicols}{2}
\begin{algorithmic}[1]
\algrestore{server-action-part-1}

    \Function{broadcast-timestamp-notification}{$T$}
        \State $msg.type\leftarrow\textsc{timestamp-notification}$
        \State $msg.\emph{g-view}\leftarrow\emph{g-view}$
        \State $msg.\emph{l-view}\leftarrow\emph{l-view}$
        \State $msg.s\leftarrow s$
        \State $msg.r\leftarrow r$
        \State $msg.\emph{txn-id}\leftarrow T.\emph{txn-id}$
        \State $msg.t\leftarrow T.t$
        \LeftComment{Notify the leaders of each shard that executes $T$}
        \For{\emph{ss}$\in$\emph{T.shards}}
            \State $rr\leftarrow$\emph{g-vec}[$ss$]$\%(2f+1)$ \LeftComment{<$ss,rr$> is the leader in $shard_{ss}$}
            \State \textsc{send-message}($msg$, <$ss,rr$>)\Comment{Send $msg$ to \emph{<ss, rr>}}
        \EndFor
    \EndFunction

\Function{send-fast-reply}{$T$, \emph{hash},\emph{result}}
    \State $msg.type\leftarrow\textsc{fast-reply}$
    \State $msg.\emph{g-view}\leftarrow\emph{g-view}$
    \State $msg.\emph{l-view}\leftarrow\emph{l-view}$
    \State $msg.s\leftarrow s$
    \State $msg.r\leftarrow r$
    \State $msg.\emph{txn-id}\leftarrow T.\emph{txn-id}$    
    \State $msg.\emph{t}\leftarrow T.\emph{t}$
    \State $msg.hash\leftarrow hash$
    \State $msg.ret\leftarrow result$
    \LeftComment{Send the reply to the coordinator (indexed by \emph{T.coord-id}) \newline\hspace*{0.5cm}which mutlicasts the txn}
    \State \textsc{send-message}($msg$, $T.\emph{coord-id}$)
\EndFunction

\Function{send-log-sync}{$T$}
    \State $msg.type\leftarrow\textsc{log-sync}$
    \State $msg.\emph{l-view}\leftarrow\emph{l-view}$
    \State $msg.s\leftarrow s$
    \State $msg.r\leftarrow r$
    \State $msg.\emph{log-pos}\leftarrow$ log position of $T$ in the leader's log list
    \State $msg.t\leftarrow T.t$
    \State $msg.\emph{txn-id}\leftarrow T.\emph{txn-id}$
    \For{\emph{rr}$\leftarrow$0 to $2f$}
        \If{$rr=r$}
        \State \textbf{continue}
        \EndIf
        \LeftComment{<$s,rr$> is a follower in $shard_{s}$}
        \State \textsc{send-message}($msg$, <$s,rr$>)
    \EndFor
\EndFunction

\Function{send-slow-reply}{$T$}
    \State $msg.type\leftarrow\textsc{slow-reply}$
    \State $msg.\emph{g-view}\leftarrow\emph{g-view}$
    \State $msg.\emph{l-view}\leftarrow\emph{l-view}$
    \State $msg.s\leftarrow s$
    \State $msg.r\leftarrow r$
    \State $msg.\emph{txn-id}\leftarrow T.\emph{txn-id}$
    \State $msg.t\leftarrow T.t$
    \State $msg.\emph{sync-point}\leftarrow \emph{sync-point}$
    \State \textsc{send-message}($msg$, $T.\emph{coord-id}$)
\EndFunction

\end{algorithmic}
\end{multicols}
\end{algorithm*}

\begin{algorithm*}[!t]
\small 
\begin{multicols}{2}
\caption{Coordinator action}
\label{algo:coord-action-complete}
\begin{algorithmic}[1]
  \Event{$receiving$ $a$ $txn$ $T$ $from$ $client$}
      \State  $T.\emph{shards}\leftarrow$ the \emph{shard-id}s of the servers that execute $T$
      \State $T.\emph{coord-id}\leftarrow$\emph{coord-id}
      \LeftComment{Refer to \S\ref{sec:timestamp-initilization} for initializing the timestamp}
      \State $T.t\leftarrow$\textsc{clock-time}()+\emph{estimated-delay}
      \For{$s\in T.\emph{shards}$}
        \For{$r\leftarrow$ 0 to $2f$}
            \State \textsc{send-message}($T$, <$s,r$>)
        \EndFor
      \EndFor
  \EndEvent
\LeftComment{Periodically, coordinator contacts view manager (\S\ref{sec:tiga-failure-recovery}) to get the current global view \emph{g-view}, and the vector \emph{g-vec}, which records each shard's local view. }
    
  \Event{$receiving$ fast-reply \emph{or} slow-reply, $msg$}
    \If{$msg.\emph{g-view}\neq\emph{g-view} \newline\hspace*{0.5cm}\textbf{ or } msg.\emph{l-view}\neq\emph{g-vec}[\emph{msg.shardId}]$} \label{algoline:view-check-start}
    \State \Return \Comment{Views mismatch, ignore the reply}
    \EndIf  \label{algoline:view-check-end}
    \State $replySet\leftarrow replySet \cup \{msg\}$
    \State $quorum\leftarrow \{mm \in replySet: mm.\emph{txn-id}=T.\emph{txn-id} \}$    
    \State  $leaderReplies\leftarrow\{\}$
    \For{\textbf{each} $shard_s$ that executes $T$}
        \If{$quorum$ has no $shard_s$'s leader's fast reply} 
            \State \Return \Comment{$T$ is not committed}
        \EndIf 
        \State Choose $mm \in quorum:$ $mm$ is $shard_s$'s leader's fast \newline\hspace*{1cm} reply
        \State $leaderReplies\leftarrow leaderReplies \cup\{ mm\}$
        \State $fastQ, slowQ\leftarrow 0,0$ 
        \For{$r$ $\leftarrow$0 to $2f$}
            \If{$quorum$ contains server $\emph{<s,r>}$'s slow-reply}
                 \State $slowQ\leftarrow slowQ+1$
                \LeftComment{Slow-reply can be used as fast-reply}
                \State $fastQ\leftarrow fastQ+1$
            \ElsIf{$quorum$ has server $\emph{<s,r>}$'s fast-reply and the \newline\hspace*{1.5cm}fast-reply has the same hash as $mm.hash$}
                \State $fastQ\leftarrow fastQ+1$
            \EndIf
       \EndFor
       \If{$fastQ <1+f+\lceil f/2 \rceil$ and $slowQ < f$}
       \State \Return \Comment{$T$ is not committed}
       \EndIf
    \EndFor
\State $tSet\leftarrow\{msg.t: msg\in leaderReplies \}$ \label{algo-line:t-agree-start}
\If{$tSet$.\textbf{size}()>1} \Comment{Leaders used different timestamps}
\LeftComment{Filter out those replies with incorrect (smaller) timestamp}
\State $invalidSet\leftarrow\{msg\in quorum: msg.t<\max tSet \}$
\State $replySet\leftarrow replySet - invalidSet$
\State \Return   \Comment{$T$ is not committed.} \label{algo-line:t-agree-end}
\EndIf
    \State $results\leftarrow\{ mm.ret: mm\in leaderReplies \}$
    \State Deliver $results$ to the client \Comment{$T$ is committed}
  \EndEvent
    
\end{algorithmic}
\end{multicols}
\end{algorithm*}

\section{Failure Recovery in \sysname}
\label{appendix:recovery}

\Para{Message drop.} Message drop does not affect \sysname's correctness. Message drop triggers the timeout at the coordinator, which resubmits the transaction with the same identifier but different timestamps. \sysname servers maintain at-most-once semantics to avoid duplicate execution of the same transaction.

\Para{Coordinator failure.} Coordinator failure does not affect \sysname's correctness. Whenever the coordinator crashes and is rebooted, the coordinator first contacts the view manager to acquire the fresh view information, i.e., \emph{<g-view, g-vec>}, and them resume its service. 

One case to note is that the coordinator can fail in the middle of multicasting transactions to multiple servers. In this case, the timestamp agreement (\circled{2} in Figure~\ref{fig:tiga-workflow}) can be blocked because some leaders receive the transaction whereas the other leaders not. To handle such case, each leader will launch a timer after it has received the first notification message from the other leaders but it has not received the corresponding transaction from the coordinator. Then when it is timeout and the transaction does not come, the leader will fetch the missing transaction from the other leaders which have sent it the notification message (because these leaders must have received the transaction).

\Para{Server failure.} There are two types of server failure in \sysname, i.e., follower failure and leader failure. Follower failure does not interrupt the service availability, but it may cause the transactions unable to be committed in the fast path, because some shards may have insufficient servers to establish a super quorum. Leader failure is more serious and it makes the shard fail to commit any transactions until (1) a new leader is elected from the followers in the same shard and (2) the new leader has recovered and executed the committed transactions before server failure. 

%We describe the details in  \S\ref{appendix:leader-election-recovery}.

% Besides, any failed server can rejoin \sysname as followers, and we describe the details in \S\ref{appendix:server-rejoin}.

% \subsection{Leader Election and Log Recovery}
% \label{appendix:leader-election-recovery}

\Para{Leader election and log recovery.}
\sysname uses the standalone view manager to detect server failure. We implement our view manager by using Viewstamped Replication protocol~\cite{viewstamp,disc17-dcr-techreport}. Algorithm~\ref{algo:recovery-cm-action} describes the action of view manager to detect \sysname server failure and issue global view change among \sysname servers.  Algorithm~\ref{algo:recovery-server-action} describes the servers' action during the global view change (i.e., selecting new leaders and recovering logs).  Here we only focus on how the view manager supports the failure recovery of \sysname servers, and omit the details on how Viewstamped Replication supports the failure recovery of our view manager. For example, when the VR protocol~\cite{viewstamp} is running in diskless mode, its correctness can be damaged by the \emph{stray messages} (i.e., messages that are sent by some servers before crash but are forgotten by the servers after they recover)~\cite{disc17-dcr-techreport, tdsn21-recovery} in the system. To tackle this, VR needs to be equipped with the crash vector technique~\cite{disc17-dcr-techreport} to preserve correctness. Here we do not discuss how to use crash vector to preserve VR correctness, and readers can refer to~\cite{disc17-dcr-techreport,nezha-tech-report} for more details of crash vector application.

\begin{algorithm*}[!t]
\small 
\caption{View manager (VMgr) action during global view change}
\begin{multicols}{2}
\label{algo:recovery-cm-action}
\textbf{Local State}: \Comment{} \\
\emph{v-view}, \Comment{the \emph{view-id} of VMgr} \\ 
\emph{v-rid}, \Comment{the \emph{replica-id} of the VMgr replica} \\ 
\emph{g-view}, \Comment{the current global view of \sysname servers} \\ 
\emph{g-vec}, \Comment{the current \emph{g-vec} of \sysname servers} \\ 
\emph{g-mode}, \Comment{the current mode for cross-shard coordination, \newline\hspace*{4cm}i.e., preventive or detective (\S\ref{sec:tiga-colocation-optimization})} \\ 
\emph{prepare-g-view}, \Comment{the prepared global view to be committed} \\ 
\emph{prepare-g-vec}, \Comment{the prepared \emph{g-vec} to be committed} \\
\emph{prepare-mode}, \Comment{the prepared mode for cross-shard coordination} \\ 
\begin{algorithmic}[1]
    \LeftComment{VMgr leader detects whether any leader servers in \sysname fails}
  \Event{$VMgr\ leader's\ failing\ to\ hear\ the\ heartbeat\ of\ some\ \newline\hspace*{0.8cm} leader\ server(s)\ in\ \sysname$}
    \LeftComment{Before issuing global view change, VMgr leader first persists the tentative \emph{g-view} and \emph{g-vec}}
        
      \State  \emph{prepare-g-view}$\leftarrow\emph{g-view}+1$
      \State  \emph{new-leader-rids}$\leftarrow\textsc{find-new-leaders()}$
      \LeftComment{Decide the new \emph{l-view}s based on the newly selected \newline\hspace*{0.5cm}leaders}
      \For{$r\leftarrow$ 0 to $2f$ }
        \State $r_{old}\leftarrow\emph{g-vec}[r]\%(2f+1)$ \Comment{Old leader's \emph{replica-id}}
        \State $r_{new}\leftarrow\emph{new-leader-rids}[r]$ \Comment{New leader's \emph{replica-id}}
        \State \emph{prepare-g-vec}[r]$\leftarrow \emph{g-vec}[r]+ (r_{new}-r_{old})\%(2f+1)$
        \LeftComment{Preventive mode is adopted when leaders can be \newline\hspace*{1cm} co-located, and detective mode is used otherwise.}
        \State \emph{prepare-mode}$\leftarrow$\textsc{preventive} or \textsc{detective}
      \EndFor
      \LeftComment{Initialize a quorum set to receive \textsc{prepare-reply}s}
      \State $\emph{prepareQuorum}\leftarrow\{\}$
      \LeftComment{Send prepare message to VMgr followers}
      \State \textsc{send-cm-prepare}()
  \EndEvent

\Event{$receiving$ \textsc{cm-prepare}, $msg$}
    \If{$msg.\emph{v-view}\neq \emph{v-view}$}
    \State \Return \Comment{Only process messages from the same \emph{v-view}}
    \EndIf
    \State $\emph{prepare-g-view}\leftarrow\emph{m.prepare-g-view}$
    \State $\emph{prepare-g-vec}\leftarrow\emph{m.prepare-g-vec}$
    \State \textsc{send-cm-prepare-reply}(\emph{msg.v-rid})
\EndEvent

\Event{\emph{$receiving$  \textsc{cm-prepare-reply}, $msg$}}
    \If{$msg.\emph{v-view}\neq \emph{v-view}$}
        \State \Return  \Comment{Views mismatch, ignore $msg$}
    \EndIf
    \If{$\emph{msg.prepare-g-view}\neq\emph{prepare-g-view}$\newline\hspace*{0.75cm}\textbf{ or }$\emph{msg.prepare-g-vec}\neq\emph{prepare-g-vec}$}
        \State \Return 
    \EndIf
    \State $\emph{prepareQuorum}\leftarrow\emph{prepareQuorum}\cup \{m\}$
    \If{$\emph{prepareQuorum}$.size()$\geq f+1$}
        \LeftComment{The prepared info has been persisted and can be used}
        \State $\emph{g-view}\leftarrow\emph{prepare-g-view}$
        \State $\emph{g-vec}\leftarrow\emph{prepare-g-vec}$
        \State $\emph{g-mode}\leftarrow\emph{prepare-g-mode}$
    \EndIf
    \LeftComment{Broadcast \textsc{view-change-req} to every \sysname server}
    \State \textsc{send-view-change-req}()
    \LeftComment{In parallel, VMgr leader also asks each VMgr follower to \newline\hspace*{0.5cm}commit the new <\emph{g-view},\emph{g-vec}> }
    \State \textsc{send-cm-commit}()
\EndEvent

\Event{\emph{$receiving$  \textsc{cm-commit}, $msg$}}
    \If{$msg.\emph{v-view}\neq \emph{v-view}$}
        \State \Return
    \EndIf
    \If{$\emph{msg.g-view}\neq\emph{prepare-g-view}$ \newline*\hspace{0.5cm} \textbf{ or } $\emph{msg.g-vec}\neq\emph{prepare-g-vec}$}
        \State \Return 
    \EndIf
    \LeftComment{VMgr follower commits the new state: \emph{<g-view, g-vec>}}
    \State $\emph{g-view}\leftarrow\emph{prepare-g-view}$
    \State $\emph{g-vec}\leftarrow\emph{prepare-g-vec}$
    \State $\emph{g-mode}\leftarrow\emph{prepare-g-mode}$

\EndEvent

\Function{find-new-leaders()}{}
    \For{$r\leftarrow 0$ to $2f$ }
        \If{$\forall s \in [0,m)$: server <$s, r$> is alive}
        \LeftComment{Servers $\emph{<*, r>}$ are co-located in the same region; we  \newline\hspace*{1cm}can choose these servers as new leaders for their shards}
        \State $\emph{new-leaders}\leftarrow [{\underbrace{r \dots r}_{m}}]$
        \State \Return \emph{new-leaders}
        \EndIf
    \EndFor
    \LeftComment{If every region has failed servers, then we first choose the \newline\hspace*{0.5cm}region with most alive servers as leaders}
    \State Choose $r\in [0,2f]$: region-$r$ has the most alive servers
    \For{$s\leftarrow$0 to $(m-1)$}
        \If{server <$s,r$> is alive}
            \State $\emph{new-leaders}[s]\leftarrow r$
        \Else 
            \State Choose $r'\in [0,2f]$: server $<s, r'>$ is alive
            \State $\emph{new-leaders}[s]\leftarrow r'$
        \EndIf 
    \EndFor
    \State \Return \emph{new-leaders}
\EndFunction

\algstore{algo:recovery-cm-action-part-1}
\end{algorithmic}
\end{multicols}
\end{algorithm*}

\begin{algorithm*}[!t]
\small 
\begin{multicols}{2}
\begin{algorithmic}
    
\algrestore{algo:recovery-cm-action-part-1}

\Function{send-cm-prepare()}{}
    \State $msg.type\leftarrow\textsc{cm-prepare}$
    \State $msg.\emph{v-view}\leftarrow\emph{v-view}$
    \State $msg.\emph{v-rid}\leftarrow\emph{v-rid}$
    \State $msg.\emph{prepare-g-view}\leftarrow\emph{prepare-g-view}$
    \State $msg.\emph{prepare-g-vec}\leftarrow\emph{prepare-g-vec}$
    \LeftComment{Broadcast the prepare message to every VMgr replica}
    \For{$rid\leftarrow 0$ to $2f$ }
    \State \textsc{send-message}($msg$, $rid$)
    \EndFor
\EndFunction

\Function{send-cm-prepare-reply(\emph{dest})}{}
    \State $msg.type\leftarrow\textsc{cm-prepare-reply}$
    \State $msg.\emph{v-view}\leftarrow\emph{v-view}$
    \State $msg.\emph{v-rid}\leftarrow\emph{v-rid}$
    \State $msg.\emph{prepare-g-view}\leftarrow\emph{prepare-g-view}$
    \State $msg.\emph{prepare-g-vec}\leftarrow\emph{prepare-g-vec}$
    \State $msg.\emph{prepare-mode}\leftarrow\emph{prepare-mode}$
    \LeftComment{Broadcast the prepare message to every VMgr replica}
    \For{$rid\leftarrow 0$ to $2f$ }
    \State \textsc{send-message}($msg$, $dest$)
    \EndFor
\EndFunction

\Function{send-cm-commit()}{}
    \State $msg.type\leftarrow\textsc{cm-commit}$
    \State $msg.\emph{v-view}\leftarrow\emph{v-view}$
    \State $msg.\emph{v-rid}\leftarrow\emph{v-rid}$
    \State $msg.\emph{g-view}\leftarrow\emph{g-view}$
    \State $msg.\emph{g-vec}\leftarrow\emph{g-vec}$
    \LeftComment{Broadcast the commit message to every VMgr replica}
    \For{$rid\leftarrow 0$ to $2f$ }
    \State \textsc{send-message}($msg$, $rid$)
    \EndFor
\EndFunction

\Function{send-view-change-req()}{}
\LeftComment{Broadcast the global view change requests to all \sysname servers}
    \State $msg.type\leftarrow\textsc{view-change-req}$
    \State $msg.\emph{g-view}\leftarrow\emph{g-view}$
    \State $msg.\emph{g-vec}\leftarrow\emph{g-vec}$
    \For{$s\leftarrow 0$ to $m-1$ }
        \For{$r\leftarrow 0$ to $2f$ }
        \State \textsc{send-message}($msg$, $\emph{<s,r>}$)
        \EndFor
    \EndFor
\EndFunction
\end{algorithmic}
\end{multicols}
\end{algorithm*}

%% TODO: Remember last-normal-view

\begin{algorithm*}[!htbp]
\small 
%\small, \footnotesize, \scriptsize, or \tiny

\begin{multicols}{2}
\caption{Server action during global view change}
\label{algo:recovery-server-action}
\textbf{Local State}: \Comment{} \\
\emph{s}, \Comment{the server's \emph{shard-id}} \\ 
\emph{r}, \Comment{the server's \emph{replica-id}} \\ 
\emph{status}, \Comment{\textsc{normal}, \textsc{recovering} or \textsc{viewchange} }  \\
\emph{l-view}, \Comment{the local view of this server} \\ 
\emph{g-view}, \Comment{the global view of this server} \\ 
\emph{g-vec}, \Comment{\emph{g-vec} records the local view of each shard} \\ 
\emph{g-mode}, \Comment{the current mode for cross-shard coordination}
\emph{sync-point}, \Comment{the point to which the server's log} \\ 
\hspace*{\algorithmicindent}\Comment{is consistent with the leader's log}\\
\emph{last-normal-view}, \Comment{The most recent \emph{l-view} in which }\\ \hspace*{\algorithmicindent}\hspace*{\algorithmicindent}\Comment{the server's \emph{status} is \textsc{normal} } \\
\emph{vQuorum}, \Comment{the set to collect \textsc{view-change} messages} \\
\emph{tQuorum}, \Comment{ the set to collect \textsc{timestamp-verification} messages}

\begin{algorithmic}[1]
  \Event{$receiving$ \textsc{view-change-req}, $msg$, $\emph{from}$ $\emph{VMgr}$}
    \If{$\emph{msg.g-view}\leq\emph{g-view}$}
        \State \textbf{return} \Comment{$msg$ is stale, ignore it}
    \EndIf
    \If{$status=\textsc{recovering}$}
    \LeftComment{Recovering server does not participate view change}
        \State \textbf{return}   
    \EndIf

    \LeftComment{Switch server status}
    \State $status\leftarrow\textsc{viewchange}$
    \LeftComment{Empty $pq$ and append these txns to log list}
    \State $\emph{pendingEntries}\leftarrow\textbf{sort}(\emph{pq})$
    \State $\emph{pq}.\textbf{clear}()$
    \State $log\leftarrow log.\textbf{append}(\emph{pendingEntries}) $
    \State $\emph{g-view}\leftarrow\emph{msg.g-view}$
    \State $\emph{g-vec}\leftarrow\emph{msg.g-vec}$
    \State $\emph{g-mode}\leftarrow\emph{msg.g-mode}$
    \State $\emph{l-view}\leftarrow\emph{g-vec}$[$s$]
    \State $\emph{leader-r}\leftarrow\emph{l-view}\% (2f+1)$ \Comment{The new leader's \emph{replica-id}}
    \LeftComment{Send \textsc{view-change} message to the new leader}
    \State \textsc{send-view-change}(\emph{<s, leader-r>})
  \EndEvent

  \Event{$receiving$ \textsc{view-change}, $msg$}
    \If{$\emph{msg.g-view}<\emph{g-view}$}
        \State \textbf{return}
    \EndIf
    \If{\emph{status}=\textsc{recovering}}
        \State \textbf{return} 
    \EndIf
     \State $status\leftarrow\textsc{viewchange}$
    \If{$\emph{msg.g-view}>\emph{g-view}$}
    \Comment{Restart a new view }
       
        \State $\emph{g-view}\leftarrow\emph{msg.g-view}$
        \State $\emph{g-vec}\leftarrow\emph{msg.g-vec}$
        \State $\emph{g-mode}\leftarrow\emph{msg.g-mode}$
        \State $\emph{l-view}\leftarrow\emph{g-vec}$[$s$]
        \State $\emph{vQuorum}\leftarrow \emptyset$
    \EndIf
    \State  $\emph{vQuorum}\leftarrow\emph{vQuorum}\cup \{m\}$
    \If{$|vQuorum|=f+1$}
        \State \textsc{rebuild-log}()
        \State \textsc{verify-timestamp-across-shards}()
        \LeftComment{Broadcast the new log list to followers}
        \State \textsc{send-start-view}() 
    \EndIf
  \EndEvent

  \Event{$receiving$ \textsc{timestamp-verification}, $msg$}
     \If{$status\neq\textsc{viewchange}$ \textbf{or} $msg.\emph{l-view}\neq \emph{l-view}$}
        \State \textbf{return}
     \EndIf
     \State $\emph{tQuorum}\leftarrow\emph{tQuorum}\cup\{msg\}$
  \EndEvent

  \Event{$receiving$ \textsc{start-view}, $msg$}
     \If{$status\neq\textsc{viewchange}$ \textbf{or} $msg.\emph{l-view}\neq \emph{l-view}$}
        \State \textbf{return}
     \EndIf
     \State $\emph{log}\leftarrow msg.log$
     \State $\emph{sync-point}\leftarrow msg.log.\textbf{length}()-1$ 
     \State $\emph{last-normal-view}\leftarrow\emph{l-view}$
     \State $status\leftarrow\textsc{normal}$
  \EndEvent

\Function{rebuild-log()}{}
\State $\emph{new-log}\leftarrow[ ]$ \Comment{Initialize a new log list to build}
\State $\emph{largest-normal-view}\leftarrow\max\{msg.\emph{lnv} | msg \in vQuorum\}$ 
\State $\emph{largest-sync-point}\leftarrow\max\{msg.sp | msg \in vQuorum \newline\hspace*{3.5cm} \textbf{and  } msg.lnv= \emph{largest-normal-view}\}$
\LeftComment{Recover Part (a) logs (refer to the third step in \S\ref{sec:tiga-failure-recovery})}
\State Choose $msg\in vQuorum: msg.\emph{sp}=\emph{largest-sync-point}$ \label{algo-line:part-1-start}
\For{$i\leftarrow 0$ to \emph{largest-sync-point} }
    \State \emph{new-log}.\textbf{append}($msg.log$[$i$])
\EndFor \label{algo-line:part-1-end}
\LeftComment{Recover Part (b) logs}
\State $t\leftarrow \max\{msg.t: msg\in \emph{new-log}\}$  \label{algo-line:part-2-log-start}
\State $\emph{candidateLogs}\leftarrow\{\}$ 
\For{$msg \in vQuorum$}
\For{$i\leftarrow \emph{msg.sync-point}+1$ to $msg.log$.\textbf{length}()-1 }
\State $\emph{candidateLogs}\leftarrow\emph{candidateLogs}\cup\{msg.log[i]\}$
\EndFor
\EndFor
\State $\emph{committedLogs}\leftarrow\{\}$
\For{$e\in\emph{candidateLogs}$}
\LeftComment{Check how many servers in \emph{vQuorum} have $e$ in their logs}
\State $Q\leftarrow\{msg\in vQuorum: e\in msg.log \}$
\If{$|Q|\geq \left \lceil{f/2}\right \rceil + 1 $}
\State  $\emph{committedLogs}\leftarrow\emph{committedLogs}\cup\{e\}$
\EndIf
\EndFor 
\LeftComment{Sort logs by their timestamps}
\State $logList\leftarrow\textbf{sort}(\emph{committedLogs})$ \Comment{This is Part (b) logs} \label{algo-line:part-2-log-end}
\State $\emph{new-log}.\textbf{append}(logList)$ \Comment{Concat two parts of logs}
\State $log\leftarrow \emph{new-log}$ \Comment{Replace the old log list}
\EndFunction

\Function{verify-timestamp-across-shards()}{}
\State \textsc{send-timestamp-verification}()
\State Wait until $|\emph{tQuorum}|=m-1$ 
\State $\emph{logSet}\leftarrow\{log[i]|i\in 0\cdots log.\textbf{length}()-1\}$
\State $\emph{allTxnIds}\leftarrow \{\} $
\For{$msg\in tQuorum$}
\For{$i\leftarrow 0$ to $msg.info$.\textbf{length}()-1 }
\State $\emph{allTxnIds}\leftarrow\emph{allTxnIds}\cup\{msg.info[i].id\}$
\EndFor
\EndFor
\State $\emph{missingTxnIds}\leftarrow allTxnIds-\{txn.\emph{txn-id}| txn\in logSet\}$
\State Fetch missing transactions from other shards based on \newline\hspace*{0.5cm}\emph{missingTxnIds}, and add to \emph{logSet}
\For{$txn\in logSet$}
\LeftComment{If multiple shards have this txn with different timestamps, \newline\hspace*{0.5cm}pick the largest one as the agreed timestamp}
\For{$mm\in tQuorum$}
\If{$\exists e\in mm.info: e.id = txn.\emph{txn-id}$}
\State $txn.t\leftarrow\max\{e.t, txn.t\}$
\EndIf
\EndFor
\EndFor
\LeftComment{Sort $logSet$ by (verified) timestamps}
\State $log\leftarrow\textbf{Sort}(logSet)$
\EndFunction

\algstore{part1}
\end{algorithmic}
\end{multicols}
\end{algorithm*}

\begin{algorithm*}[!t]
\small
\begin{multicols}{2}
\begin{algorithmic}[1]
\algrestore{part1}

\Function{send-view-change(\emph{<ss,rr>})}{}
\LeftComment{Send \textsc{view-change} message to server \emph{<ss,rr>}}
    \State $msg.type\leftarrow\textsc{view-change}$
    \State $msg.\emph{g-view}\leftarrow\emph{g-view}$
    \State $msg.\emph{l-view}\leftarrow\emph{l-view}$
    \State $msg.\emph{g-mode}\leftarrow\emph{g-mode}$
    \State $msg.\emph{lnv}\leftarrow\emph{last-normal-view}$
    \State $msg.log\leftarrow log$
    \State $msg.\emph{sp}\leftarrow\emph{sync-point}$
    \State \textsc{send-message}($msg$, \emph{<ss,rr>})
\EndFunction

\Function{send-timestamp-verification()}{}    
\For{$ss\leftarrow 0$ to $m-1$ }
    \If{$ss=s$}
    \State \textbf{continue} \Comment{No need to send the message to itself}
    \EndIf
    \State $msg.type\leftarrow\textsc{timestamp-verification}$
    \State $msg.\emph{g-view}\leftarrow\emph{g-view}$
    \State $msg.\emph{l-view}\leftarrow\emph{l-view}$
    \State $msg.\emph{s}\leftarrow\emph{s}$
    \State $msg.\emph{r}\leftarrow\emph{r}$
    \State $msg.info\leftarrow []$
    \For{$i\leftarrow 0$ to $log$.\textbf{length}()-1 }
        \State $txn\leftarrow logs[i]$
        \If{$txn$ $involves$ $shard_s$}
           \State $\emph{msg.info}.$\textbf{append}(\{$id:txn.\emph{txn-id}$, $t:txn.t$\})
        \EndIf
    \EndFor
    \LeftComment{Calculate \emph{replica-id} of $shard_ss$'s leader}
    \State $rr\leftarrow \emph{g-vec}[ss]\% (2f+1)$
    \State \textsc{send-message}(($msg$, \emph{<ss,rr>})
\EndFor
\EndFunction

\Function{send-start-view()}{} 
    \State $msg.type\leftarrow\textsc{start-view}$
    \State $msg.\emph{g-view}\leftarrow\emph{g-view}$
    \State $msg.\emph{l-view}\leftarrow\emph{l-view}$
    \State $msg.\emph{g-mode}\leftarrow\emph{g-mode}$
    \State $msg.\emph{s}\leftarrow\emph{s}$
    \State $msg.\emph{r}\leftarrow\emph{r}$
    \State $msg.\emph{log}\leftarrow\emph{log}$
    \For{$rr\leftarrow 0$ to $2f$}
        \If{$rr=r$}
        \State \textbf{continue}
        \EndIf
        \State \textsc{send-message}(($msg$, \emph{<s,rr>})
    \EndFor
\EndFunction

\end{algorithmic}
\end{multicols}
\end{algorithm*}

% \subsection{Server Rejoins}
% \label{appendix:server-rejoin}

\Para{Server rejoins.} A failed server can rejoin the its shard as a follower. After running the rejoining procedure (described in Algorithm~\ref{algo:recovery-server-action}), a server switches from \textsc{recovering} status to \textsc{normal} status with the proper log list, and then work as a follower to continue processing transactions.

% No need for crash vector

% In general, the algorithm follows the approach used by viewstamped replication (VR)~\cite{viewstamp}. 

%  It is worth noting that when VR is running without stable storage, then its correctness might be damaged by the \emph{stray messages} (i.e., messages that are sent by some servers before crash but are forgotten by the servers after they recover)~\cite{disc17-dcr-techreport, tdsn21-recovery} in the system. Therefore, we use the crash vector technique~\cite{disc17-dcr-techreport,vldb23-nezha} to detect stray messages and preserve the correctness

% However, the prior works~\cite{disc17-dcr-techreport,tdsn21-recovery} have observed that VR's correctness can be damaged by \emph{stray messages}  
% The prior works~\cite{disc17-dcr-techreport,vldb23-nezha} have discussed the use of crash vectors to avoid the misleading effect of \emph{stray messages}. Here we omit it in our algorithm description for simplicity.

\begin{algorithm*}[!t]

\small 
%\small, \footnotesize, \scriptsize, or \tiny

\begin{multicols}{2}
\caption{Server action during rejoining \sysname}
\label{algo-replica-rejoin}
\textbf{Local State}: \Comment{} \\
\emph{s}, \Comment{the server's \emph{shard-id}} \\ 
\emph{r}, \Comment{the server's \emph{replica-id}} \\ 
\emph{status}, \Comment{\textsc{normal}, \textsc{recovering} or \textsc{viewchange} }  \\
\emph{l-view}, \Comment{the local view of this server} \\ 
\emph{g-view}, \Comment{the global view of this server} \\ 
\emph{g-vec}, \Comment{\emph{g-vec} records the local view of each shard} \\ 
\emph{g-mode}, \Comment{the current mode for cross-shard coordination} \\
\emph{sync-point}, \Comment{the point to which the server's log} \\ 
\hspace*{\algorithmicindent}\Comment{is consistent with the leader's log}\\
\emph{last-normal-view}, \Comment{The most recent \emph{l-view} in which }\\ \hspace*{\algorithmicindent}\hspace*{\algorithmicindent}\Comment{the server's \emph{status} is \textsc{normal} } \\
\begin{algorithmic}[1]
  \Event{$status=$\textsc{recovering}}
    \State $msg'\leftarrow$\textsc{inquire-view-manager()}
    \State $\emph{g-view}\leftarrow msg'.\emph{g-view}$
    \State $\emph{g-vec}\leftarrow msg'.\emph{g-vec}$
    \State $\emph{g-mode}\leftarrow msg'.\emph{g-mode}$
    \State $\emph{l-view}\leftarrow \emph{g-vec}[s]$
    \LeftComment{Compute the leader server's \emph{replica-id} of $shard_s$}
    \State $\emph{leader-r}\leftarrow \emph{g-vec[s]}\ \%\  (2f+1)$
    \State \textsc{state-transfer}(\emph{<s, leader-r>})
  \EndEvent

\Function{inquire-view-manager()}{}
    \State $msg.type$$\leftarrow$\textsc{inquire-req}
    \State $msg.s\leftarrow s$
    \State $msg.r\leftarrow r$
    \State Send the inquiry $msg$ to any replica of the view manager (if the replica is a follower, it will forward the inquiry to the leader of the view manager.)
    \State Wait until receiving reply $msg'$ containing \emph{<g-view},\emph{g-vec>} 
    \State \Return $msg'$
\EndFunction

\Function{state-transfer}{$\emph{<ss,rr>}$}
    \State $msg.type\leftarrow $\textsc{state-transfer-req}
    \State $msg.\emph{g-view}\leftarrow \emph{g-view}$
    \State $msg.\emph{l-view}\leftarrow \emph{l-view}$
    \State $msg.s\leftarrow s$
    \State $msg.r\leftarrow r$
    \State \textsc{send-message}($msg$, $\emph{<ss,rr>}$)
    \State Wait until $status=\textsc{normal}$
    \State \Return
\EndFunction

   \Event{\emph{$receiving$} state-transfer-req, $msg$}
    \If{$status\neq\textsc{normal}$}
    \State \Return
    \EndIf
    \If{$\emph{g-view}\neq \emph{msg.g-view} $ \textbf{or} $\emph{l-view}\neq\emph{msg.l-view} $}
    \State \Return
    \EndIf
    \State $msg'.type\leftarrow \textsc{state-transfer-rep}$
    \State $msg'.log\leftarrow log$
    \State $msg'.\emph{g-view}\leftarrow  \emph{g-view}$
    \State $msg'.\emph{l-view} \leftarrow  \emph{l-view}$
    \State $msg'.sp\leftarrow  \emph{sync-point}$
    \State \textsc{send-message}($msg',\emph{<msg.s, msg.r>}$)
  \EndEvent
 
    \Event{\emph{$receiving$} state-transfer-rep, $msg$}
    \If{$status\neq\textsc{recovering}$}
    \State \Return
    \EndIf
    \If{$\emph{g-view}\neq \emph{msg.g-view} $ \textbf{or} $\emph{l-view}\neq\emph{msg.l-view} $}
    \State \Return
    \EndIf
    \State $log\leftarrow msg.log$
    \State $\emph{last-normal-view}\leftarrow\emph{l-view}$
    \State $\emph{sync-point}\leftarrow msg.sp$
    \State $status\leftarrow \textsc{normal}$
    \Comment{Rejoin as a \textsc{normal} follower}
  \EndEvent

  % \algstore{replica-rejoin}
    
\end{algorithmic}
\end{multicols}
\end{algorithm*}

\newpage
\section{Correctness Proof of \sysname}
\label{appendix:correctness-proof}

% \newtheorem{theorem}{Theorem}
% \newtheorem{lemma}{Lemma}

% Strict serializability implies both serializability and linearizability~\cite{jepson-strict serializability, linearizability}. Therefore, we decompose the correctness proof of \sysname into two steps. 

\sysname guarantees two correctness properties, i.e., per-shard linearizability and strict serializability across shards. 
\begin{itemize}
    \item (Per-shard linearizability) All committed transactions always satisfy linearizability in each shard. 
    \item (Strict Serializability)  The execution results for all committed transactions across multiple shards are strictly serializable.
\end{itemize}

% Combing Step 1 and Step 2, we have proved that \sysname guarantees strict serializability for all the committed transactions. 

\subsection{Proof of Per-Shard Linearizability} 

\begin{lemma}[Durability]
\label{lemma:durability}
If a transaction $T$ is committed in $shard_i$ under the global view $gv_1$, then $T$ can always be recovered in any follow-up views $gv_2\geq gv_1$. 
\end{lemma}

First of all, a minority ($\leq f$) of  follower failure does not interrupt service availability of \sysname, so it does not trigger global view change. As a result, there is no violation of durability. We focus on the cases when the leader fails or encounters network partition. In both cases, the view manager will launch a global view change and leads to the election of new leaders and the reconstruction of log lists in the new global view. Based on which path is used to commit $T$ in $shard_i$, we discuss two different cases.

\Para{Case-1: $T$ is committed in the fast path.} According to the super quorum check in the fast path (\S\ref{sec:tiga-qc-fast}), $T$ can be committed through the fast path if the leader and $f+\lceil f/2 \rceil$ followers have the \textbf{same} log lists containing $T$. "\textbf{Same} log lists" indicates that $T$ has the same timestamp on these $1+ f+\lceil f/2 \rceil$ servers. Because at most $f$ servers simultaneously fail in $shard_i$, then based on quorum intersection, among the $f+1$ servers that participate in the recovery, there are at least  $1+ f+\lceil f/2 \rceil -f = \lceil f/2 \rceil+1$ servers containing $T$. Therefore, $T$ can be recovered from any $f+1$ servers that are selected to rebuild the log list (refer to line~\ref{algo-line:part-2-log-start}--\ref{algo-line:part-2-log-end} in Algorithm~\ref{algo:recovery-server-action}).

\Para{Case-2: $T$ is committed in the slow path.} According to the quorum check in the slow path (\S\ref{sec:tiga-qc-slow}), Servers' advancing their sync-points \emph{happens before} sending the  slow-replies. In other words, when $T$ is committed in the slow path, there have been at least $f+1$ servers whose sync-points surpasses the position of $T$ in the servers' log lists. Based on quorum intersection, among the $f+1$ server that participate in the recovery, there is at least $(f+1)+(f+1)-(2f+1)=1$ server's sync-point that surpasses the position of $T$ in the server's log list. Since the new leader will pick the server holding largest sync-point among the participating servers, and copy all the log entries from the corresponding server up to sync-point, so $T$ will be included in the recovered log list (refer to line~\ref{algo-line:part-1-start}--\ref{algo-line:part-1-end} in Algorithm~\ref{algo:recovery-server-action}).  

Combing Case-1 and Case-2, we have proved that the committed transactions can always be recovered after any view change (due to server failure or network partition).

\begin{lemma}[Consistency]
\label{lemma:consistency}
If a transaction $T$ is committed in $shard_i$ under a global view $gv_1$, then its execution result will remain the same in any follow-up views $gv_2\geq gv_1$. 
\end{lemma}

(1) Given a committed transaction $T$ in the global view $gv_1$, we can also prove that any transaction $T'$ executed before $T$ on the leader is also committed in $gv_1$. More specifically, if $T$ is committed through the fast path, then there are $f+\lceil f/2 \rceil +1$ servers (including the leader) have consistent log lists up to $T$. Since the log list up to $T'$ is a sub-sequence, these  $f+\lceil f/2 \rceil +1$ servers also have the consistent log list up to $T'$, so $T'$ is also a committed transaction through the fast path. If $T$ is committed through the slow path, then the recovered sync-point, which surpasses the log position of $T$, will also surpass the log position of $T'$. Therefore, $T'$ is also a committed transaction in the slow path.

(2) Because the timestamp is a property of the transaction, the timestamp will also be recovered if the committed transaction is recovered in the new view (Durability). Therefore, all the committed transactions in the global view $gv_1$ will remain the same timestamp in the new view $gv_2$. Since the leader executes the transactions based on the timestamp order, then for any transaction $T'$, which is committed in the view $gv_1$ and executed before $T$ (i.e., $T'$ has a smaller timestamp), the transaction $T'$ will still be recovered with the same timestamp and executed before $T$. 

Combining (1) and (2), we conclude that given a transaction $T$ committed in the global view $gv_1$, all the transactions executed before $T$ by the leader in $gv_1$ will are also committed and be executed before $T$ by the new leader in the follow-up global view $gv_2$.  

(3) Therefore, the only case that can cause the execution inconsistency to $T$ is that, there exists another conflicting transaction $T''$, which has not been executed by the leader in the global view $gv_1$, but is recovered on the new leader in $gv_2$ with a smaller timestamp than $T$, making $T''$ executed before $T$ in the new global view $gv_2$. However, we will prove by contradiction that such cases can never happen.

\Para{Case-1: $T$ is committed in the fast path.} Assume there exists another $T''$ with a smaller timestamp than the committed transaction $T$, but has not been executed by the leader in the old global view $gv_1$. In order for $T''$ to be recovered in the new view $gv_2$, $T''$ must exist in the log lists of at least $\lceil f/2 \rceil +1$ servers which have survived from $gv_1$ to $gv_2$. However, if these $\lceil f/2 \rceil +1$ servers contains $T''$ during the global view $gv_1$, then their log lists will be inconsistent with the leader in the view $gv_1$ because the leader's log list does not contain $T''$ ahead of $T$. As a result, the coordinator can only obtain at most $(2f+1)-(\lceil f/2 \rceil +1)=f+\lfloor f/2 \rfloor$ fast-replies containing the same hash (i.e., indicating the servers have the same log lists). Thus, $T$ cannot be committed in the fast path, which contradicts the assumption of Case-1. 

\Para{Case-2: $T$ is committed in the slow path.} Still assume there exists such a $T''$ with a smaller timestamp than $T$, but has not been executed by the leader in the old global view $gv_1$. Since $T$ is committed in the slow path, $T$ and the other committed transactions before $T$ are recovered by using the largest sync-point from the remaining $f+1$ servers that participate in the recovery. If $T''$ is recovered after the sync-point, then $T''$ has a larger timestamp than $T$ because transactions are appended to log lists according to their timestamp order. On the other hand, if $T''$ is also recovered prior to the sync-point, then it indicates there exists at least one follower in the previous global view $gv_1$, whose log list contains $T''$ even after it has ensured the log list is consistent with the leader's log list. Further, it can be derived that the leader's log list also contains $T''$ with a smaller timestamp than $T$. As a result, the leader in the global view $gv_1$ should have executed $T''$ before $T$, which contradicts the assumption that the old leader has not executed $T''$.

Combining Case-1 and Case-2, we have proved that there does not exist such a transaction $T''$ which has not been executed before a committed transaction $T$ in an old global view but is executed before $T$ in a new global view.

Finally, combining (1), (2) and (3), we have proved the consistency property in each shard of \sysname.

\begin{theorem}[Per-Shard Linearizability] 
\label{thm:linearizability}
All committed transactions always satisfy linearizability in each shard. 
\end{theorem}

According to~\cite{nsdi19-curp}, the linearizability property can reworded as: Given two committed transactions $T_1$ and $T_2$, if the execution effect of $T_1$ is observed by $T_2$, then no contrary observation can occur afterwards (i.e., it should not appear to revert or be reordered).

Based on the design of \sysname, for any shard in \sysname, if $T_2$ can observe $T_1$'s execution effect before $T_2$ is executed, then $T_2$ must have a larger timestamp than $T_1$. 

Based on the proved durability property (Lemma~\ref{lemma:durability}), $T_1$ and $T_2$ will be recovered in the new global view with the same timestamps as in the previous global view that they have been committed, so their execution will not be reordered, i.e., $T_2$ will still observe $T_1$'s execution effect, rather than that $T_1$ observes $T_2$'s execution effect. 

Based on the proved consistency property (Lemma~\ref{lemma:consistency}), $T_1$'s execution result will remain unchanged in the new global view, so $T_2$ will continue to observe the \emph{same} execution effect (result) of $T_1$ and there is no contrary observation. Therefore, linearizability is guaranteed in each shard of \sysname.

% Step 1 has already been proved in Nezha~\cite{vldb23-nezha,nezha-tech-report} (refer to Appendix B in~\cite{nezha-tech-report}). Each shard is one Nezha instance containing $2f+1$ replicas. \cite{nezha-tech-report} has proved that the committed transactions in Nezha can be survived when at most $f$ replicas fail and linearizability is preserved after recovery. The linearizable order is the transaction's agreed timestamp order. 

% Based on \sysname's design, only leaders will execute the transactions, so we continue to prove Step 2: the transactions' execution in the leader's shard satisfy strict serializability. 

% Formally, strict serializability can be decomposed into two requirements~\cite{jepson-strict-serializability,linearizability,sigmod95-clocc}, i.e., serializability and linearizability (i.e., real-time ordering). We first prove by contradiction that the execution of transactions by \sysname's leader servers satisfies serializability. Based on that, we continue to prove the serializable execution also meets the linearizability requirement.

\subsection{Proof of Strict Serializability}. 
\label{appendix:proof-strict-serializability}
We decompose the proof of strict serializability into two steps. First, we prove \sysname guarantees serializability for all transactions and the serializable order is the order of transactions timestamp order. Second, we continue to prove the timestamp order is strictly serializable, i.e., transactions' real-time order does not contradict the serializable order.   

\begin{lemma}[Serializability]
\label{lemma:serializability}
The execution of all committed transactions in \sysname is serializable.
\end{lemma}

We use the serializability graph~\cite{berstein-chapter,bernstein-book} to prove serializability. Serializability graph considers every transaction as one vertex in the graph, and adds directional edges between conflicting (non-commutative) transactions to represent the \emph{happened-before} order: Given two conflicting transactions $T_1$ and $T_2$, if $T_1$ is executed earlier than $T_2$ on one leader server, then there is an edge from $T_1$ to $T_2$. According to the Serializability Theorem~\cite{berstein-chapter}, the execution of transactions is serializable if the serializability graph is acyclic. We prove it by contradiction.

% Here the exeuction is completed after timestamp agreement

Suppose there is a cycle in the serializability graph representing \sysname's execution. Without the loss of generality, we assume a pair of conflicting transactions are involved in the cycle, denoted as $T_1$ and $T_2$. Thus the serializability graph can be described as $T_1\rightarrow\cdots T_2\rightarrow\cdots T_1$. 

According to \sysname's protocol design, each leader executes the transactions following the order of their timestamps after timestamp agreement (\S\ref{sec:timestamp-agreement}).
From $T_1\rightarrow\cdots T_2$, we know that there must exist one leader server, denoted as $leader_1$, which executes $T_1$ earlier than $T_2$, so $T_1$'s timestamp is smaller than $T_2$'s timestamp on $leader_1$, denoted as $t_{1}^1<t_{2}^1$. But on the other hand, from $T_2\rightarrow\cdots T_1$, we know there is another leader server, denoted as Leader-2, where $T_2$'s timestamp is smaller than $T_1$'s timestamp, denoted as $t_{2}^2<t_{1}^2$ (the subscripts indicate transaction identifier and the superscripts indicate the shard identifier). 

For any committed transaction, the coordinator considers it committed after the coordinator has verified that all shards are using the same timestamp (Line~\ref{algo-line:t-agree-start}-\ref{algo-line:t-agree-end} in Algorithm~\ref{algo:coord-action-complete}).\footnote{The coordinator might know this agreement earlier than leaders: At this point, leaders themselves might be still uncertain whether they already used the agreed timestamp if they have not completed timestamp agreement.}
Therefore, we have $t_{1}^1=t_{1}^2$ and $t_{2}^1=t_{2}^2$. Then we come to the contradiction: both $t_{1}^1<t_{1}^2$ and $t_{1}^1>t_{1}^2$ hold simultaneously. 

Therefore, the assumption is not true. The serializability graph is acyclic, which indicates the execution by \sysname's leader servers is serializable, which follows the transactions' timestamp order after timestamp agreement.

\begin{theorem}[Strict Serializability]  The execution of all committed transactions in \sysname is strictly serializable. In other words, the execution of all committed transactions respects their real-time ordering. 
\end{theorem}

\Para{Notation.} For every transaction $T_i$, we use $\textsc{start}_i$ to represent the real time that $T_i$ is started, and use $\textsc{complete}_i$ to represent the real time that $T_i$ is completed. $T_i$ involves a set of shards for joint execution, so we use $\emph{ShardSet}_i$ to represent the \emph{shard-id}s of $T_i$'s involved shards. For an involved shard $\emph{shard}_s$, we use $\textsc{exec}_{i}^{s}$ to represent the real time when $T_i$ is optimistically executed on $\emph{shard}_s$'s leader $L_s$ ($s\in\emph{ShardSet}_i$), and we use $\textsc{deq}_{i}^{s}$ to represent the real time when $T_i$ dequeued from $L_s$'s queue. Some transactions might be re-executed if its first optimistic execution is invalid and revoked. Here we do not consider the invalid execution, and only consider the $T_i$'s final optimistic execution that is eventually committed. In this execution, $T_i$ used the agreed timestamps across all leaders, $T_i.t=T_i.t_{agreed}$. Thus, we have $\forall s\in\emph{ShardSet}_i: \textsc{start}_i<\textsc{exec}_{i}^{s}<\textsc{complete}_i$.

From Lemma~\ref{lemma:serializability}, we know \sysname has a valid serializability schedule according to the transactions' agreed timestamp order. Therefore, given two transactions in the serializable schedule, i.e., $T_i$ and $T_j$ ($i<j$ and $T_i.t_{agreed}<T_j.t_{agreed}$), we represent the dependency relationship between them as $T_i\rightarrow T_{i+1}\rightarrow\cdots \rightarrow T_{j-1}\rightarrow T_{j}$, with each two consecutive transactions conflict on some data items. To prove by contradiction, we assume $T_i$ and $T_j$'s real-time ordering violates the serializable schedule, i.e. $\textsc{complete}_j < \textsc{start}_i$.

Based on \sysname's protocol design:

(1) $T_i$ can be dequeued from $L_s$'s queue only after timestamp agreement succeeds and $L_s$ confirms all leaders have used the same timestamp for $T_i$. Besides, on each leader, $T_i$'s optimistic execution happens before timestamp agreement, so we have $\forall s_1,s_2\in\emph{ShardSet}_i : \textsc{exec}_{i}^{s_1} < \textsc{deq}_{i}^{s_2}$

(2) On any leader $L_s$, transactions can only be executed when there is no conflicting transactions ahead of it in the queue. Consider two consecutive transactions $T_i\rightarrow T_{i+1}$, and $s\in\emph{ShardSet}_i \cap \emph{ShardSet}_j$,  since $T_i.t_{agreed}<T_{i+1}.t_{agreed}$ , $T_{i+1}$ can only be executed after $T_i$ has dequeued, then we have $\textsc{exec}_{i}^{s}<\textsc{deq}_{i}^{s}<\textsc{exec}_{i+1}^{s}$.

Since each two consecutive transactions in the schedule conflict on some data items, there are at least 1 common shard shared by the two transactions, we use $s_{i, i+1}$ to represent any common shard that is shared by $T_i$ and $T_{i+1}$, so we have 

$s_{i,i+1}\in \emph{ShardSet}_i \cap \emph{ShardSet}_{i+1}$,\\
$s_{i+1,i+2}\in \emph{ShardSet}_{i+1} \cap \emph{ShardSet}_{i+2}$,\\
$\cdots$,\\
$s_{j-1,j}\in \emph{ShardSet}_{j-1} \cap \emph{ShardSet}_{j}$

Combing (1) and (2), we have $\textsc{exec}_{i}^{s_{i,i+1}} < \textsc{deq}_{i}^{s_{i,i+1}} < \textsc{exec}_{i+1}^{s_{i, i+1}} < \textsc{deq}_{i+1}^{s_{i+1,i+2}} < \textsc{exec}_{i+2}^{s_{i+1,i+2}}<\cdots <\textsc{exec}_{j}^{s_{j-1,j}} $

However, since we assume $\textsc{complete}_j < \textsc{start}_i$, then we have $\textsc{exec}_{j}^{s_{j-1,j}}<\textsc{complete}_j<\textsc{start}_i <\textsc{exec}_{i}^{s_{i,i+1}}$, which contradicts with the derived relation $\textsc{exec}_{i}^{s_{i,i+1}} <\textsc{exec}_{j}^{s_{j-1,j}} $ based on \sysname's design. Therefore, our assumption is not correct, there does not exists such $T_i$ and $T_j$, whose real-time ordering violates the serializable schedule, i.e., the schedule is strictly serializable.

\section{\sysname's Incremental Hash}
\label{appendix:incremental-hash}

\sysname's incremental hash is computed based on the current log list and is included in servers' fast-replies. The coordinator can compare the hash values from different servers' fast-replies. If the hash values are the same, then the coordinator knows the two servers have the same log list. 

The incremental hash is computed as follows. 

First, for every log entry, a hash can be calculated by concating the \emph{client-id}, \emph{txn-id}, and timestamp as a string, and then converting the string into a hash value. In \sysname's implementation, we use \texttt{SHA1} to do this, but the choice of hash function is orthogonal to \sysname's protocol design: \texttt{SHA1} can be replaced by the other alternatives (e.g., \texttt{SHA256}) for stronger collision resistance.

Second, the server aggregates the hashes of the log entries. Initially, the server maintains a zeroed hash value $h$. Since we use \texttt{SHA1} in \sysname's implementation, $h$ is composed of 80 bits and its initial value is 80 bits of 0. For every log entry $e$ appended to the log list, we use the logical operation, exclusive-or $\oplus$, to aggregate its hash $\texttt{SHA1}(e)$ with $h$, i.e., $$h\leftarrow h\ \oplus\ \texttt{SHA1}(e)$$
On the other hand, during log synchronization (\S\ref{sec:tiga-qc-slow}), some log entries might be removed from the log list, which also leads to a hash update. Due to the nature of exclusive-or operation, removing the log entry $e$ from the log list does the same operation as the aggregation, i.e., $h\leftarrow h\ \oplus\ \texttt{SHA1}(e)$.

The benefit of using incremental hash is to avoid recomputing the hash value from scratch every time, because adding or removing a log entry only incurs one exclusive-or operation. 

In \sysname, we further extend the incremental hash approach to support commutativity optimization. First of all, we realize that read-only transactions do not modify the server state. Therefore, the hash does not need to encode read-only transactions. Then, for write transactions, we consider commutativity when updating the hash. Instead of maintaining one single hash value, each server maintains a table of per-key hashes. For every newly appended write transaction, the sever will XOR its hash to update the corresponding per-key hashes in the table according to the transaction's read-set and write-set. For example, if a transaction $T$ is appended to the log list and $T$ needs to access two keys $x$ and $y$, then in the table, both $x$ and $y$'s hashes will aggregate $\texttt{SHA1}(T)$, i.e., $hash_x\leftarrow hash_x \oplus \texttt{SHA1}(T)$  and $hash_y \leftarrow hash_y \oplus \texttt{SHA1}(T)$. Meanwhile, when sending fast-reply for $T$, the server only encodes the $x$'s and $y$'s hashes instead of the hashes for all log entries. Specifically, the server first concat the key with the $hash$ as a string, and then convert the per-key string into a \texttt{SHA1} hash, and finally aggregate them into the final hash, i.e., $$h_T\leftarrow \texttt{SHA1}(\emph{<}x, hash_x \emph{>}) \oplus \texttt{SHA1}(\emph{<}y, hash_y \emph{>})$$

\section{Optimizations in Slow Reply}
\label{appendix:optim-slow-reply}
In \sysname's workflow (\circled{5}-\circled{6} in Figure~\ref{fig:tiga-workflow}), the followers send a slow reply for each entry after the follower's sync-point has surpassed the entry, i.e., the follower has confirmed that its log is consistent with the leader's log up to this entry. Such design cuts down the length of the slow path and allows the transaction to be committed only 1 message delay later if the transaction fails to be committed in the fast path. However, we find two issues in real implementation. 

First, the slow-replies can be redundant and increase the load for coordinators. Even when the transaction is committed in the fast path~\circled{4}, the servers are unaware of that, so they continue to send the unnecessary slow-replies. Processing these slow-replies adds more burden to the coordinators, especially when the coordinator machine is not powerful.

Second, when implementing \sysname, we find many ordinary RPC libraries do not support ``server push''~\cite{server-push} (i.e., the server can  proactively send messages to the client instead of passively replying the the client's request), including the RPC library used by Janus and \sysname.

Therefore, we choose not to let servers proactively send slow-replies for each transaction. Instead, the coordinator should actively ask for slow-replies from followers when the coordinator finds the followers' fast-replies contain different hash values from the leader. Besides, we also notice that slow-replies can be batched, especially when the coordinator maintains many outstanding transactions. Instead of requesting the slow-replies one by one, we let the coordinators periodically (e.g., every \SI{10}{\milli\second} when \sysname works in WAN) inquire each follower's <\emph{g-view}, \emph{l-view}> and sync-points. 

We use an example to explain how the periodic inquiry works. Consider the coordinator multicasts a transaction to a targeting shard but fails to commit it in the fast path because the leader and followers have inconsistent log lists. However, followers will later receive the cross-region synchronization message (\circled{5} in Figure~\ref{fig:tiga-workflow}), and then modify their log lists, and also advance their sync-points to indicate to which point their log lists have become consistent with the \emph{leader}'s log list. In the leader's fast-reply, we let the leader include its \emph{view-id} and the \emph{log-id} assigned to this entry (transaction). Through the periodic inquiry, the coordinator finally confirms the transaction is committed in the targeting  shard if (1) At least $f$ followers in this shard have the same view as their leader, and (2) these followers have advanced their sync-points to be larger than the \emph{log-id} in the leader's fast-reply.

\begin{figure*}[!t]
    \centering
    \includegraphics[width=0.75\textwidth]{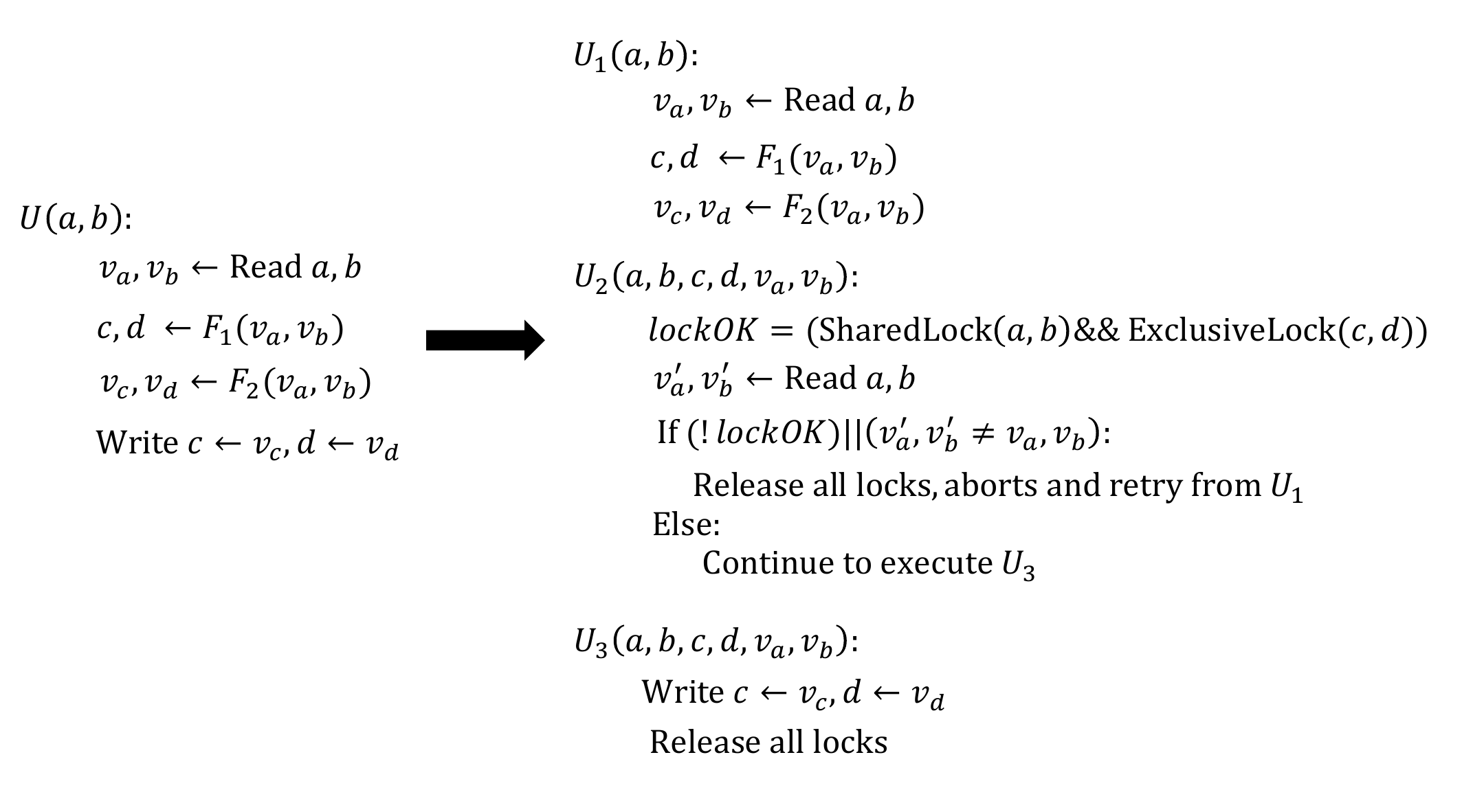}
    \caption{Example of decomposing dependent transaction}
    \label{fig:deptxn-example}
\end{figure*}

\section{Decomposing Dependent Transactions}
\label{appendix:decomposition}

\subsection{Common Existence of One-Shot Transactions}
\label{appendix:bench}
\sysname targets improving the performance for OLTP systems to process one-shot transactions. One-shot transaction is a constrained category of transactions, but commonly exists in many real-world workloads. 
We have examined 12 benchmarks widely used and publicized in literature. Table~\ref{table:common-bench} summarizes the characteristics of these benchmarks. We find that almost all the transactions included in these workloads are one-shot transactions, which verifies the common existence of one-shot transactions in practice.

\begin{table*}[!t]
\caption{Common Benchmarks}
\label{table:common-bench}
\begin{threeparttable}
\begin{center}
\renewcommand{\arraystretch}{1.2}
\begin{tabular}{ccccccc}
\hline
    Benchmark & Domain & Tables & Columns & Types of Txns & Read Ratio & One-Shot?  \\
    \hline
    AuctionMark~\cite{auctionMark-bench} & Online auctions & 16 & 125 & 9 & 55.0\% & Yes \\
    JPAB~\cite{jpab-bench} &  Java performance bench & 7 & 68 & 4 & 25.0\% & Yes\\
    LinkBench~\cite{sigmod13-linkbench} &  Social network & 3 & 7 & 10 & 69.05\% & Yes\\ 
    SEATS~\cite{Seats} & Online airline ticketing & 10 & 189 & 6 & 45.0\% & Yes\\ 
    SIBench~\cite{tods-sibench} & Snapshot isolation  & 1 & 2 & 2 & 50.0\% & Yes \\ 
    SmallBank~\cite{smallbank} & Banking system & 3 & 6 & 6 & 15.0\% & Yes \\ 
    TATP~\cite{tatp-bench}  & Caller location system & 4 & 51 & 7 & 40.0\% & Yes \\ 
    TPC-C~\cite{tpcc} & data warehouse & 9 & 92 & 5 & 8\% & Yes \\ 
    Twitter~\cite{vldb13-oltpbench} & Social network & 5 & 18 & 5 & 0.9\% & Yes \\ 
    Voter~\cite{voter} & Phone-based electon system & 3 & 9 & 1 & 92.2\% & Yes\\ 
    YCSB~\cite{ycsb-workload} & NoSQL benchmark & 1 & 11 & 6 & 50.0\% & Yes \\ 
    TAOBench~\cite{vldb22-taobench} & Social network & 2 & 5 & 8 & 99.8\% & >99.99\%~\tnote{\color{green!80!black} 1} \\
    \hline \\

\end{tabular}
    \begin{tablenotes}
      \small
      \item[1] In TAOBench, only the \texttt{edge\_add} transaction are not one-shot because it needs check the eligibility before insertion. However, such transactions only occupy <0.01\% according to \texttt{workload\_o.json}, and such transactions can be also processed by \sysname using the decomposition technique described in \S\ref{appendix:decomposition}.
    \end{tablenotes}
    
\end{center}
\end{threeparttable}
\end{table*}

\subsection{Example of Decomposing Transactions}
\label{appendix:example-decomposition}
% \begin{figure}[!htbp]
%     \centering
%     \includegraphics[width=0.24\textwidth]{figs/deptxn-example.pdf}
%     \caption{Example of dependent transaction}
%     \label{fig:deptxn-example}
% \end{figure}

% \begin{figure}[!t]
%     \centering
%     \includegraphics[width=0.4\textwidth]{figs/decomposition-deptxn.pdf}
%     \caption{Decomposing dependent transaction}
%     \label{fig:decomposition-deptxn}
%     % \vspace{-0.4cm}
% \end{figure}

Although \sysname targets one-shot transactions, it can also support handling dependent transactions by using the decomposition technique. Figure~\ref{fig:deptxn-example} (left side) illustrates a typical dependent transaction. We use $a$, $b$, $c$, $d$ to represent different keys distributed across shards, and use $v_a$, $v_b$, $v_c$, $v_d$ to represent their values.

% ~\footnote{Here we only consider the first-order dependent transactions because (1) the higher-order dependent transactions rarely appear in the real-world workloads~\cite{vldb10-txn-determinisim}, and (2) our solution can be further extended to handle higher-order dependent transactions.}

Different from one-shot transactions, the read set (i.e., $a$ and $b$) and write set (i.e., $c$ and $d$) of the transaction are not completely known in advance. In this example, the write set (the keys and/or the values) depends on the values of the read set and some functions ($F_1$ and $F_2$). Therefore, this transaction $U(a, b)$ cannot be executed in one shot. In order for \sysname to execute such transactions in strict serializability, we decompose the transaction into three one-shot transactions, i.e., $U_1$, $U_2$, $U_3$, shown on the right side of Figure~\ref{fig:deptxn-example}. 

% \begin{equation*}
% \begin{split}
% U_1&(a, b): \\
%     & v_a, v_b \leftarrow\texttt{Read } a, b\\
%     &  c, d = \texttt{F}(v_a, v_b) \\\\
% U_2&(a, b, c, d, v_a, v_b):\\
%     & lockOK = (\texttt{SharedLock}(a,b) \ \&\&\  \texttt{ExclusiveLock}(c,d))\\
%     &  v_a', v_b' \leftarrow\texttt{Read } a, b\\
%     & \texttt{If } (!lockOK) \ || \  v_a', v_b' \neq  v_a, v_b \\
%     & \texttt{\qquad Release all locks} \\
%     & \texttt{\qquad Return failure} \\
%     & \texttt{Else } \\
%     & \texttt{\qquad Return sucess} \\
% &\slash*\texttt{ If $U_2$ succeeds, then execute $U_3$;}\\ 
% &\texttt{otherwise, abort and retry $U_1$} *\slash \\\\
% U_3&(a, b, c, d, v_c, v_d):\\
%     & \texttt{Write } c\leftarrow v_c, d\leftarrow v_d \\
%     & \texttt{Release all locks} \\
% \end{split}
% \end{equation*}

The decomposition technique enables \sysname to handle dependent transactions because each of the three transactions, $U_1$, $U_2$ and $U_3$, are one-shot transactions. However, the decomposition inevitably introduces aborts: If $U_2$ fails because some other transactions come in between $U_1$ and $U_2$, causing $U_2$'s lock failure or dirty read (i.e., $v_a$ and $v_b$), then the transaction will be aborted and we need to retry from $U_1$ after releasing all locks. Fortunately, as previous works~\cite{vldb10-txn-determinisim,osdi16-janus} observed, real-life OLTP workloads rarely involve key-dependencies on frequently updated data, leading to very few occurrence of such aborts (e.g., $U_2$ fails).

\balance

% \newpage
% \input{coverletter}
\end{document}